\documentclass[aps,pra,twocolumn,notitlepage,superscriptaddress]{revtex4-1}

\usepackage{graphicx,color}
\usepackage{dcolumn}
\usepackage{bm}
\usepackage{amsmath,amssymb,mathrsfs}
\usepackage{url}
\usepackage{hyperref}

\newcommand{\Choi}{\mathsf{Choi}}
\newcommand{\NS}{{\mathsf{NoSig}}}
\newcommand{\Bi}{{\mathsf{Bi}}}

\newcommand{\R}{\mathbb{R}}
\newcommand{\C}{\mathbb{C}}

\newcommand{\set}[1]{\mathsf{#1}}
\newcommand{\grp}[1]{\mathsf{#1}}
\newcommand{\spc}[1]{\mathcal{#1}}

\def\d{{\rm d}}

\def\>{\rangle}
\def\<{\langle}
\def\kk{\>\!\>}
\def\bb{\<\!\<}
\newcommand{\st}[1]{\mathbf{#1}}
\newcommand{\bs}[1]{\boldsymbol{#1}}

\newcommand{\map}[1]{\mathcal{#1}}
\newcommand{\Tr}{\operatorname{Tr}}

\newcommand{\op}[1]{\operatorname{#1}}

\newcommand{\Chan}{{\mathsf{Chan}}}
\newcommand{\BiChan}{{\mathsf{BiChan}}}
\newcommand{\SpanBiChan}{{\mathsf{SpanBiChan}}}
\newcommand{\Map}{{\mathsf{Map}}}

\newtheorem{theo}{Theorem}

\newtheorem{lemma}{Lemma}
\newtheorem{prop}{Proposition}
\newtheorem{cor}{Corollary}
\newtheorem{defi}{Definition}

\def\Proof{{\bf Proof.~}}
\def\qed{$\blacksquare$ \newline}

\usepackage{qcircuit}

\begin{document}
    \title{ { Quantum operations with  indefinite  time direction}}

    \author{Giulio Chiribella}
    \affiliation{QICI Quantum Information and Computation Initiative, Department of Computer Science, The University of Hong Kong, Pokfulam Road, Hong Kong}
    \affiliation{Department of Computer Science, University of Oxford, Wolfson Building, Parks Road, Oxford, UK}
    \affiliation{Perimeter Institute for Theoretical Physics, 31 Caroline Street North, Waterloo,  Ontario, Canada}
    \author{Zixuan Liu}
    \affiliation{QICI Quantum Information and Computation Initiative, Department of Computer Science, The University of Hong Kong, Pokfulam Road, Hong Kong}
    
\begin{abstract}
The fundamental dynamics of quantum particles is neutral with respect to the arrow of time. And yet, our experiments are not: we observe quantum systems evolving from the past to the future, but not the other way round.    A fundamental question is  whether it is possible to conceive  a broader set of operations that probe quantum processes in the backward direction, from the future to the past, or more generally,  in a combination of the forward and    backward directions.    Here we introduce a mathematical framework for  these operations, showing that some of them 
cannot be interpreted as  random mixtures of operations that probe  processes  in a definite direction.   
As a concrete example, we construct an operation, called the quantum time flip, 
that  probes an unknown dynamics in a quantum superposition of the forward and backward directions. 
This operation exhibits  an information-theoretic  advantage over  all operations with 
 definite direction.  
 It  can realised probabilistically using quantum teleportation, and can be  reproduced experimentally with photonic systems.    More generally,  we introduce a set of multipartite operations that  include indefinite  time direction as well as indefinite causal order,  providing a framework for potential extensions of quantum theory.   
 
     \end{abstract}

  \maketitle

The experience of time flowing in a definite direction, from the past to the future, is deeply rooted in our thinking.     At the microscopic level, however, the laws of Nature seems to be indifferent  to the distinction between past and future.   Both in classical and quantum mechanics, the fundamental equations of motion are reversible, and  changing the sign of the time coordinate  (possibly together with the sign of some other parameters) still yields a valid dynamics. For example,   
the CPT theorem  in quantum field theory \cite{luders1954equivalence,pauli1955niels} implies that an evolution backwards in time is indistinguishable from  an evolution forward in time in which  the  charge and parity of all particles have been inverted.   
An asymmetry between past and future emerges in  thermodynamics, where the second law postulates an increase of entropy in the forward time direction.  But even the time-asymmetry of thermodynamics can be reduced to  time-symmetric laws at the microscopic level \cite{halliwell1996physical}, {\em e.g.}
 by postulating a low entropy initial state \cite{wald2006arrow}.

While the  microscopic world  
is time-symmetric, the way in which we interact with it 
  is not.  As a matter of fact, we operate only in the forward time direction:  in ordinary experiments, we initialise  physicals system at a given moment,  let them evolve forward in time, and perform measurements at a later moment. 
 Still, this asymmetry in the structure of our experiments does not feature in the dynamical laws  themselves. 
  This fact suggests that, rather than being fundamental, time asymmetry  may be   specific to the  way  in which  ordinary agents, such as ourselves, interact with other physical systems \cite{maccone2009quantum,rovelli2017time,di2020quantum,hardy2021time}.

 An  intriguing possibility is that, at least in principle,   some other type of agent could  perform experiments 
 in the opposite direction, by initialising the state of physical systems in the future, and by observing their evolution backward in time.   This possibility is implicit in a variety of frameworks wherein  pre-selected and post-selected quantum states are treated on the same footing \cite{aharonov1964time,aharonov1990superpositions,aharonov2002two,abramsky2004categorical,hardy2007towards,oeckl2008general,svetlichny2011time,lloyd2011closed,genkina2012optimal,oreshkov2015operational,silva2017connecting}.      Building on these frameworks,  one  can  even conceive agents with the ability to deterministically pre-select certain systems and to deterministically post-select others, thus observing physical processes in an arbitrary combination of the forward and backward direction.    
 Such  agents     may or may not exist in reality, but can  serve as a useful fiction to  shed light  on 
 the operational  significance of  the constraint of a fixed time direction, by contrasting  the information-theoretic capabilities  associated to  alternative ways to operate in time.  

 Here  we establish a  mathematical framework for  operations that use quantum devices in arbitrary combinations of the forward and backward direction.
We first characterise  the set of   bidirectional quantum processes, that is, processes that could in principle be accessed in both directions. 
    We  then construct a set of operations that use bidirectional processes, and we show that some of these operations cannot be obtained as random mixtures of operations that probe the processes of interest in  a definite direction.  As a concrete example, we introduce an operation, called the quantum time flip, that uses processes in a coherent superposition of the forward and backward directions.  The potential of the quantum time flip is illustrated by a game where  a referee challenges a player to discover a hidden  relation between two black boxes implementing two unknown unitary gates.  As it turns out, a player with the ability to query the boxes in a coherent superposition of directions  can identify the correct relation with no error, while every player who can only access  the two boxes in a definite time  direction  will  have an error probability  of at least 11\%, even if the player is able to combine the two boxes in an indefinite order   \cite{chiribella2009beyond,oreshkov2012quantum,chiribella2013quantum}. 

{ Our work initiates  the exploration 
 of a new type of quantum operations that are not constrained to a single time direction, and provides a rigorous framework for analysing   their information-theoretic power.} It also   allows for multipartite operations where both the time direction and the causal order are indefinite,  and  rises the open question whether these operations are physically accessible in new regimes, such as quantum gravity, or whether  they are prevented by some yet-to-be-discovered 
  mechanism.

 \section{Results}

 {\bf Bidirectional devices  and their characterisation.}   We start by identifying  the largest set of quantum devices that  are in principle compatible with two alternative modes of operation: either in the forward time direction, or in the backward time direction.   
 
 \begin{figure}
	\includegraphics[width=0.4\textwidth]{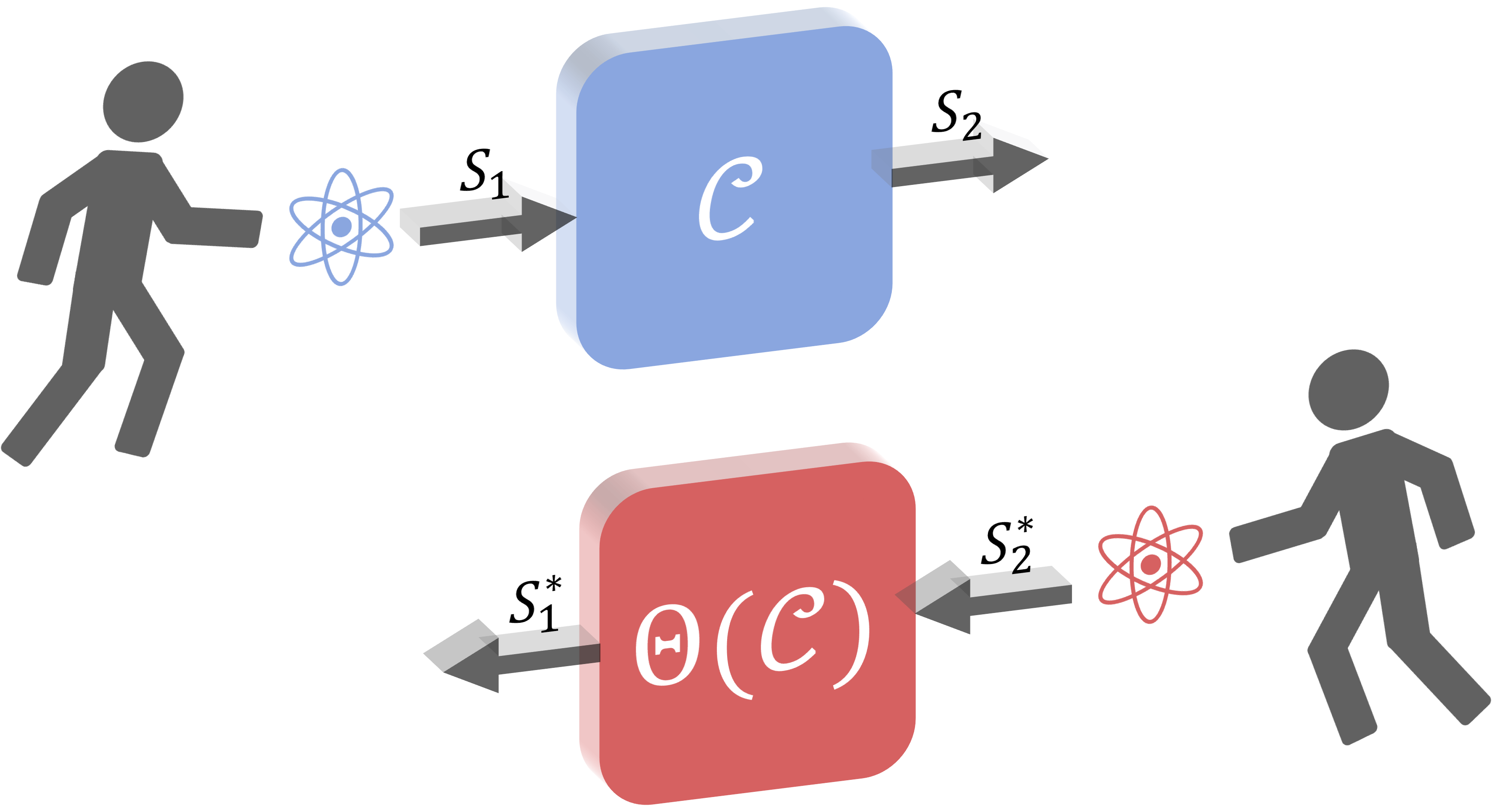}
	\caption{\label{fig:agents} {\bf Bidirectional devices.}  A bidirectional device is in principle compatible with two alternative modes of operation. In the forward mode (top), an agent prepares an input system at time $t_1$ and obtains an  output system at time $t_2\ge t_1$. In the backward mode (bottom),   a hypothetical agent could prepare an input at time $t_2$ and obtain an  output at time $t_1$. These two modes of using the device correspond to two different input-output transformations $\map C$ and $\Theta (\map C)$, respectively.}
\end{figure}

     Consider  a process that takes place   between two times $t_1$ and $t_2\ge t_1$, corresponding to two events, such as the entry of a system into a Stern-Gerlach apparatus, and its exit from the same apparatus.   Ordinary agents can  interact with the process in the forward time direction: they can   deterministically pre-select state of an incoming system   $S_1$  at time $t_1$,  and later they can measure an outgoing system  $S_2$ at time $t_2$.    The overall input-output transformation from time $t_1$ to time $t_2$ is described by a quantum channel  $\map C$, that is, a trace-preserving, completely positive (CPTP) map transforming density matrices of system $S_1$ into density matrices of system $S_2$ \cite{heinosaari2011mathematical}.    Now,  imagine a hypothetical  agent that operates in the opposite time direction, by deterministically post-selecting the system at time $t_2$ and performing measurements  at time $t_1$, as illustrated in Figure \ref{fig:agents}.     For such a backward-facing agent, the role of the input and output systems is exchanged, and the two systems at time $t_1$ and $t_2$ may even appear to be different from $S_1$ and $S_2$, e.g. they may have opposite charge and opposite parity.  In the following  we denote the systems observed by the backward-facing agent as $S_1^*$ and $S_2^*$, and we only assume that they have the same dimensions of $S_1$ and $S_2$, respectively. 
  If the overall input-output transformation observed by the backward-facing agent is still described by a  valid quantum channel (CPTP map), we call  the process  {\em bidirectional}.


 
 To determine whether a given process is bidirectional, one has to specify  a map $\Theta$, converting  the channel   $\map C$ observed by the forward-facing agent into the corresponding channel $\Theta (\map C)$ observed by the backward-facing agent.   We call the map  $\Theta$ an {\em input-output inversion}.   
  The set of bidirectional processes is then defined as  the set of  all quantum channels    $\map C$ with the property that also  $\Theta (\map C)$ is a quantum channel.    In the following, the  set of bidirectional channels will be denoted by $\sf B  (S_1\to S_2)$. 
    
We  now characterise all the possible input-output inversions satisfying four natural requirements. Specifically, we require that  the map $\Theta$ be 
\begin{enumerate}
       \item {\em order-reversing:}  $\Theta  (\map D \,  \map C  )    =   \Theta  (\map C) \,   \Theta  (\map  D)$  for every pair of channels $\map C  \in   \set B  (S_1  \to S_2)$ and $\map D  \in \set B  (S_2\to S_3)$,           \item  {\em identity-preserving}:  $\Theta(\map I_S)  =  \map I_{S^*}$, where $\map I_S$  ($S_*$) is the identity channel on system $S$ ($S^*$).    
       \item   { {\em distinctness-preserving:}  
         if  $\map C  \not  =   \map D$, then  $\Theta (\map C)  \not  = \Theta (\map D)$, }
           \item {\em compatible with random mixtures}:    $\Theta  \left(  p\, \map     C  +    (1-p) \, \map  D    \right)   = p\, \Theta (\map C)  + (1-p)\,  \Theta (\map D)$ for every pair of channels $\map C$ and $\map D$ in $ \set B  (S_1  \to S_2)$,  and for every probability $p\in  [0,1]$.
            \end{enumerate}

\begin{figure}
	\includegraphics[width=0.45\textwidth]{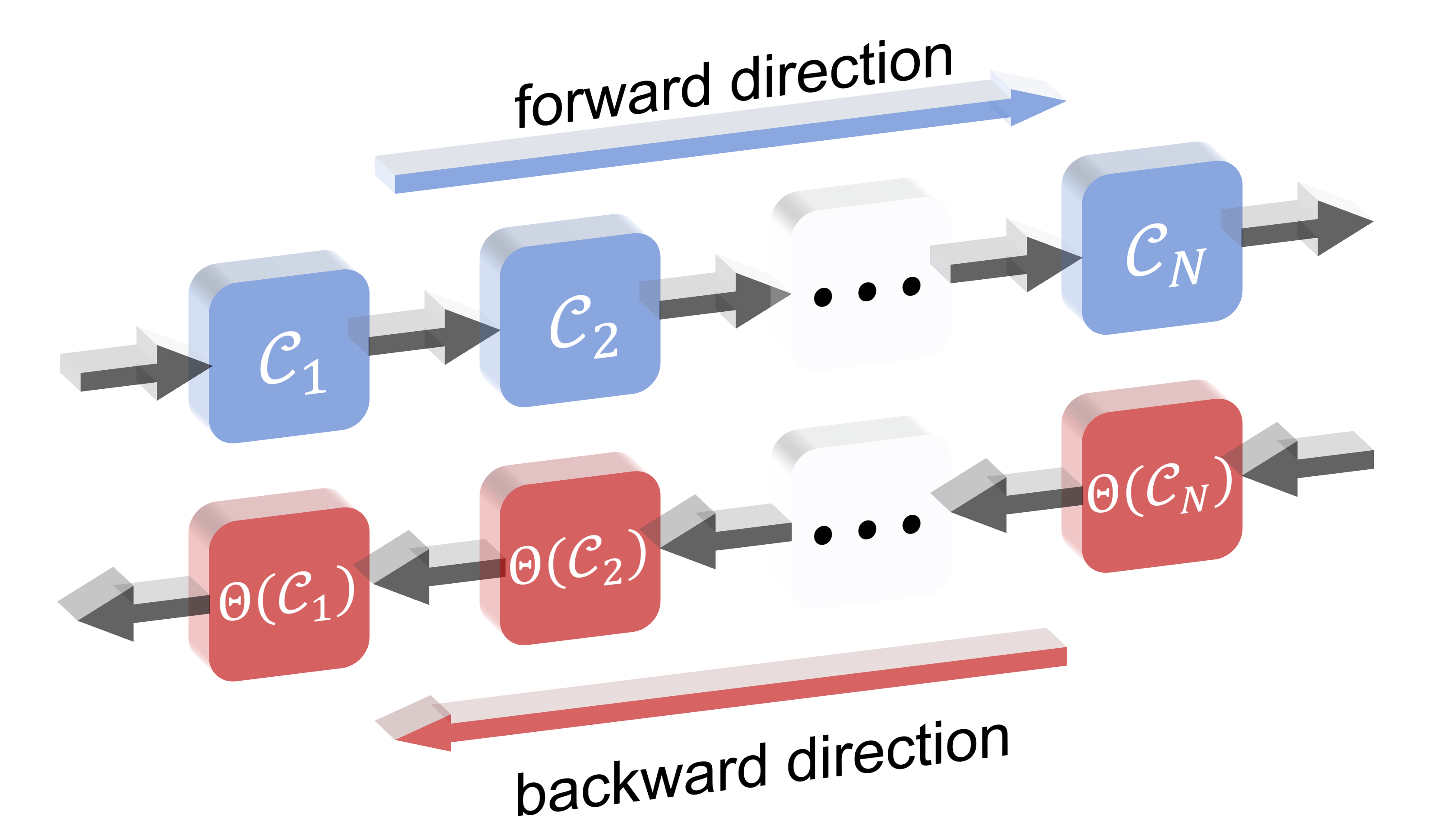}
	\caption{\label{fig:orderreversing} {\bf The order-reversing condition.}  If a system experiences a sequence of processes $\map C_1,  \dots, \map C_N$ in the forward-time representation (in blue), then
	the system should experience the opposite sequence $\Theta (  \map C_N),  \dots,  \Theta (\map C_1)$  in the backward-time representation (in red).}
\end{figure}

Requirement 1, illustrated in Figure \ref{fig:orderreversing}, is  the most fundamental:  for every sequence of processes, the order in which a backward-facing agent sees the processes  should be  the opposite of the order in which a forward-facing agent sees them.  Requirement 2 is also quite fundamental:  if the forward-facing agent  does not see any change in the system,   then also the backward-facing agent should not see any change.     
 Requirement 3  is a weak form of symmetry: processes that appear distinct to a forward-facing agent should appear distinct also to a backward-facing agent.  A stronger requirement would have been to require that applying  $\Theta$ twice   should bring every process back to itself.  This condition is stronger than our Requirement 3, because it implies not only that  $\Theta$  must be invertible, but also that $\Theta$  is its own inverse.    Finally, Requirement 4  is that if   a process has probability $p$ to be $\map C$ and probability $1-p$ to be $\map D$ for the forward-facing agent, then, for the backward-facing agent the process will have probability   $p$ to be $\Theta (\map C)$ and probability $1-p$ to be $\Theta(\map D)$.

 Our notion of input-output inversion is closely related with the notion of time-reversal in quantum mechanics \cite{wigner1959group,messiah1965quantum}  and in  quantum thermodynamics \cite{campisi2011colloquium}.  It is worth stressing, however, that  input-output inversion is more general than time-reversal, because it can include combinations of time-reversal with other symmetries, such as charge conjugation and parity inversion (see Appendix \ref{app:unitary} for more discussion).    Moreover,  the input-output inversion can also describe situations that do not involve time-reversal, including, for example, the use of optical devices where the roles of the input and  output modes can be exchanged, as discussed later in the paper.

 In the following, we will focus on the scenario where the systems $S_1$ and $S_2$ have the same dimension.   We will assume that all unitary dynamics are bidirectional, that is, that  the set   $\sf B (S_1\to S_2)$ contains all possible unitary channels.    For unitary channels, Requirements 1-3 completely determine the action of the input-output inversion. 
   Specifically,  we show that the input-output inversion must either  be unitarily equivalent to the adjoint $\theta (U):  =  U^\dag$, or to  the transpose $\theta  (U):=  U^T$ (Appendix \ref{app:unitary}).

 For general quantum channels, we show that the set of bidirectional processes coincides with  the set of  {\em bistochastic channels}  \cite{landau1993Birkhoff,mendl2009unital}, that is, channels $\map C$ with a Kraus representation $\map C  (\rho)    =  \sum_i   C_i \rho  C_i^\dag$ satisfying both conditions      $\sum_i   C_i^\dag C_i  =  I_{S_1}$ and $\sum_i C_iC_i^\dag =I_{S_2}$ (see Methods).   
Also in this case  we find  that, up to unitary equivalence, 
 there exist only two possible choices of input-output inversion:  the adjoint  $\map C^\dag$, defined by $\map C^\dag (\rho)   :=  \sum_{ i}  C_i^\dag  \rho  C_i$, and the transpose $\map C^T$, defined by $\map C^T (\rho)   =  \sum_{ i}  C_i^T  \rho  \overline C_i$, with $\overline C_i  :  = (C_i^T)^\dag  $.

For two-dimensional quantum systems  the adjoint and transpose are unitarily equivalent,
 and therefore  the input-output inversion  is essentially unique. For higher dimensional systems, however,  the adjoint and the transpose  exhibit a fundamental  difference:  unlike the transpose, the adjoint   does not generally produce quantum channels (CPTP maps) when applied locally to to the dynamics of bipartite quantum systems (see Methods).   Technically, the difference is that the adjoint is not a completely positive map on quantum channels. In the terminology of  Refs. \cite{chiribella2008transforming,chiribella2009theoretical,chiribella2013quantum,bisio2019theoretical}, the adjoint is not an admissible supermap on quantum channels.   


\medskip

{\bf Quantum operations with indefinite time direction.}
The standard operational framework of quantum theory describes  sequences of operations performed in the forward  time direction.  We now define a more general type of operations, which  use quantum devices in arbitrary  combinations of the forward and backward direction. 
  { Our framework is based 
  on
   the framework of quantum supermaps \cite{chiribella2008transforming,chiribella2009theoretical,chiribella2013quantum,bisio2019theoretical}, a mathematical framework  describes candidate  operations that could in principle be performed on a set of quantum devices.    In general, a quantum supermap from an input set of quantum channels $\set B$ to an output set of quantum channels $\set B'$ is a  map that preserves convex combinations, and can act locally on the dynamics of  composite systems, transforming any extension of a channel in $\set B$ into an extension of a channel in $\set B'$~\cite{chiribella2013quantum}. }
  
    The possible operations on bidirectional devices correspond to   quantum supermaps transforming  bistochastic channels into ordinary channels (CPTP maps).     Some of these supermaps use the devices in the forward direction:   they are  of the form $\map S_{\rm fwd} (\map C)  = \map B (\map C\otimes \map I_{\rm aux}) \map A$, where $\map C$ is the bistochastic channel describing the device of interest, and   $\map A$ and $\map B$ are two fixed channels, possibly involving an auxiliary system $\rm aux$ \cite{chiribella2008transforming}.   Other  supermaps  could be realised by using the device is  the backward direction: they are of the form  $\map S_{\rm bwd} (\map C)  = \map B' ( \Theta (\map C)\otimes \map I_{\rm aux'}) \map A'$, where $\map A'$ and $\map B'$ are two fixed channels and $\Theta$ is (unitarily equivalent to) the transpose.

  A complete characterization of the possible supermaps acting on bistochastic channels is provided in Appendix \ref{app:bidirectional}.   As we will see in the following, the set of these supermaps contains   operations that are neither of the forward type nor of the backward type, nor of any random mixture of these two types.  We call these transformations {\em quantum operations with indefinite time direction}.     
These operations are the analogue for the time direction of the operations with  indefinite causal order~\cite{chiribella2009beyond,oreshkov2012quantum,chiribella2013quantum}, also known as causally inseparable operations \cite{oreshkov2012quantum,araujo2015witnessing,oreshkov2016causal}. 

 In Appendix \ref{app:multipartite},  we extend our construction from operations on a single bistochastic channel to more general multipartite operations,   described by  quantum supermaps $\map S$ that transform a list of bistochastic channels $(\map C_1, \map C_2, \dots,  \map C_N)$ into an ordinary channel $\map S  (\map C_1, \map C_2, \dots,  \map C_N)$. 
  This general type of supermaps can exhibit both indefinite  time direction and indefinite causal order, and provide a broad framework for potential extensions of quantum theory.   
  
\medskip

{\bf The quantum time flip.}  {We now introduce a concrete example of operation with indefinite time direction, called the quantum time flip. This operation is  a analogue of  the quantum SWITCH~\cite{chiribella2009beyond,chiribella2013quantum}, previously introduced in the study of  indefinite causal order.}   
 The quantum time flip takes in input a bidirectional device, and produces as output  a controlled channel \cite{aharonov1990superpositions,oi2003interference,chiribella2019quantum,abbott2020communication,dong2019controlled}, which acts as $\map C$ if a  control qubit is initialised in the state $|0\>$, and as $\Theta  (\map C)$ if the control qubit is initialised in the  state $|1\>$.    For a fixed set of Kraus operators ${\bf C}   =  \{ C_i \}$,  we consider the controlled channel 
   $\map F_{\bf C}$  of the form   $\map F_{\bf  C}      (\rho)=\sum_i    F_{i}  \rho  F_{i}^\dag $, with
  \begin{align}\label{timeflip}
  F_{i} :=  C_i  \otimes |0\>\<0|  +    \theta (C_i)  \otimes   |1\>\<1|  \,  \, ,
\end{align}
where  the map $\theta:  C_i \mapsto \theta (C_i)$ is either unitarily equivalent to the adjoint  or to the transpose.   In passing, we observe that the channel $\map F_{\bf C}$ is itself bistochastic,
 and therefore  it also admits an input-output inversion.  

It is worth stressing that {\em (i)} $\map F_{\bf C}$ is a valid quantum channel (CPTP map) if and only if the input channel $\map C$ is bistochastic, and {\em (ii)}  the definition of  $\map F_{\bf C}$ is independent of the Kraus representation if and only if the map $\theta$  is unitarily equivalent to the transpose (Appendix \ref{app:supermap}).  When these two conditions are satisfied,  we show that the map $\map F:  \map C \to \map F_{\bf C}$  satisfies all the requirements of a valid quantum supermap. We call this supermap the {\em quantum time flip} and we will write the controlled channel as $\map F (\map C)$.

{ The quantum time flip is an example of an operation with indefinite time direction:  
   it is impossible to decompose it as a random mixture $ \map F  =  p \,  \map S_{\rm fwd}  +  (1-p)  \,  \map S_{\rm bwd}$ where $p$ is a probability, and $\map S_{\rm fwd}$ ($\map S_{\rm bwd}$) is a forward (backward) supermap.       In Appendix \ref{app:incompatible} we show that, if such decomposition existed,  then there would exist an ordinary quantum circuit that  transforms a completely unknown unitary gate $U$ into its transpose $U^T$, a task that  is known to be impossible \cite{chiribella2016optimal,quintino2019probabilistic}. 
  We also show that the quantum time flip cannot be realised in a definite time direction even if one has access to two copies of the original channel $\map C$.  Remarkably, this stronger no-go result holds even if the two copies of the channel $\map C$ are combined in an indefinite order: as long as all copies of the channel are used in the same time direction, there is no way to reproduce the action of the quantum time flip.

 }

\medskip

 \begin{figure}
	\includegraphics[width=0.45\textwidth]{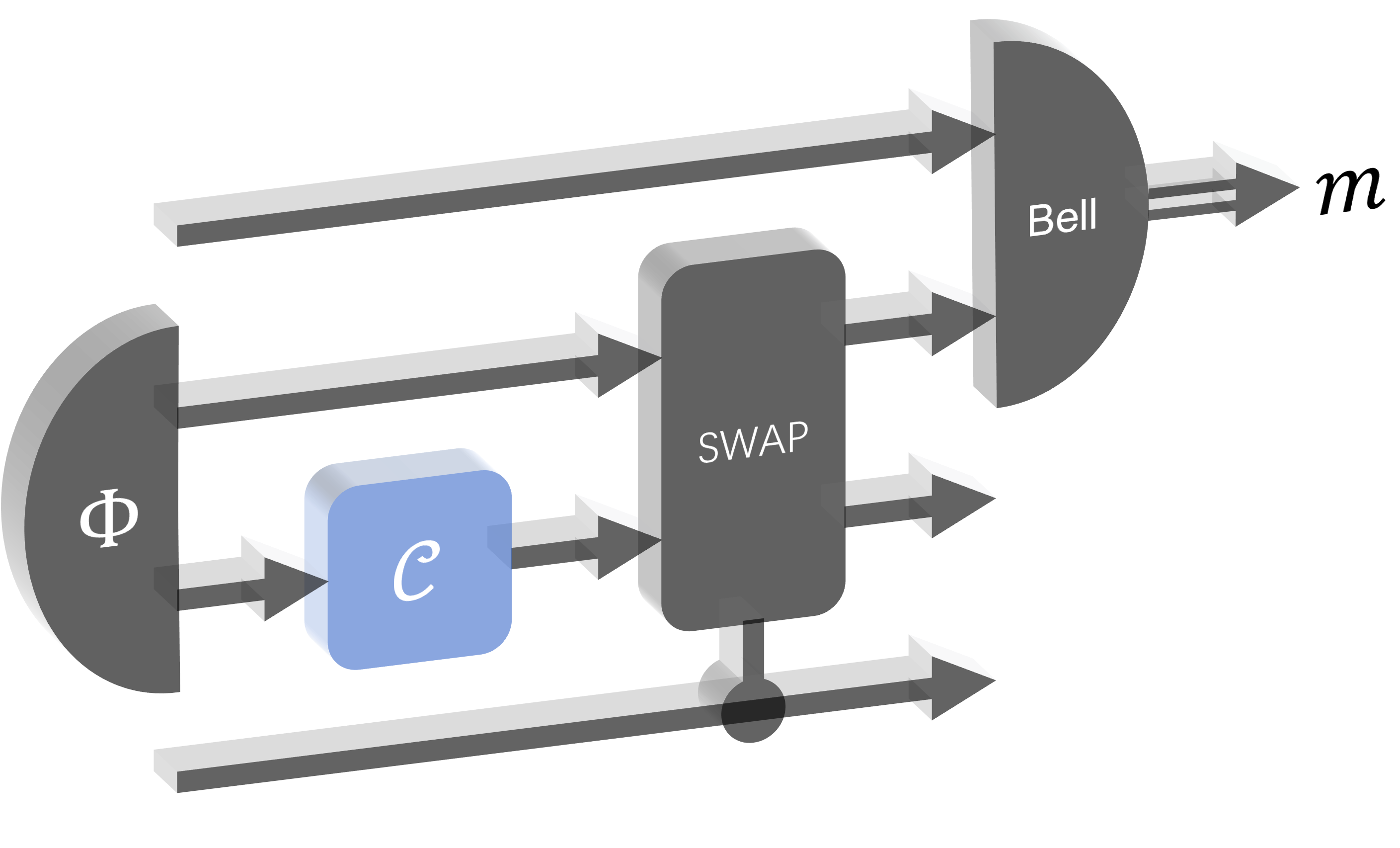}
	\caption{\label{fig:circuit} {\bf Probabilistic realisation  of the quantum time flip.}  An unknown channel $\map C$ is applied locally on a maximally entangled state, which then undergoes a controlled swap operation and is used as a resource for quantum teleportation.    The probabilistic realisation of the  quantum time flip is  heralded by  a specific value of the  outcome $m$ of the Bell measurement in the teleportation protocol.}
\end{figure}

{ {\bf Realisation of the quantum time flip through teleportation.}  }
  We have seen that the quantum time flip cannot be  perfectly realised by any quantum circuit with a definite time direction.  This  no-go result concerns perfect realisations, which reproduce the quantum time flip with unit probability and without error.   On the other hand,  the  quantum time flip  can be realised with non-unit probability   in an ordinary quantum circuit, using  quantum teleportation \cite{bennett1993teleporting}.

The setup is depicted in  Figure \ref{fig:circuit}.   An unknown bistochastic channel  $\map C$  is applied on one side of { a  maximally entangled state,  say the canonical Bell state  $|\Phi\>=  \sum_{i=1}^{d}  \,  |i\>\otimes |i\>/\sqrt{d}$}, and the output is used as a resource for quantum teleportation.    The transpose is realized by swapping the two copies of the  system:     for example, when  the channel $\map C$ is unitary, the application of the channel to the Bell state $|\Phi\>$ yields another maximally entangled state $|\Phi_U\>  :=  (   I\otimes  U) |\Phi\>$, where $U$ is a unitary matrix, and swapping the two entangled systems   produces the state $|\Phi_{U^T}\>$,  where the unitary $U$ is replaced by its transpose $U^T$.    Coherent control of the choice between the forward channel $\map C$ and the backward channel $\Theta (\map C)$  is realized by adding control to the swap.   Finally, a Bell measurement is performed and the outcome corresponding to the projection on the state $|\Phi\>$ is post-selected.  When this outcome occurs, the circuit reproduces  the quantum time-flipped channel  $\map F(\map C)$, as shown in the following in the unitary case.    

  Let us denote  by $|\phi\>_{S}$ the initial state of the target system and by $|\psi\>_{C}=  \alpha \, |0\>_{C}  +  \beta\,  |1\>_{C}$ the initial state of the control qubit. Then,  the joint state of the target and control after the controlled swap  is  $  \alpha\,   |\phi\>_S  \otimes |\Phi_U\>  \otimes |0\>  +  \beta  \,   |\phi\>_S  \otimes |\Phi_{U^T}\>  \otimes |1\>$.  When the  Bell measurement is performed, the target system and the control are collapsed to one of the states  $\alpha   \,   U   U_m   |\phi\>_S  \otimes |0\>_C   + \beta\,    U^T   U_m |\phi\>_S \otimes |1\>_C  $, where $m \in  \{1,\dots, d^2\}$ is the measurement outcome and $\{U_m\}_{m=1}^{d^2}$ are the unitaries associated to the Bell measurement.    For the outcome corresponding to the state $|\Phi\>$,  one obtains   the overall state transformation  $  |\phi\>_S  \otimes |\psi\>_C  \mapsto  \alpha   \,   U  |\phi\>_S  \otimes |0\>_C   + \beta\,    U^T  |\phi\>_S \otimes |1\>_C  $, corresponding to the time-flipped channel $\map F(\map C)$.    More generally, each outcome  of the Bell measurement  gives rise to  a conditional transformation that uses the gate $U$ in an indefinite  time direction.   This fact  is not in contradiction with the definite time direction of the overall setup in  Figure \ref{fig:circuit}:  averaging over all outcomes of the Bell measurement   yields an overall operation that uses the gate $U$ in a well-defined  direction (the forward one).

In the teleportation setup, the quantum time flip is realised probabilistically. 
However,  in principle the quantum time flip {could} also be implemented deterministically and without error by some agent who is not constrained to operate in a well-defined time direction. { For example,    Figure \ref{fig:circuit} shows that an agent with the ability to deterministically pre-select a Bell state, and to deterministically post-select the outcome of a Bell measurement would be able to deterministically achieve the quantum time flip.  Note that not all circuits built from deterministic pre-selections and deterministic post-selections are compatible with quantum theory.  In this respect, the framework of quantum  operations with indefinite time direction  provides a  candidate  criterion for determining  which postselected circuits can be allowed and which ones should be forbidden. 

\medskip

 {\bf An information-theoretic advantage of the quantum time flip.}    We now introduce a game where the quantum time flip offers an advantage over arbitrary setups  with fixed time direction.   The structure of the game is similar to that of another game, previously introduced by one of us to  highlight  the advantages  of the quantum SWITCH~\cite{chiribella2012perfect}. However, the variant introduced here highlights fundamental diffference:  in this variant of the game,   no perfect win can be achieved by the quantum SWITCH, or by any of the processes with indefinite causal order considered so far  in the literature.

 The game involves a referee, who challenges a player to discover a property of two black boxes.  The referee promises that the two black boxes  implement two unitary gates $U$ and $V$   satisfying either the condition  {$UV^T   =  U^T V$, or the condition $UV^T  =  -  U^T  V $.}  
 The goal of the player is to discover which of these two alternatives holds.   
 
A player with access to the quantum time flip can win the game with certainty.   The winning strategy is to apply the quantum time flip to both gates,  exchanging the roles of $|0\>$ and $|1\>$  in the control for gate $V$.  
    In this strategy,  one time flip generates the gate $  S_U  =    U  \otimes |0\>\<0|  + U^T \otimes |1\>\<1|$, while the other generates the gate $ S_V  =   V^T  \otimes |0\>\<0|  +  V \otimes |1\>\<1|$.   The strategy is to prepare the target and control systems in the product state   $|\psi\> \otimes|+\>$, where $|\psi\>$ is arbitrary,  and  $|\pm\>  :=  (|0\>  \pm |1\>)/\sqrt 2$.  Then, the target and control are sent first through the gate $S_V$ and then through the gate $S_U$,  obtaining the state
\begin{align}
 \nonumber S_U S_V  (  |\psi\>\otimes |+\>)   = &  
\left[   \frac{ UV^T   +  U^T  V }2    |\psi\>   \right]  \otimes |+\>\\
& +  
 \left[   \frac{ UV^T   -  U^T  V }2      |\psi\>   \right]  \otimes |-\> \, .  \label{state}
 \end{align}
If $U$ and $V$   satisfy the condition $UV^T   = U^T V $, then the second term in the sum vanishes, and the control qubit ends up in the state $|+\>$. Instead, if the gates satisfy the condition $UV^T   = -U^T V$, then the first term vanishes, and the control qubit ends up in the state $|-\>$.   Hence, the player can measure the control qubit in the  basis $\{|+\>, |-\>\}$, and figure out exactly which condition is satisfied.   

 Overall, the transformation of the gate pair $(U,V)$ into the controlled-gate $S_U S_V$ is an example of a bipartite supermap with indefinite time direction, of the type discussed in Appendix \ref{app:multipartite}.  A player that implements this supermap  can in principle win the game with certainty.

The situation is different for players who can only probe the two unknown gates in a definite time direction.   In  Appendix \ref{app:bound} we show that every such player will have a probability { of at least 11\% to lose the game.  Remarkably, this limitation applies not only to strategies that use the two gates $U$ and $V$ in a fixed order, but also to all strategies where the relative order of $U$ and $V$ is indefinite. 
}  }

\medskip

\begin{figure}
	\includegraphics[width=0.45\textwidth]{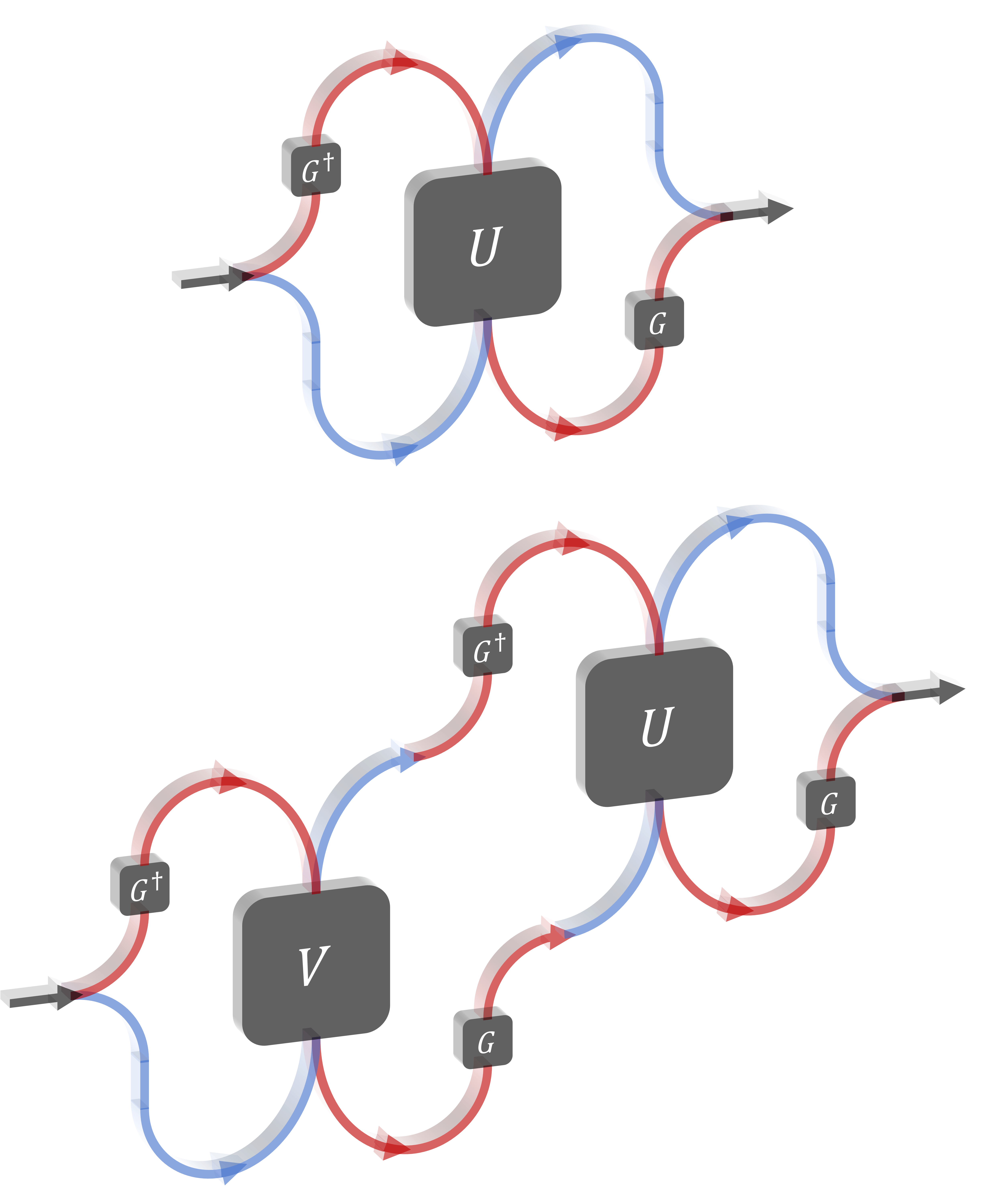}
	\caption{\label{fig:loop} {\bf  Photonic realisation of the superposition of a process and its input-output inverse.}             { Using a beamsplitter,} a single photon is coherently routed  along two paths, one (in blue) traversing an unknown waveplate from top to bottom, and the other (in red) traversing it from bottom to top.    Along one path, the photon polarisation experiences a unitary gate $U$, while on the other path it experiences the transpose gate $U^T$, up to a change of basis $G$ that is undone by placing suitable polarisation rotations before and after the waveplate.
	 The two paths are finally recombined in order to allow for an interferometric measurement on the control qubit  (top image).   
    By concatenating two setups with the above structure, one can  implement  the winning strategy  in Eq. (\ref{state})   (bottom image).  
   }
\end{figure}

{\bf Photonic realisation of the suporposition of a process and its input-output inverse.}  A coherent superposition of a unitary process  and its input-output inverse  can be  realised with polarisation qubits, using the interferometric setup  illustrated in Figure \ref{fig:loop}.   In this setup, the control qubit is the path of a single photon.  
   A beamsplitter puts the photon in a coherent superposition of two paths, which lead to an unknown polarisation rotator from two opposite spatial directions, respectively.     Along one path, the passage through the polarisation rotator induces an unknown  unitary gate $U$. Along the other path, the passage through the polarisation rotator induces  the unitary gate $G U^T  G^\dag$, where $G$ is a fixed unitary gate depending on the choice of basis used for representing polarisation states (in the standard representation of the Poincar\'e sphere, $G$ is the Pauli matrix $Z=  |0\>\<0| - |1\>\<1|$).  By undoing the unitary gate $G$, one can then obtain a quantum process with coherent control over the gates $U$ and $U^T$, as described by Eq.~(\ref{timeflip}). 


Note that the above realisation is  not in contradiction with our no-go result on the realisation of the quantum time flip in a quantum circuit with a fixed direction of time. The no-go result states that it is impossible to build the controlled unitary gate $U \otimes |0\>\<0|  +  U^T  \otimes |1\>\<1|$ starting from an unknown and uncontrolled gate $U$ as the initial resource.  However, it does not rule out the existence of a device that directly implements the controlled gate $U \otimes |0\>\<0|  +  U^T  \otimes |1\>\<1|$ in the first place. { Such devices do exist in nature, as shown above, and the unitary $U$ appearing in them can be either known or unknown.}   A similar situation arises in the implementation of  other controlled gates, which cannot be constructed from their uncontrolled version \cite{nakayama2014universal,araujo2014quantum,chiribella2016optimal,thompson2018quantum},   but  can be directly realised in various experimental setups  \cite{zhou2011adding,friis2014implementing}.

\section{Discussion}

{ In this work we defined a framework for  quantum operations  with indefinite time direction. 
  This class of operations is broader than the set of operations considered so far in the literature, and in the multipartite case it includes  all  known operations with definite and indefinite causal order.   
  Quantum operations with both indefinite time direction and indefinite causal order provide a  framework for describing  the interactions of an agent with the fundamentally  time-symmetric dynamics of quantum theory, and for composing local processes into more complex structures.  This higher order framework  is expected to contribute to the study of quantum gravity scenarios,  as envisaged by Hardy~\cite{hardy2007towards}. These applications, however, are beyond the scope of the present paper, and remain as a direction for future research.      }

The characterization of the bidirectional quantum channels provided in this paper  reveals an interesting connection with  thermodynamics.  We showed that the the set of bidirectional quantum processes coincides with the set of  bistochastic channels. On the other hand, bistochastic channels   can also be characterised  as  the largest set of entropy non-decreasing processes: any entropy non-decreasing process must transform the maximally mixed state into itself, and therefore be bistochastic; {\em vice-versa}, every bistochastic channel is entropy non-decreasing  \cite{gour2015resource}.  Combining these two characterizations, we conclude that the processes admitting a time-reversal are exactly those that are compatible with the non-decrease of entropy 
both in the forward and in the backward time direction.  
This conclusion is  remarkable, because no entropic  consideration was included in the derivation of our results. A promising direction for future research is to  further investigate the role of input-output inversion in the search of axiomatic principles  for   quantum thermodynamics \cite{chiribella2017microcanonical,krumm2017thermodynamics}.  

Finally, another interesting direction is to explore generalisations of quantum thermodynamics to the scenario where  agents are not constrained to operate in a definite time direction.  {A first  step in this direction has been recently taken by  Rubino, Manzano, and Brukner \cite{rubino2020time},  who explored thermal machines using a coherent superpositions of  forward and backward processes.   
Their notion of backward process is different from ours, in that it is defined in terms of the joint unitary evolution of the system and an environment, rather than the dynamics of the  system alone.  
Due to the dependence on the environment,  the superposition of forward and backward processes  considered in \cite{rubino2020time} cannot be  interpreted as the result of an operation performed solely on the original channel.      An interesting direction of future research is to explore the thermodynamic power of the operations  introduced in our work,  combining them with the  insights of Ref. \cite{rubino2020time} and with similar insights arising from the research on indefinite causal order \cite{Felce20, Guha20, Simonov21}.   }

\section{Methods}

{\bf Characterisation of the input-output inversions.}
The foundation of our framework is the characterisation of the bidirectional quantum devices.  
  The logic of our argument is the following:  first, we observe that the input-output inversion must be linear in its argument (Appendix \ref{app:linear}).   Hence, the input-output inversion of unitary gates uniquely determines the time-reversal of every channel in the linear space generated by the unitary channels. 
 The linear span of the unitary channels is characterised by the following theorem from \cite{mendl2009unital}, for which we provide a new, constructive proof in Appendix \ref{app:bisto}. 
  \begin{theo}\label{theo:bisto}
 The linear span   of the set of unitary channels coincides with the linear span of the set of bistochastic channels. 
  \end{theo}

Theorem \ref{theo:bisto} implies that the input-output inversion of bistochastic channels is uniquely determined by the input-output inversion of unitary channels.      In particular, it implies that, up to changes of basis, there are only two possible choices of input-output inversion of bistochastic channels: either the adjoint, or the transpose.   

Interestingly, the adjoint and the transpose exhibit a fundamental difference when applied to the local dynamics of a subsystem.     Suppose that a composite system $S\otimes E$ undergoes a joint evolution with the property that the reduced evolution of system $S$ is  bistochastic. 
  Then, one may want to apply the input-output inversion only on the $S$-part of the evolution, while leaving the $E$-part unchanged.    In Appendix \ref{app:CP} we show that, when the dimension of system $S$ is larger than two,  the  local application of the input-output inversion generates valid quantum evolutions if and only if the input-output inversion is described by the transpose. In contrast, if the input-output inversion is described by the adjoint, then there is no consistent way to define its local action on the dynamics of a subsystem.
 
 \medskip 

{\bf Characterisation of the bidirectional channels.}  
We now show that the set of channels with an input-output inversion  satisfying Requirements (1-4) coincides with the set of  bistochastic channels.    The key of the argument is the following result: 
\begin{theo}\label{theo:allchannels}
If a channel $\map C$ admits an input-output inversion satisfying Requirements 1,2, and 4, then its input-output inversion  $\Theta (\map C)$ is a bistochastic channel.   \end{theo}
{ The proof is provided in Appendix \ref{app:allchannels}.  Theorem \ref{theo:allchannels}, combined with Requirement 3 (the input-output inversion  maps distinct channels into distinct channels), implies that  only bistochastic channels can admit an input-output inverse. Indeed, if a non-bistochastic channel had an input-output inversion, then the time reversal should coincide with the input-output inversion of a bistochastic channel, in contradiction with Requirement 3. }

In Appendix \ref{app:nobij} we show that, even if Requirement 3 is dropped,  defining a non-trivial input-output inversion satisfying requirements 1,2, and 4  is impossible for every system of dimension $d>2$.   For $d=2$, instead, an input-output inversion satisfying conditions (1-3) can be defined on all channels, but it maps all channels into bistochastic channels, in agreement with Theorem~\ref{theo:allchannels}.

\medskip

\section*{Data Availability}    The authors declare that the data supporting the findings of this study are available within the paper and in the supplementary information files. \\

\section*{Acknowledgments}   We acknowledge discussions with  L Maccone, Y Mo, BH Liu,  H Kristj\'ansson, A Vanrietvelde, M Christodoulou, A Di Biagio, E Aurell, K \.Zyczkowski, MT Quintino, and X Zhao.   This work was supported by the National Natural Science Foundation of China through grant 11675136, by the Hong Kong Research Grant Council through grant 17307719 and though the Senior Research Fellowship Scheme SRFS2021-7S02, by the Croucher Foundation,  and by the John Templeton Foundation through grant  61466, The Quantum Information Structure of Spacetime  (qiss.fr).  Research at the Perimeter Institute is supported by the Government of Canada through the Department of Innovation, Science and Economic Development Canada and by the Province of Ontario through the Ministry of Research, Innovation and Science. The opinions expressed in this publication are those of the authors and do not necessarily reflect the views of the John Templeton Foundation.

\medskip 

\section*{Author Contributions}    Both authors contributed substantially to the research presented in this paper and to  the preparation of  the manuscript. 

 \section*{Competing Interests}    The authors declare no competing interests.

    \appendix

\section{Input-output inversion  of unitary dynamics and its relation with time-reversal
}\label{app:unitary}

Here we characterise the action of the input-output inversion on the  set of unitary evolutions. 
 Using such characterisation, we will then discuss the relation between the notion of input-output inversion and the notion of  time-reversal  in quantum mechanics \cite{wigner1959group,messiah1965quantum}  and in  quantum thermodynamics \cite{campisi2011colloquium}.

\subsection{Input-output inversion of unitary dynamics}

Here we characterise the action of the possible input-output inversions on the set of unitary evolutions.  For this part of the paper, we will only use Requirements 1 (order reversal), 2 (identity preservation),  and 3 (distinctness preservation). 

First, note that Requirements 1 and 2 together imply that the map $\Theta$ transforms unitary channels into unitary channels:  
\begin{lemma}\label{lem:unitarypreserving}
Every  input-output inversion  $\Theta$,  satisfying  Requirements 1 and 2 in the main text must map unitary channels into unitary channels. 
\end{lemma} 

\Proof Recall that our standing assumption is that all unitary channels are bidirectional, that is, they are in the domain of the map $\Theta$.  Now,  a channel $\map C$  with input $S_1$ and output $S_2$   is unitary if and only if  there exists another channel $\map D$, with input $S_2$ and output $S_1$,  such that $\map D  \circ \map C =  \map I_{S_1}$ and $\map C \map \circ \map D = \map I_{S_2}$, where $S_1$ ($S_2$) is the input (output) of channel $\map C$, and  $\map I_{X}$ is the identity channel on system $X  \in  \{S_1,S_2\}$.     If $\map C$ is a unitary channel, then, applying the   map $\Theta$ on both sides of the two equalities, one obtains  $\Theta (\map D  \circ \map C) = \Theta( \map I_{S_1})$ and $\Theta(\map C \map \circ \map D) = \Theta (\map I_{S_2})$. Using Requirements 1 and 2, one then gets   $\Theta (\map C)   \circ \Theta( \map D) =  \map I_{S^*_1}$ and $\Theta(\map D) \circ \Theta( \map C) =\map I_{S_2^*}$, which imply that $\Theta (\map C)$ is a unitary channel.   (In passing, we observe that the above proof applies to any map $\Theta$ that is defined on a set of channels $\sf B$ with the property that, for every unitary channel $\map C$ in $\sf B$, its inverse $\map D$ is also in $\sf B$.)   
 \qed

Now, every unitary channel $\map U$ can be written in  the form $\map U (\rho)  =   U \rho  U^\dag$, for some unitary matrix $U  $
 in the special unitary group $\grp {SU}  (d)$.   Since the map $\Theta$ maps unitary channels into unitary channels, it induces a map $\theta$  from   $\grp {SU}  (d)$ to itself.   For the map $\theta$, Requirements 1-3  in the main text amount to the conditions 
 \begin{align}
\label{thetaorderrev}\theta (UV)   &  =    \theta(V) \theta (U)  \qquad \forall U,V\in \grp{SU}  (d)\\
\label{thetaunitpres}\theta (I)   &  =  I \\
\label{thetainj}    U  &  \not  =  V   \quad \Longrightarrow \theta (U) \not  =  \theta (V)   \quad \forall U,V\in \grp{SU}  (d) \, .
 \end{align}   
 
 We now show that the map $\theta$ must be unitarily equivalent to the adjoint or to the transpose.


\begin{lemma}\label{lem:unitary}
Let $\theta: \grp {SU}  (d) \to \grp {SU}  (d)$ be a map  satisfying the conditions in Eqs. (\ref{thetaorderrev})-(\ref{thetainj}).  Then, one has either   $\theta (U)  =   V U^\dag  V^\dag$ or $\theta(U) =  V U^T  V^\dag$, where $V \in  \grp {SU}  (d)$ is a fixed unitary operator. 
\end{lemma}
\Proof  Let $\theta$ be a time-reversal on $\grp{SU} (d)$. Define the transformation $\alpha:  \grp {SU}  (d)  \to \grp {SU}  (d) $ as    
 $\alpha  (U)  : =   \theta (U^\dag)$.
  By construction, $\alpha$ is a representation of the group $\grp{SU} (d)$, that is, it satisfies the condition $\alpha (U_1U_2)   =  \alpha (  U_1) \alpha (U_2)$ for every pair of matrices $U_1$ and $U_2$ in $\grp{SU} (d)$.  
  
The classification of the representations of $\grp{SU} (d)$ implies that,   up to unitary equivalences,  there exist only three representations in dimension $d$ \cite{fulton2013representation}: the trivial representation $\alpha (U)  =  I , \, \forall U$, the defining representation  $\alpha (U)   =  U , \, \forall U$, and the conjugate representation $\alpha (U)   = \overline U , \, \forall U$. 
 
 Now, the definition of $\alpha$  implies  the relation   $  \theta  (U)   =    \alpha  (U^\dag)  $.   Hence, there are only three possibilities, up to unitary equivalence:  {\em (i)} $  \theta  (U)  =  I \, ,\forall U$,    {\em (ii)} $ \theta  (U)  =  U^\dag \, ,\forall U$, and  {\em (iii)} $  \theta  (U)  =  \overline {U}^{\dag} \equiv U^T \, ,\forall U$.   The first possibility   $  \theta  (U)  =  I \, ,\forall U$ is ruled out by Eq. (\ref{thetainj}).  \qed

\subsection{Relation with time-reversal of unitary dynamics}

 The classic  notion of time-reversal in quantum mechanics dates back to Wigner \cite{wigner1959group}. In this formulation, time-reversal corresponds to  a symmetry  of the state space. By Wigner's theorem, state space symmetries are described either by operators that are either unitary or anti-unitary (see e.g.  \cite{Uhl16}).  For the time-reversal symmetry, the canonical choice is to take a  anti-unitary operator, motivated by physical considerations such as the preservation of the canonical commutation relations  under the transformation $X \mapsto  X$, $P \mapsto  -  P$ \cite{messiah1965quantum}, or the requirement that the energy be bounded from below both in the forward-time picture and in the backward-time picture \cite{weinberg1995quantum,roberts2017three}. 
  In the following, we will first provide some remarks that are valid both for unitary and anti-unitary operators, and then we will specialise them to the canonical  choice, namely the anti-unitary case.  
 
Let $A$ be an  operator (either unitary or anti-unitary)  that maps    generic pure states $|\psi\>$ into the corresponding time-reversed states  $|\psi_{\rm rev}\>  =  A |\psi\>$.  The time-reversal of states then induces a time-reversal of unitary evolutions.  The latter  is determined by the condition that, if a forward-time evolution $U$ transforms the state  $|\psi\>$ into the state $|\psi'\>$, then the corresponding backward-time evolution $U_{\rm rev}$  must transform the state  $|\psi'_{\rm rev}\>$ into  the state $|\psi_{\rm rev}\>$,   for every possible initial state $|\psi\>$.   This condition amounts to the equation $U_{\rm rev}  A  U  |\psi\>   =   A  |\psi\>  \, ,\forall |\psi\>$, or equivalently, to the equation
    \begin{align}
   \label{timerev}
  U_{\rm rev}    =   A   U^\dag A^{-1} \, ,
  \end{align}
  where $A^{-1}$ is the inverse of $A$.  
 This equation is known in quantum control and quantum thermodynamics, where it corresponds to the so-called microreversibility principle in the special case of autonomous ({\em i.e.} non-driven) systems with Hamiltonian invariant under time-reversal (cf. Eq. (40) of \cite{campisi2011colloquium}).

   Let us now focus on the canonical case where $A$ is an anti-unitary operation.    Eq. (\ref{timerev})   can be made explicit by recalling that every antiunitary operator  $A$    can be decomposed  as $  V  K  $, where $V$ is a unitary operator, and $K:  |\psi\> \mapsto  |\overline \psi\>$ is the complex conjugation in a given basis \cite{Uhl16}.  
  Using the relations $V^{-1}  =  V^\dag$ and $K^{-1}  =  K$,   one then obtains $A^{-1}  =   K^{-1}  V^{-1}   =    K  V^\dag$, and therefore 
  \begin{align}
 \nonumber 
   U_{\rm rev}  &  =   V  \,  (  K  U^\dag  K)   \,   V^\dag  \\
  \label{trevU}  &  =   V  U^T  V^\dag \, ,
  \end{align}    
  where $U^T$ denotes the transpose of $U$ in the given basis.

Eq. (\ref{trevU}) shows that the transformation of unitary evolutions due to the  canonical time-reversal  is unitarily equivalent to the transpose.   This transformation   corresponds to one of the two  possible forms of an input-output inversion allowed by our Lemma \ref{lem:unitary}. 

One can also consider non-canonical choices of time-reversal, such as the one advocated by Albert \cite{albert2000time} and Callender \cite{callender2000time}, who argued that, in certain systems, time-reversal should leave quantum states unchanged.   This choice corresponds to setting $A$ equal to the identity operator, which, inserted into Eq. (\ref{timerev}), gives the time-reversed dynamics $U_{\rm rev}  =  U^\dag$.  More generally, if one were to choose   the operator $A$ to be a generic unitary,   one would get the time-reversed dynamics $U_{\rm rev}  =   A  U^\dag  A^\dag$. This choice corresponds to the second option in our Lemma \ref{lem:unitary}.     

\subsection{Other order-reversing symmetries:  CT,  PT, and CPT.}

Our characterisation of the input-output inversions is not specifically about time-reversal symmetry, but more generally about any symmetry that reverses the order of time evolutions, cf.  Eqs. (\ref{thetaorderrev})-(\ref{thetainj}).  As such, it also applies to other combination of the time-reversal symmetry with other order-reversing symmetries, such as the combinations of time-reversal (T), with parity inversion (P) and charge-conjugation (C).    In other words,  all the combinations CT, PT, and CPT are possible order-reversing symmetries. The two options allowed by Lemma \ref{lem:unitary} cover the possible cases that may arise in these scenarios.   For example, Ref. \cite{skotiniotis2013quantum} argued that the  full CPT symmetry corresponds to a unitary transformation  $V$ at the state space level.  In this case, the same argument used in the derivation of Eq. (\ref{timerev}) implies that the action of the CPT symmetry on the dynamics is given by the mapping $U\mapsto  V U^\dag V^\dag$, in agreement with the second option in Lemma \ref{lem:unitary}.  

\subsection{Relation with time-reversal in non-unitary case}

So far we discussed the  input-output inversion of unitary dynamics, and its relation with time-reversal and other order-reversing symmetries.      In all these cases, the input-output inversion  can  be interpreted as an inversion of the system's trajectory in state space, corresponding to the intuitive idea of ``playing a movie in reverse''~\cite{sachs1987physics,albert2000time}: if  a system  transitions from $|\psi\>$ to $|\psi'\>$ in the forward time direction, then it  transitions from $|\psi'\>$ to $|\psi\>$ (up to unitary transformations  and/or complex conjugations)  in the backward time direction.    

The extension to non-unitary processes, discussed in the main text,   comes with  a key difference. For non-unitary processes, the reversal of state space trajectories is generally impossible, unless one includes the environment into the picture. Nevertheless, the notion of input-output inversion introduced in the main text is still valid, and can be defined without specifying the details of the interaction with the environment.     We now discuss the relation of our notion of input-output inversion with the notions of time-reversal of non-unitary evolution considered in the literature, in particular in the works of  Crooks  \cite{crooks2008quantum},   Oreshkov and Cerf \cite{oreshkov2015operational}, Aurell, Zakrzewski, and \.Zyczkowski \cite{aurell2015time}, and, more recently, in  Ref.  \cite{chiribella2020symmetries}).

 In Crooks' formulation  \cite{crooks2008quantum},   time-reversals constructed from the adjoint using a non-linear procedure.  
   Specifically,  the time-reversal of a quantum channel $\map C$ coincides with Petz' recovery map $\map C_{\rm Petz}$~\cite{petz1988sufficiency,hayden2004structure}, defined as    $\map C_{\rm Petz} (\rho)  : =   \rho_0^{1/2}\,   \map C^\dag\,  (\rho_0^{-1/2} \, \rho \, \rho_0^{-1/2}) \, \rho_0^{1/2}$  where  $\rho_0$ is any quantum state such that $\map C (\rho_0)  =  \rho_0$, and $\map C^\dag$ is the adjoint of channel $\map C$ (see the main text for the explicit definition).  This  procedure  can be applied to arbitrary channels, but in general it is non-linear, due to the dependence on the state $\rho_0$.     The non-linearity implies that the time-reversal of a mixture of channels is generally not equal to the mixture of their time-reversals, in violation of Requirement 4 in the main text.  More importantly, this time-reversal is generally not order-reversing, and Requirement 1 in the main text is generally violated:  for two generic channels $\map C$ and $\map D$, it is often not the case that $(\map C \circ \map D)_{\rm rev}  =  \map D_{\rm rev} \circ \map C_{\rm rev}$.  
     Requirements 1 and 4 are restored  if one restricts the time-reversal to  bistochastic channels, and chooses $\rho_0$ to be the maximally mixed state. In this case, the Petz recovery map coincides with the definition the adjoint map $\map C^\dag$, and is in agreement with the classification provided in the main text.  
     
An extension of the approach by Crooks was proposed in Ref.  \cite{chiribella2020symmetries}, following up on a suggestion by Andreas Winter.  There,  one defines   a fixed  reference state for every system, and defines the time-reversal only on the subset of channels $\map C$ satisfying the condition $\map C (\rho_{S_1})  =  \rho_{S_2}$, where $\rho_{S_1}$ and $\rho_{S_2}$ are the fixed reference states of the systems $S_1$ and $S_2$ corresponding to the input and output of channel $\map C$, respectively.  On this subset of channels, the time-reversal is defined as the Petz recovery map  $\map C_{\rm Petz} (\rho)  : =   \rho_{S_1}^{1/2}\,   \map C^\dag\,  (\rho_{S_2}^{-1/2} \, \rho \, \rho_{S_2}^{-1/2}) \, \rho_{S_1}^{1/2}$, or as the  variant  of the Petz recovery map where the adjoint $\map C^\dag$ is  replaced by the transpose $\map C^T$.     This definition satisfies all the Requirements 1-4 in the main text. However, it does not assign a time-reversal to every unitary evolution, unless $\rho_1$ and $\rho_2$ are set to the maximally mixed state.   Depending on the purpose, this may or may not be an issue. For example, it may  be interesting to consider time reversals that are defined only on a subset of unitary evolutions, {\em e.g.} the evolutions that preserve the Hamiltonian of the system. More generally,  extending the results of the present paper to the scenario where only a subgroup of the unitary group admits a time reversal is an interesting direction of future research.

Oreshkov and Cerf \cite{oreshkov2015operational} considered symmetries in an extended framework for quantum theory where arbitrary postselection are allowed.  Their main result is an extension of Wigner's theorem, where the allowed symmetries are described by  invertible  operators that are  either linear or anti-linear.   
  In their framework,  the time-reversal is   defined as an operation that  transforms states into measurement operators, and vice-versa.  This formulation does not explicitly specify how the time-reversal should be defined on quantum channels, but, since the time-reversal operation on states is generally non-linear (due to the presence of postselection), it is natural to expect that any time-reversal of quantum channels based on it would also be non-linear, thereby  violating our Requirement 4.   

Finally, Aurell, Zakrzewski, and \.Zyczkowski \cite{aurell2015time} define  the time-reversal  of general quantum channels in terms of a special decomposition,  whereby each channel is decomposed as $\map C  =    \map V_1  \map C_{\rm ess}  \map V_2^\dag$, where $\map V_1$ and $\map V_2$ are unitary channels (generally depending on $\map C$), and $\map C_{\rm ess}$ is a (generally non-unitary) quantum channel, called the ``essential map''.   The time-reversal is then defined as the channel $\map C_{\rm rev}   =   \map V_2    \map C_{\rm ess} \map V_1^\dag$.     This notion of time-reversal is an involution on the set of quantum channels, and for unitary channels it coincides with the adjoint.  On the other hand, for general non-unitary channels it is not an order-reversing operation, nor a linear one: like Crook's time-reversal, this choice of time reversal generally violates Requirements 1 and 4 in the main text.   Aurell, Zakrzewski, and \.Zyczkowski also consider other possible notions of time-reversals, defined as involutions on the set of quantum channels.  For this broader definition, it is possible to show that  the time-reversal of unitary dynamics cannot be extended to a time-reversal of arbitrary  quantum operations~\cite{chiribella2020symmetries}, and it is conjectured that an extension to the set of all quantum channels is also impossible.



{
\section{Characterisation of the operations on bidirectional quantum devices
}\label{app:bidirectional}\subsection{Definition}
 
A basic way to interact with a bidirectional quantum device is described by a particular type of quantum supermap  \cite{chiribella2013quantum} that transforms bistochastic channels into ordinary channels (CPTP maps).        

 Hereafter, we will denote by $L(\spc H,\spc K)$ the set of linear operators on a generic Hilbert space $\spc H$ to another generic Hilbert space $\spc K$, and we will use the shorthand notation $L(\spc H):  =  L(\spc H,\spc H)$.  Also, we will denote by  $\Map(S_{\rm i},S_{\rm o})$ the set of linear maps from $L(\spc H_{S_{\rm i}})$ to $L(\spc H_{S_{\rm o}})$,  by    $\Chan(S_{\rm i},S_{\rm o})$ the set of all quantum channels with input system $S_{\rm i}$ and output system $S_{\rm o}$, and by  $\Bi\Chan(S_{\rm i},S_{\rm o})$ the subset of all bistochastic channels.

A quantum supermap on bistochastic channels is  a linear map $\map S:  \Map  (A_{{ \rm i}}   \,  ,   \,  A_{{ \rm o}} ) \to  \Map ( B_{{ \rm i}},  B_{{ \rm o}}  )$, where $A_{\rm i}$   ($ A_{\rm o})$  is the  input (output) of the bistochastic channel on which  $\map S$ acts, and    $B_{\rm i}$   ($ B_{\rm o})$ is the  input (output) of the channel produced by $\map S$.  The map $\map S$ is required to transform channels into channels even when acting locally on part of a composite process.    Explicitly, this means that the map $(\map S \otimes \map I_{E_{\rm i}  E_{\rm o}})  \,  (  \map  M)$ must be a valid quantum channel whenever $\map M   \in  \Map  (A_{{ \rm i}}  E_{\rm i}\,  ,   \,  A_{{ \rm o}}  E_{\rm o}  ) $ is a channel that extends a bistochastic no-signalling channel, that is, a channel $\map M$ is such that the reduced channel $\map M_\sigma$, defined by 
\begin{align}
\map M_\sigma  (\rho)   =  \Tr_{E_{\rm o}}  [  \map  M(  \rho \otimes \sigma)]  \, ,\end{align}
belongs to $\NS (  A_{{ \rm i}}  , A_{ {\rm o}} )$  for every density matrix  $\sigma \in  L(  E_{\rm i})$.

\subsection{Choi representation}
An equivalent way to represent quantum supermaps   is to use the Choi representation \cite{choi1975completely}.    A generic linear map  $\map M   :  L(\spc H_{S_{\rm i}}) \to  L(\spc H_{S_{\rm o}})$ is in one-to-one correspondence with its  Choi operator  ${\rm Choi} ({\spc M})  \in L  (\spc H_{S_{\rm o}}   \otimes \spc H_{S_{\rm i}})$, defined by   
\begin{align}
\nonumber {\rm Choi}({\map M})   &:  =    \sum_{m,n}  \map M  (|m\>\<n| )\otimes |m\>\<n|\\
  &    =   (\map M\otimes \map I_{S_{\rm i}} )   (|I_{S_{\rm i}}\kk \bb I_{S_{\rm i}}|)  \, ,  \label{choief}
  \end{align}
  where the second equality uses the  double-ket notation~\cite{royer1991wigner,dariano2000bell} 
  \begin{align}\label{doubleket}|A\kk :  =  \sum_{m,n} \, \< m|  A  |n\>  \,  |m\>\otimes |n\> \,,
  \end{align} for  a generic  operator  $A \in  L(\spc H_{S_{\rm i}},  \spc H_{S_{\rm o}})$.

Now, a supermap  $\map S:  \Map  (A_{{ \rm i}}   \,  ,   \,  A_{{ \rm o}} ) \to  \Map ( B_{{ \rm i}},  B_{{ \rm o}}  )$  is itself a linear map, and, as such, is in one-to-one correspondence with a linear operator  $S   \in   L(  \spc H_{      B_{\rm o} }  \otimes \spc H_{  B_{{\rm i}} }  \otimes  \spc H_{A_{{ \rm o}}}   \otimes \spc H_{A_{ {\rm i}}})$.  The correspondence is specified by the relation
\begin{align}
\nonumber &\Choi   ( \map S  (\map N))  \\ 
& =  \Tr_{A_{{ \rm i}}   A_{ {\rm o}}    }  [     (  I_{B_{\rm o}   B_{{\rm i}}}  \otimes \Choi (\map N) )^T\,  S ] \, . \label{choicond2}            
\end{align}   
In this representation, the requirement that $\map S$ be applicable locally on part of a larger process is equivalent to the requirement  that the operator $S$ be positive \cite{chiribella2008transforming,chiribella2013quantum}.   The requirement that $\map S$ transforms any bistochastic  channel into a quantum channel is equivalent to the  condition
\begin{align}\label{choicond1}
\Tr_{B_{\rm o}} \Tr_{A_{{ \rm i}}   A_{ {\rm o}}  }
 [     (  I_{B_{\rm o}   B_{{\rm i}}}  \otimes  N)^T\,  S ]  
= I_{B_{\rm i}}  
\quad \forall j  \, \,,
\end{align} 
where $N$ is an arbitrary  Choi operators of a bistochastic channel in $\Bi \Chan (  A_{{ \rm i}}  ,  A_{ {\rm o}})$.

The normalisation conditions  (\ref{choicond1}) can be put in a more explicit form by decomposing the operator $S$ into orthogonal operators, each of which is  either proportional to the identity  or traceless on some of the Hilbert spaces, in a similar way as it was done in \cite{oreshkov2012quantum} for the characterisation of the operations  with definite time direction.   Using this fact, in the following we provide a complete characterisation.

\subsection{Characterisation of the supermaps from bistochastic channels to channels}

Here we characterise the Choi operators of the supermaps transforming bistochastic channels into channels.  
First, note that the Choi operator $N$ of any bistochastic channel $\map N  \in \BiChan  (A_{\rm i} ,   A_{\rm o})  $ with $A_{\rm i} \simeq A_{\rm o}$ satisfies the conditions 
\begin{align}
\Tr_{A_{\rm i} }  [  N]=   I_{A_{\rm o} }  \quad{\rm and}\quad \Tr_{A_{\rm o} }  [  N]=   I_{A_{\rm i} }  \, .
 \end{align}
 As a consequence, the operator $N$ can be decomposed as 
 \begin{align}N  =   \frac{I_{A_{\rm i}} \otimes I_{A_{\rm o} }}{d}   +   T\,,
 \end{align} 
 where  $T$ is operator such that 
 \begin{align}\label{doubletraceless}
\Tr_{A_{\rm i} }  [  T]=   0  \quad{\rm and}\quad \Tr_{A_{\rm o} }  [  T]=   0  \, .
 \end{align}
and $d$ is the dimension of systems $A_{\rm i}$ and $A_{\rm o}$. 
Choosing $T=  0$, the condition~(\ref{choicond1}) becomes 
\begin{align}\label{condizioneuno}
\frac{\Tr_{A_{\rm i}  A_{\rm o} B_{\rm o}}  [  S]}{d}  =  I_{B_{\rm i}} \, .
\end{align}
Choosing an arbitrary $T$, instead, we obtain
\begin{align}\label{condizionedue}
{\Tr_{A_{\rm i}  A_{\rm o} B_{\rm o}}  [ (T_{A_{\rm i}A_{\rm o}}  \otimes I_{B_{\rm i}B_{\rm o}}) S]} =  0 \, .
\end{align}
The combination of conditions~(\ref{condizioneuno}) and~(\ref{condizionedue}) is equivalent to the original condition~(\ref{choicond1}).   We will now cast condition~(\ref{condizionedue}) in a more explicit form.   Condition~(\ref{condizionedue}) is equivalent to the requirement that $S$ be orthogonal (with respect to the Hilbert-Schmidt product) to all operators of the form $T_{A_{\rm i}A_{\rm o}}  \otimes   J_{B_{\rm i}} \otimes I_{B_{\rm o}}$, where $J_{B_{\rm i}}$ is an arbitrary operator on $\spc H_{B_{\rm i}}$ and $T_{A_{\rm i}A_{\rm o}}$ is an arbitrary operator satisfying Eq.~(\ref{doubletraceless}).   In other words, $S$ must be of the form 
\begin{align}
\nonumber S    =     &
 I_{A_{\rm i} }\otimes I_{\rm A_{\rm o}} \otimes M_{B_{\rm i}  B_{\rm o}} \\
   \nonumber  &+ I_{A_{\rm i}}  \otimes  K_{ A_{\rm o}   B_{\rm i} B_{\rm o}}      +        I_{A_{\rm o}}  \otimes  L_{ A_{\rm i}   B_{\rm i}  B_{\rm o}}  \\
&  + W_{  A_{\rm i}  A_{\rm o}  B_{\rm i}    B_{\rm o}} \, ,  \label{pfui}
\end{align}
  where  $M_{B_{\rm i}  B_{\rm o}}$ is an arbitrary operator on $\spc H_{B_{\rm i}}\otimes \spc H_{B_{\rm _o}}$, and 
  the remaining  operators on the right hand side satisfy the relations 
  \begin{align}
\nonumber  \Tr_{A_{\rm o}}    [K_{ A_{\rm o}   B_{\rm i}  B_{\rm o}}]    =  0  \qquad   &&\\
\nonumber  \Tr_{A_{\rm i}}   [ L_{ A_{\rm i}   B_{\rm i}  B_{\rm o}}  ]  =  0 \qquad &&\\
\nonumber  \Tr_{A_{\rm i}}     [W_{  A_{\rm i}  A_{\rm o}  B_{\rm i}    B_{\rm o}} ]    =0  \qquad  && \\
\nonumber   \Tr_{A_{\rm o}}     [W_{  A_{\rm i}  A_{\rm o}  B_{\rm i}    B_{\rm o}} ]    =0\qquad &&\\
  \Tr_{B_{\rm o}}     [W_{  A_{\rm i}  A_{\rm o}  B_{\rm i}    B_{\rm o}} ]    =0   \, .\quad~\, &&   \label{relations}
  \end{align}
The last step is to express  the operators in the right hand side of Eq.~(\ref{pfui}) in terms of the partial traces of $S$.    Explicitly, we have 
\begin{align}
\nonumber
\nonumber M_{B_{\rm i}  B_{\rm o}}  & = \frac {\Tr_{{A_{\rm i}}  {A_{\rm o}}   }[S]}{  d^2  }   \\
\nonumber K_{ A_{\rm o}   B_{\rm i}  B_{\rm o}  }   &  =  \frac {\Tr_{{A_{\rm i}}    }[S]}{  d    }      -    I_{A_{\rm o}} \otimes M_{B_{\rm i}  B_{\rm o}} \\
 L_{ A_{\rm i}   B_{\rm i}  B_{\rm o}  }   &  =  \frac {\Tr_{{A_{\rm o}}    }[S]}{  d   }      -    I_{A_{\rm o}} \otimes M_{B_{\rm i}  B_{\rm o}} 
\end{align} 
while $W_{  A_{\rm i}  A_{\rm o}  B_{\rm i}    B_{\rm o}} $ is a generic operator satisfying the last three relations in Eq.~(\ref{relations}). 
  
Inserting the above relations into Eq.~(\ref{pfui}), we obtain  
\begin{align}
\nonumber S    =     &
    \frac{ I_{A_{\rm i}} }{  d  }  \otimes \Tr_{A_{\rm i}}  [S]      +        \frac{ I_{A_{\rm o}} }{  d  }  \otimes     {\Tr_{{A_{\rm o}}   }[S]}    \\
\nonumber &  -  \frac{ I_{A_{\rm i}}}{d}    \otimes \frac{I_{A_{\rm o}} }{    d  }   \otimes {\Tr_{{A_{\rm i}}  {A_{\rm o}}   }[S]}  \\
&  + W_{  A_{\rm i}  A_{\rm o}  B_{\rm i}    B_{\rm o}} \, ,  \label{pfui2}
\end{align}
or equivalently,  

\begin{align}
\nonumber &S    -     
    \frac{ I_{A_{\rm i}} }{  d   }  \otimes \Tr_{A_{\rm i}}  [S]      -       \frac{ I_{A_{\rm o}} }{  d   }  \otimes     {\Tr_{{A_{\rm o}}   }[S]}   \\
 \nonumber    &  \qquad  \qquad  +  \frac{ I_{A_{\rm i}}}{d}    \otimes \frac{I_{A_{\rm o}} }{    d }   \otimes {\Tr_{{A_{\rm i}}  {A_{\rm o}}   }[S]}  \\
 &\quad =  W_{  A_{\rm i}  A_{\rm o}  B_{\rm i}    B_{\rm o}} \, .  \label{pfui3}
\end{align}
In other words, the left hand side of the equation should be  an operator that satisfies the last three conditions of Eq.~(\ref{relations}).   The first two conditions are automatically guaranteed by the form of the right hand side of Eq.~(\ref{pfui3}), while the third  condition reads
\begin{align}
\nonumber  \Tr_{B_{\rm o}} [S]  = &\frac{ I_{A_{\rm i}} }{  d   }  \otimes \Tr_{A_{\rm i}  B_{\rm o}}  [S]      +        \frac{ I_{A_{\rm o}} }{  d  }  \otimes     {\Tr_{{A_{\rm o}}  B_{\rm o}  }[S]}    \\
 &  -  \frac{ I_{A_{\rm i}}}{d}    \otimes \frac{I_{A_{\rm o}} }{    d  }   \otimes {\Tr_{{A_{\rm i}}  {A_{\rm o}}  B_{\rm o} }[S]}  \, . \label{condizioneue2}
\end{align}
Summarising, we have shown that the normalisation of the supermap $\map S$ is expressed by the two conditions~(\ref{condizioneuno}) and~(\ref{condizioneue2}). As an example, it can be easily verified that the Choi operator of the quantum time flip, provided in Eq.~(\ref{choiflip}) satisfies conditions~(\ref{condizioneuno}) and~(\ref{condizioneue2}).    In fact, the quantum time flip satisfies these conditions even when the roles of $A_{\rm i}$ and $A_{\rm o}$ are exchanged, showing that the quantum time flip is a supermap from bistochastic channels to bistochastic channels. }

\section{Multipartite quantum operations with no definite time direction
}\label{app:multipartite}\subsection{General definition}
 Here we provide the definition of the set of multipartite operations with indefinite time direction. 
    The definition adopts  the framework of   \cite{chiribella2013quantum}, which defines general quantum supermaps from a subset of quantum channels to another.   In our case, the input and output set are as follows:  
 \begin{itemize}
 \item {\bf Input set:}  the set of $N$-partite no-signalling bistochastic channels, defined as the set of $N$-partite quantum channels of the form  
 \begin{align}\label{binosig}
 \map N   =  \sum_j \,  c_j \,   
  \map A_{1, j}    \otimes  \map A_{2, j} \otimes \cdots \otimes  \map A_{N, j}\, .      
 \end{align}    
 where each $c_j$ is a real coefficient,  each $\map C_{i,j}$ is a bistochastic channel.     
  We denote the set of channels of this form as $\Bi\NS (  A_{1{ \rm i}},   A_{1 {\rm o}} \, |\,   A_{2 \rm i}  , A_{2 {\rm o}}\, | \, \cdots  \,  |  A_{N \rm i} ,  A_{N {\rm o}})$ where  $A_{n \rm i}$   ($ A_{n \rm o})$ is the  input (output) of channel $\map A_{n,  j}$, for every possible $j$. 
  \item {\bf Output set:} the set $\Chan  (B_{\rm i},  B_{\rm o})$, consisting of ordinary channels from system $B_{\rm i}$ to system $B_{\rm o}$. 
\end{itemize}
A quantum supermap on no-signalling bistochastic channels is then defined as a linear map $\map S:  \Map  (A_{1{ \rm i}} A_{2{ \rm i}} \cdots  A_{N{ \rm i}}\,  ,   \,  A_{1{ \rm o}} A_{2{ \rm o}} \cdots  A_{N{ \rm o}}  ) \to  \Map ( B_{{ \rm i}},  B_{{ \rm o}}  )$, where $B_{\rm i}$   ($ B_{\rm o})$ is the  input (output) of the channel produced by $\map S$.  The map $\map S$ is required to transform channels into channels even when acting locally on part of a composite process.    Explicitly, this means that the map $(\map S \otimes \map I_{E_{\rm i}  E_{\rm o}})  \,  (  \map  M)$ must be a valid quantum channel whenever $\map M   \in  \Map  (A_{1{ \rm i}} A_{2{ \rm i}} \cdots  A_{N{ \rm i}}  E_{\rm i}\,  ,   \,  A_{1{ \rm o}} A_{2{ \rm o}} \cdots  A_{N{ \rm o}}  E_{\rm o}  ) $ is a channel that extends a bistochastic no-signalling channel, that is, $\map M$ is such that the reduced channel $\map M_\sigma$, defined by 
\begin{align}
\map M_\sigma  (\rho)   =  \Tr_{E_{\rm o}}  [  \map  M(  \rho \otimes \sigma)]  \, ,\end{align}
belongs to $\Bi\NS (  A_{1{ \rm i}}  , A_{1 {\rm o}} \, |\,   A_{2 \rm i} ,  A_{2 {\rm o}}\, | \, \cdots  \,  |  A_{N \rm i} ,  A_{N {\rm o}})$  for every density matrix  $\sigma \in  L(  E_{\rm i})$.

Quantum supermaps on bistochastic no-signalling channels describe the most general way in which $N$ bidirectional quantum processes can be combined together.   
In general, this combination can be incompatible with a definite direction of time, and, at the same time, incompatible with a definite ordering of the $N$ channels.  

Here we provide three examples for $N=2$.   To specify a supermap $\map S$, we specify its action on the set of product channels $\map A_{1}  \otimes \map A_{2}$, which---by definition---are a spanning set of the set of bipartite bistochastic no-signalling channels.
The first supermap, $\map S_1$, is defined as 
\begin{align}
\nonumber \map S_1  (\map A_{1}  \otimes \map A_{2})   (\rho) &:=  \sum_{m,n}  S_{1mn}   \rho S_{1mn}^\dag\\
S_{1mn}  &: =  A_{1m} A_{2n}^T  \otimes |0\>\<0|  +     A_{1m}^T A_{2n}  \otimes |1\>\<1| \, ,\end{align}
where $\{A_{1m}\}$ and $\{A_{2,n}\}$  are Kraus operators of channels $\map A_1$ and $\map A_2$, respectively.    
  This supermap can be generated by applying two independent quantum time flips to channels $\map A_1$ and $\map A_2$, respectively, and by exchanging the roles of the control states $|0\>$ and $|1\>$  in the second time flip.  
  This supermap describes the  winning strategy for the game defined in the main text. The supermap is incompatible with a definite time direction,  but is compatible with a definite causal  order between the two black boxes corresponding to channels $\map A_1$ and $\map A_2$ (channel $\map A_1$, or its transpose $\map A_1^T$ always acts after channel $\map A_2$ or its transpose $\map A_2^T$).

The second supermap, $\map S_2$, is the quantum SWITCH
~\cite{chiribella2009beyond,chiribella2013quantum}, defined as
\begin{align}
\nonumber \map S_2  (\map A_{1}  \otimes \map A_{2})   (\rho) &:=  \sum_{m,n}  S_{2mn}   \rho S_{2mn}^\dag\\
S_{2mn}  &: =  A_{1m} A_{2n}  \otimes |0\>\<0|  +      A_{2n} A_{1m} \otimes |1\>\<1| \, .\end{align}  
Note that the quantum SWITCH here is defined only on the set of bistochastic no-signalling channels. Interestingly, however, this definition determines the action of the quantum SWITCH  on arbitrary channels (and on arbitrary linear maps as well): the reason is that the set of bistochastic no-signalling channels includes the set of all products of unitary channels, and it is known that the quantum SWITCH is uniquely determined by its action on such channels \cite{dong2021quantum}.  
Finally, note that  the order of the channels $\map A_1$ and $\map A_2$ in the quantum SWITCH is indefinite, but each channel is used in the forward time direction.  

A third supermap, $\map S_3$,  is a combination of the quantum time flip and the quantum SWITCH, and is defined as follows: 
\begin{align}
\nonumber \map S_3  (\map A_{1}  \otimes \map A_{2})   (\rho) &:=  \sum_{m,n}  S_{3mn}   \rho S_{3mn}^\dag\\
S_{3mn}  &: =  A_{1m} A_{2n}  \otimes |0\>\<0|  +      A_{2n}^T A_{1m}^T \otimes |1\>\<1| \, .\end{align}  
This supermap  describes a coherent superposition of the process $\map A_1 \circ \map A_2$ and its time reversal  $\Theta(\map A_1  \circ \map A_2 )   = \map A_2^T \circ  \map A_1^T$. 
Such supermap   is incompatible with both a definite time direction and with a definite causal order.

\subsection{Choi representation}
An equivalent way to represent quantum supermaps  on bistochastic no-signalling channels is to use the Choi representation, thus obtaining a generalisation of the notion of process matrix~\cite{oreshkov2012quantum}, originally defined for supermaps that combine processes in an indefinite order, while using each process in a definite time direction.    

Since $\map S$ is a linear map, it is in one-to-one correspondence with a linear operator  $S   \in   L(  B_{\rm o}   B_{{\rm i}}   A_{1{ \rm o}}   A_{1 {\rm i}}      A_{2 \rm o}   A_{2 {\rm i}}\, \, \cdots  \,   A_{N \rm o}   A_{N {\rm i}}  )$.  The correspondence is specified by the relation
\begin{align}
\nonumber &\Choi   ( \map S  (\map N))  \\ 
& =  \Tr_{A_{1{ \rm i}}   A_{1 {\rm o}}      A_{2 \rm i}   A_{2 {\rm o}}\, \, \cdots  \,   A_{N \rm i}   A_{N {\rm o}} }  [    (   I_{B_{\rm i}   B_{{\rm o}}}  \otimes  \Choi (\map N)  )^T  \, S    ] \, .            
\end{align}   
In this representation, the requirement that $\map S$ be applicable locally on part of a larger process is equivalent to the requirement  that the operator $S$ be positive  \cite{chiribella2008transforming,chiribella2013quantum}.     The requirement that $\map S$ transforms any bistochastic no-signalling channel into a quantum channel is equivalent to the  condition
\begin{align}
\Tr_{B_{\rm o}} \Tr_{A_{1{ \rm i}}   A_{1 {\rm o}}      A_{2 \rm i}   A_{2 {\rm o}}   \cdots  \,   A_{N \rm i}   A_{N {\rm o}}} [ ( I_{B_{\rm o}   B_{{\rm i}}}  \otimes N  )^T  \,  S      ] = I_{B_{\rm i}}  
\quad \forall j  \, \,,
\end{align} 
where $N$ is  the Choi operator of an arbitrary bistochastic no-signalling channel in  $\Bi\NS (  A_{1{ \rm i}}  ,  A_{1 {\rm o}} \, |\,    \, A_{2 \rm i}   , A_{2 {\rm o}}\, | \, \cdots  \,  |  A_{N \rm i}  ,  A_{N {\rm o}})$.

\section{The quantum time flip supermap
}\label{app:supermap}

Here we show that  that the quantum time flip is a well-defined transformation of  bistochastic channels, that is, it is a valid quantum supermap \cite{chiribella2008transforming,chiribella2009theoretical,chiribella2013quantum}.   


First,  we observe  that the quantum time flip  transformation   is well defined:

\begin{prop}\label{prop:welldefined}
The transformation  $\map F:  \map C \mapsto  \map F_{\bf C}$ defined in the main text   is independent of the choice of Kraus operators ${\bf C}  = \{C_i\}$ used for channel  $\map C$. \end{prop}

The proof uses the Choi  isomorphism  \cite{choi1975completely} and the double-ket notation in Eq. (\ref{doubleket}).  Using this notation, the Choi operator of a quantum channel   $\map C$  can be written as 
\begin{align}\label{chuno}
{\sf Choi}  (  \map C)    =  \sum_i  |C_i\kk \bb  C_i  |   \,  . 
\end{align} 

{\bf Proof  of Proposition \ref{prop:welldefined}.}  The definition in the main text implies that the map $\map F_{\bf  C}$ has  Choi operator
\begin{align}\label{chdue}
 {\sf Choi}  ( \map F_{\bf C})      =   \sum_i    |F_i\kk\bb F_i| \, ,
\end{align}
with $F_i  =  C_i\otimes |0\>\<0|  +  \theta (C_i)  \otimes |1\>\<1|$. Explicitly, one has
\begin{align}  
|F_i\kk&  =     |C_i\kk  \otimes |0\>\otimes |0\>   +   |\theta (C_i)\kk  \otimes |1\>\otimes |1\>  \, .
\end{align}
When $\theta (C_i) =  C_i^T$,  we have 
\begin{align}  
\nonumber |F_i\kk&  =     |C_i\kk  \otimes |0\>\otimes |0\>   +   |C_i^T\kk  \otimes |1\>\otimes |1\>  \\
  &  =  V   |C_i\kk \, , \label{piropiro}
\end{align}
with
\begin{equation}
    V  =   I^{\otimes 2}    \otimes |0\> \otimes |0\> +   {\tt SWAP}   \otimes |1\> \otimes |1\> \, .
\end{equation}
Combining Eqs. (\ref{chuno}), (\ref{chdue}), and (\ref{piropiro}),  we then obtain 
\begin{align}\label{comppos}
 {\sf Choi}  ( \map F_{\map C})   =   V  {\sf Choi}  (\map C) V^\dag  \, . 
\end{align}
This equation implies that  (the Choi operator of)  $\map F_{\bf C}$ depends only on (the Choi operator of) $\map C$, and not on the specific choice of Kraus operators for $\map C$ used to define the Kraus operators $\map F_{\bf C}$.  \qed  

\medskip 

Next we observe that  the map $\map F$ is  {\em completely positive}, in the sense that the induced map $\widehat{\map F}: {\sf Choi}  ( \map  C)  \mapsto  {\sf Choi}  ( \map F_{\map  C}) $ is completely positive.  Complete positivity is immediate  from Eq. (\ref{comppos}).   Operationally, complete positivity means that the supermap $\map F$ can be applied locally to one part of a larger quantum evolution~\cite{chiribella2008transforming,chiribella2009theoretical,chiribella2013quantum}.  




Since the induced map $\widehat{\map F}$ is completely positive, it also has  a positive Choi operator.  Specifically, the operator is 
\begin{align}\label{choiflip}
 {\sf Choi}  (\widehat{\map F})   &  =  |V \kk \bb V| \, ,
\end{align} 
with 
\begin{align}
\nonumber 
\nonumber |V\kk   &  =   |I^{\otimes 2}  \kk  \otimes |  0\>\otimes   |0\>     +   |{\tt SWAP}  \kk  \otimes |1\>\otimes |1\>    \\
 \nonumber  &  =  |I\kk_{A_{\rm i}B_{\rm it}} \otimes |I\kk_{A_{\rm o}B_{\rm ot}}   \otimes |  0\>  \otimes |0\>\\
  &\qquad  +  |I\kk_{A_{\rm i}B_{\rm ot}} \otimes |I\kk_{A_{\rm o}B_{\rm it}}   \otimes |  1\>  \otimes |1\> \, ,
\end{align}
where  $A_{\rm i}$  ($A_{\rm o}$) is the input (output) system of the process $\map C$ on which the quantum time flip acts, and $B_{\rm it}$  ($B_{\rm ot}$) is the input (output) target system of the process $\map F  (\map C)$ produced by the quantum time flip.  

Finally, note that the quantum time flip maps bistochastic channels into bistochastic channels as one can check immediately from the Kraus representation $F_i =  C_i  \otimes |0\>\<0|  +  \theta(C_i)  \otimes |1\>\<1|$.  

Summarising, the quantum time flip $\map F$ is a well-defined, completely positive supermap transforming bistochastic channels into bistochastic channels. 

{
\section{The quantum time flip is incompatible with a definite time direction
}\label{app:incompatible}\subsection{Basic proof}

Here we show that the quantum time flip  cannot be decomposed as $\map F   =  p \, \map S_{\rm fwd} +  (1-p)  \, \map S_{\rm bwd}$, where $p\in  [0,1]$ is a probability and $\map S_{\rm fwd}$ ($\map S_{\rm bwd}$) is a  supermap corresponding to a quantum circuit that uses  the input  channel in the forward  (backward) direction.     

The proof proceeds by contradiction.  Let us consider the application of the quantum time flip to a unitary channel $\map U$.  Since the output channel $\map F(\map U)$ is unitary, and since unitary channels are extreme points of the convex sets of quantum channels,  the condition  $\map F (\map U)   =  p \, \map S_{\rm fwd}  (\map U) +  (1-p)  \, \map S_{\rm bwd}  (\map U)$  implies   $\map F (\map U)   =  \map S_{\rm fwd}  (\map U)   =  \map S_{\rm bwd}  (\map U)$.  Now, the condition $\map F (\map U)   =  \map S_{\rm fwd}  (\map U)$ implies the equality  
\begin{align}
  \nonumber \map  U^T  (\rho)  &  =  \map F (\map U)    (  \rho \otimes  |1\>\<1|)  \\
 \nonumber   &  =  \map S_{\rm fwd} (\map U)    (  \rho \otimes  |1\>\<1|) \\
 \nonumber   &  =  \map A  \circ  (   \map U\otimes \map I_{\rm aux})  \circ \map B   (  \rho\otimes |1\>\<1|) \\
    &  =  \map A\circ  (   \map U\otimes \map I_{\rm aux})  \circ \map B_1  (\rho) \, ,\label{contraiction}
\end{align} 
where $\rm aux$ is an auxiliary quantum system, $\map A $ and $\map B$ are suitable quantum channels, and $\map B_1$ is the quantum channel defined by $\map B_1 (\rho) :  =  \map B   (\rho\otimes |1\>\<1|)$. 

Equation~(\ref{contraiction}) should hold for all unitary channels $\map U$.  But this is a contradiction, because it is known that no quantum circuit can implement the transformation $\map U \mapsto \map U^T$~\cite{chiribella2016optimal,quintino2019probabilistic}. 

\subsection{Strenghtened proof with two copies of the input channel}

We show that  the  time-flipped channel $\map F (\map U)$ cannot be generated by an quantum process that uses two copies of a generic unitary channel $\map U$ in a definite time direction. This impossibility result holds even for processes that combine the two copies of the channel $\map U$ in an indefinite causal order.  Our result highlights a difference between the quantum time flip and the quantum SWITCH, as the quantum SWITCH of two unitary gates can be reproduced by ordinary circuits if two copies of each unitary gate  are provided~\cite{chiribella2009beyond,chiribella2013quantum}.  

Let us consider  operations that transform a pair of input channels into a single output channel.  These operations were defined in  \cite{chiribella2013quantum}, which we briefly summarise in the following.

An operation on a pair of channels can be described by a quantum supermap $\map S:  (\map A,  \map B)  \mapsto  \map S (\map A,\map B)$ that is linear in both arguments.  Let us denote by   $A_{\rm i}$  ($A_{\rm o}$) the input (output) system of channel $\map A$, and by  $B_{\rm i}$  ($B_{\rm o}$) the input (output) system of channel $\map B$, and by $C_{\rm i}$  ($C_{\rm o}$) the input (output) system of channel  $ \map S (\map A,\map B)$.  

The normalisation condition for the supermap $\map S$ is that the map $\map S (\map A,  \map B)$ should be a quantum channel  for every  $\map A  \in \Chan(A_{\rm i},A_{\rm o})$  and $\map B  \in  \Chan(B_{\rm i},B_{\rm o})$.    Linearity implies that the supermap $\map S$ can be extended to  a supermap $\widetilde{\map S}$ that is  well-defined on every bipartite channel of the form $\map N   =  \sum_j  \,  c_j \,    \map A_j\otimes \map B_j$, where each $c_j$ is a  real coefficients, and each $\map A_j$ ($\map B_j$) is a channel in $\Chan(A_{\rm i},A_{\rm o})$   ($\Chan(B_{\rm i},B_{\rm o})$).    The set of such channels $\map N$  coincides with  the set of no-signalling channels with respect to the bipartition  $A_{\rm i}  A_{\rm o}$  vs  $B_{\rm i}  B_{\rm o}$. This   set will be denoted by $\NS  (A_{\rm i}  , A_{\rm o}|  B_{\rm i} , B_{\rm o}) $.    The relation between the bilinear supermap $\map S$ and its extension $\widetilde {\map S}$ is given by the equality  $\widetilde{\map S}  (\map A\otimes \map B):  =  \map S (\map A,\map B)$, valid for every pair of channels $\map A$ and $\map B$.    In the following, we will focus on the map $\widehat{\map S}$.  
 

A general supermap with indefinite causal order is   a linear map $\widetilde{\map S}:  \Map  (A_{\rm i}B_{\rm i},  A_{\rm o} B_{\rm o}) \mapsto  \Map  (C_{\rm i},C_{\rm o})$ transforming no-signalling channels in $   \NS  (A_{\rm i}  ,  A_{\rm o}|  B_{\rm i}  , B_{\rm o})$ into ordinary channels in $\Chan  (C_{\rm i}  ,C_{\rm o})$.   Besides normalisation, the map   $\widehat{\map S}$ is required to be well-defined when acting locally on part of a larger process, that is, to satisfy the condition  
$(\widetilde{\map S} \otimes \map I_{D_{\rm i} D_{\rm o}})     (\map M)     \in \Chan  ( C_{\rm i} D_{\rm i}  ,  C_{\rm o} D_{\rm o})$ for every channel  $\map M \in   \Chan  (A_{\rm i}   B_{\rm i}  D_{\rm i} ,  A_{\rm o} B_{\rm o}  D_{\rm o})$ that extends a no-signalling channel, that is, any channel  $\map M$   such that the channel $\map M_\sigma:    \rho   \mapsto   \map M_\sigma (\rho ):  =  \Tr_{D_{\rm o}}  [\map M (\rho  \otimes \sigma)]$ is no-signalling for every state $\sigma$ of system $D_{\rm i}$~\cite{chiribella2013quantum}.  

The set of all supermaps $\widetilde{\map S}$  from  no-signalling channels in $  \NS  (A_{\rm i}  ,A_{\rm o}|  B_{\rm i}    , B_{\rm o})$ into ordinary channels in $\Chan  (C_{\rm i}  ,C_{\rm o})$ can be used to describe all the ways in which two generic quantum channels $\map A$ and $\map B$ can be combined, either in a definite or in an indefinite order.   In the special case where the channels $\map A$ and $\map B$ are bistochastic,  the above supermaps correspond to operations that use both channels in the  forward time direction.    To emphasise this fact, we use the notation 
\begin{align}
\map S_{\rm fwd} (\map A,  \map B)   :  =  \widetilde{\map S}  (\map A,  \map B) \,,
\end{align} 
where $\map A$ and $\map B$ are arbitrary bistochastic channels.   Operations that use  channels $\map A$ and $\map B$ in the backward time direction can be defined similarly as 
\begin{align}\map S_{\rm bwd} (\map A,  \map B)   :  =  \widetilde{\map S}  (\Theta  (\map A), \Theta( \map B)) \,,
\end{align} 
where $\Theta  $ is the input-output inversion,  defined in terms of the transpose.    


We now show that the quantum time flip cannot be reproduced  by a forward supermap $\map S_{\rm fwd}$, nor by a backward supermap  $\map S_{\rm bwd}$, nor by a random mixture of these two types of maps.   

 \begin{theo}\label{theo:swa}
It is impossible to find  supermaps $\map S_{\rm fwd}$ and  $\map S_{\rm bwd}$, and a probability $p\in  [0,1]$ such that    $\map F   (\map U)=  p \, \map S_{\rm fwd}  (\map U ,  \map U) +  (1-p)  \, \map S_{\rm bwd}  (\map U,\map U)$ for every unitary channel $\map U$. 
\end{theo}    

The proof consists of three steps.   First, note that, since $\map F(\map U)$ is a unitary channel and unitary channels are extreme points of the convex set of quantum channels, the condition  $\map F   (\map U)=  p \, \map S_{\rm fwd}  (\map U ,  \map U) +  (1-p)  \, \map S_{\rm bwd}  (\map U,\map U)$ implies  $\map F   (\map U)=  \map S_{\rm fwd}  (\map U ,  \map U) =   \map S_{\rm bwd}  (\map U,\map U)$ for every unitary channel $\map U$.  Hence, to prove the theorem it is enough to prove that the quantum time flip cannot be reproduced by a forward supermap.   

Second, note that  the condition $\map F   (\map U)=  \map S_{\rm fwd}  (\map U ,  \map U)$ implies that there exists a forward supermap implementing the transformation $\map U\otimes \map U \mapsto \map U^T$ where $\map U$ is an arbitrary unitary gate. 

Third, we prove the following lemma:  
\begin{lemma}  No forward supermap can implement the transformation  $\map U\otimes \map U \mapsto \map U^T$ where $\map U$ is an arbitrary unitary gate. 
  \end{lemma}

\Proof  
The similarity between the output channel $\map S_{\rm fwd}   (\map U ,  \map U)$  and the target gate $\map U^T$ can be measured by the average  fidelity between their Choi operators, given by 
\begin{align}
D_U  =   ( \map S   (\map U ,  \map U)  \otimes \map I  )    (|I\kk\bb I|)
 \quad  {\rm and} \quad   E_U  =      |U^T\kk\bb U^T| \, .  \end{align}         
Explicitly, the average fidelity is given by 
\begin{align}\label{fiU}
F  =  \int \d   U    \,   \frac{\Tr [  D_{U}   E_U  ]}{d^2}   \, .
\end{align}

In the following we will show that the fidelity $F$ is strictly smaller than 1 for every forward supermap. Specifically, we will show that the fidelity is upper bounded by $5/6$ for qubits, and by $6/d^2$ for higher-dimensional quantum systems.  The derivation of the bounds is inspired by  the semidefinite programming techniques developed in~\cite{chiribella2016optimal}, although no knowledge of semidefinite programming is needed in the proof.

The first step in the derivation is to write down the supermap $\map S_{\rm fwd}$ in the Choi representation.  Choi operators of quantum supermaps are also known as process matrices~\cite{oreshkov2012quantum}.

The fidelity can be rewritten by introducing the Choi operator of the supermap $\map S$, hereafter  denoted by $S$.   
The Choi operator $S$ is a positive operator on the tensor product space $\spc H_{A_{\rm i}}  \otimes  \spc H_{A_{\rm o}}  \otimes  \spc H_{B_{\rm i}}  \otimes\spc H_{B_{\rm o}}  \otimes\spc H_{C_{\rm i}}  \otimes\spc H_{C_{\rm o}} $, and the action of the supermap $\map S_{\rm fwd}$ on a pair of channels $\map A$ and $\map B$ is given by  
\begin{align}\label{SAB}
\map S_{\rm fwd} (  \map A,\map B)   =  \Tr_{A_{\rm i}, A_{\rm o}, B_{\rm i} ,B_{\rm o}}  [   ({\sf Choi}_{\map A}  \otimes  {\sf Choi}_{\map B}\otimes I_{C_{\rm i}C_{\rm o}})^T \,  S ] \, ,
\end{align}
where ${\sf Choi}_{\map A}$  an ${\sf Choi}_{\map B}$ are the Choi operators of $\map A$ and $\map B$, respectively, and $  \Tr_{A_{\rm i}, A_{\rm o}, B_{\rm i} ,B_{\rm o}}$ denotes the partial trace over the Hilbert space $\spc H_{A_{\rm i}}  \otimes  \spc H_{A_{\rm o}}  \otimes  \spc H_{B_{\rm i}}  \otimes\spc H_{B_{\rm o}}$.

Combining Eqs.~(\ref{fiU}) and~(\ref{SAB}), we obtain
\begin{align}\label{fisomega}
F  &=  \Tr [ S \, \Omega ] \\
\nonumber \Omega   &  =  \frac {\int \d U  \,  |\overline U\kk\bb \overline U|_{A_{\rm i},A_{\rm o}} \otimes   |\overline U\kk\bb \overline U|_{B_{\rm i},B_{\rm o}}  \otimes   |U^T\kk\bb U^T|_{C_{\rm i},C_{\rm o}}}{d^2} \, ,
\end{align}   
where the subscripts label the Hilbert spaces of the input/output systems. 

Note that the operator $\Omega$ satisfies the commutation relation 
\begin{align}
[\Omega,  \overline  U_{A_{\rm i}} \otimes  \overline  V_{A_{\rm o}}  \otimes \overline U_{B_{\rm i}}\otimes \overline V_{B_{\rm o}} \otimes V_{C_{\rm i}} \otimes  U_{C_{\rm o}}  ] =  0  \, ,
\end{align}
for every pair of unitary operators $U$ and $V$.  
Now, recall that the Hilbert space $\C^d \otimes \C^d \otimes \C^d$ can be decomposed into irreducible subspaces for the representation $\overline U\otimes \overline U\otimes U$ as
\begin{align}\label{isotypic}
\C^d \otimes \C^d \otimes \C^d    =   \bigoplus_j   \, (  \spc R_j\otimes \spc M_j) \, ,
\end{align}
where $\spc R_j$ is a representation space, where the representation $\overline U\otimes \overline U\otimes U$ acts irreducibly, and $\map M_j$ is a multiplicity space, which is invariant under the action of the representation $\overline U\otimes \overline U\otimes U$.   Here there are 3 possible irreducible representations,  of which one has dimension $d$, and the other two have dimensions $d (d_\pm  -1)$, respectively, where $d_\pm  =  d  (d  \pm  1)/2$ is the dimension of the symmetric/antisymmetric subspace of $\C^{d} \otimes \C^{d}$.   The $d$-dimensional representation, denoted by $j_0$, has multiplicity $m_{j_0}  =2  \equiv  \dim  (\spc M_{j_0})$, while all the other representations have multiplicity $m_j  =  1  \equiv  \dim   (\spc M_j)$. 

Using the decomposition~(\ref{isotypic}) and Schur's lemma,  we obtain the  expression 
\begin{align}
\Omega  =  \frac 1 {d^2}  \,  \bigoplus_j    \frac  {  P_j^{A_{\rm i} B_{\rm i} C_{\rm o}} \otimes P_j^{A_{\rm o} B_{\rm o} C_{\rm i}}  \otimes |I_{\spc M_j}\kk\bb I_{\spc M_j}|  }  {d_j}   \, .
\end{align}  

For quantum systems of dimension $d  >2$, we  now show that no quantum process with indefinite causal order can achieve  fidelity higher than $6/d^2$.  
To prove this bound, we define the quantum state 
\begin{align}
\rho_{A_{\rm o} B_{\rm o} C_{\rm i}}   :=  \frac 16   \,  \bigoplus_j   \, m_j  \, \frac{P_j^{A_{\rm o} B_{\rm o} C_{\rm i}}  \otimes I_{\spc M_j}^{A_{\rm o} B_{\rm o} C_{\rm i}} }{d_j}\, .    
\end{align}
Note that we have 
\begin{align}
 \frac 6{d^2} \,   \rho_{A_{\rm o} B_{\rm o} C_{\rm i}} \otimes I_{A_{\rm i} B_{\rm i} C_{\rm o}}
  \ge  \, \Omega\, , 
\end{align}
and therefore 
\begin{align} 
\nonumber F   & = \Tr [  S \, \Omega] \\
\label{uffa0}&    \le  \frac 6{d^2} \,  \Tr [ 
S   \,  (\rho_{A_{\rm o} B_{\rm o} C_{\rm i}} \otimes I_{A_{\rm i} B_{\rm i} C_{\rm o} }) \,  ] \, . 
\end{align}
Now, expand the state $\rho_{A_o B_o C_i}$ as an affine combination 
\begin{align}\label{uffa1}
\rho_{A_{\rm o} B_{\rm o} C_{\rm i}}    =  \sum_{k=1}^K   \, c_k   \,   ( \alpha_k  \otimes \beta_k \otimes \gamma_k)  \, ,
\end{align}
where $\alpha_k $,  $\beta_k$, and $\gamma_k$ are density matrices of systems $A_{\rm o}$, $B_{\rm o}$, and $C_{\rm i}$, respectively, and $(c_k)_{k=1}^K$ are real coefficients summing up to 1.  Define the quantum channels $\map A_k$ and $\map B_k$ with Choi operators ${\sf Choi}_{\map A_k}  =  I_{A_{\rm i}} \otimes \overline \alpha_k$  and ${\sf Choi}_{\map B_k}  =   I_{B_{\rm i}}  \otimes \overline \beta_k$, and note that one has 
\begin{align}\label{uffa2}
\Tr[  S\,   (\rho_{A_{\rm o} B_{\rm o} C_{\rm i}} \otimes I_{A_{\rm i} B_{\rm i} C_{\rm o}})   ]  = \sum_{k=1}^K c_k  \,  \Tr[ D_k \,   (\gamma_k\otimes I_{C_{\rm o}}) ] \, ,
\end{align}
where 
\begin{align}
D_k:  =  \Tr_{A_{\rm i}A_{\rm o} B_{\rm i} B_{\rm o}}   [ S\,    (   {\sf Choi}_{\map A_k} \otimes  {\sf Choi}_{\map B_k}  \otimes I_{C_{\rm i} C_{\rm o}} )^T]
\end{align}
is the Choi operator of the channel $\map D_k :  =   \map S  (\map A_k,  \map B_k)$, as per Eq.~(\ref{SAB}).   Since the channel $\map D_k$ is trace-preserving, its Choi operator satisfies the condition 
\begin{align}\label{uffa3}
 \Tr[ D_k \,   (\gamma_k\otimes I_{C_{\rm o}}) ]   = 1   \qquad \forall  k\in  \{1,\dots,  K\} \, .
\end{align}
Combining~Eqs.~(\ref{uffa0}),~(\ref{uffa2}), and~(\ref{uffa3}), we finally obtain    
\begin{align}
\nonumber F    & \le  \frac 6{d^2} \,  \Tr [ 
S   \,  (\rho_{A_{\rm o} B_{\rm o} C_{\rm i}} \otimes I_{A_{\rm i} B_{\rm i} C_{\rm o}}) \,  ]    \\
\nonumber &   =  \frac 6{d^2} \,     \sum_{k=1}^K  \,  c_k \,  \Tr[ D_k \,   (\gamma_k\otimes I_{C_o}) ]  \\ 
\nonumber &  =  \frac 6{d^2} \,     \sum_{k=1}^K  \,  c_k  \\
&  =  \frac 6{d^2} \, .  
\end{align}
Hence, no process with locally definite time arrow can achieve fidelity larger than $6/d^2$.  

Let us consider now the  $d=  2$ case.  In this case, we define two states 
\begin{align}
\rho_{ +,  A_{\rm o}B_{\rm o}  C_{\rm i}}   & =  \alpha\,   \frac{  Q_+}{2}   +     (1-\alpha)  \,  \frac{Q_{+,\perp}}{4} \\
\rho_{-, A_{\rm o}B_{\rm o}  C_{\rm i}}   & =  \alpha\,   \frac{  Q_-}{2}   +     (1-\alpha)  \,  \frac{Q_{+}}{2 }  \, ,
\end{align}
where $\alpha  =   3/5$,   $Q_\pm   =  \sum_{n  \in\{ 0,1\}}   |\Phi_n\>\<\Phi_n|_{A_{\rm o}B_{\rm o}C_{\rm i}}$ with $|\Phi_n\>_{A_{\rm o}B_{\rm o}C_{\rm i}}  =  ( |n\>_{A_{\rm o}}  |I\kk_{B_{\rm o} C_{\rm i}}    +  |I\kk_{A_{\rm o}  C_{\rm i}}  |n\>_{B_{\rm o}}  )/\sqrt{2 (\d \pm1)}$,  and $Q_{+,  \perp}    :  =     P_{+,  A_{\rm o}B_{\rm o}} \otimes I_{C_{\rm i}}  -  Q_{\pm,  A_{\rm o}B_{\rm o}C_{\rm i}}$, where $P_+$ is the projector on the symmetric subspace of $\spc H_{A_{\rm o}} \otimes \spc H_{B_{\rm o}}$.

Direct inspection shows that the states $\rho_{ +,  A_{\rm o}B_{\rm o}  C_{\rm i}}  $ and $\rho_{ -,  A_{\rm o}B_{\rm o}  C_{\rm i}}  $ have the same marginals on systems $A_{\rm o} C_{\rm i}$ and $B_{\rm o}  C_{\rm i}$.  In formula,
\begin{align}
\nonumber 
\Tr_{A_{\rm o}}[\rho_{ +,  A_{\rm o}B_{\rm o}  C_{\rm i}}]  & = \Tr_{A_{\rm o}}[\rho_{ -,  A_{\rm o}B_{\rm o}  C_{\rm i}}]   =:  \sigma_{B_{\rm o}  C_{\rm i}}\\
\Tr_{B_{\rm o}}[\rho_{ +,  A_{\rm o}B_{\rm o}  C_{\rm i}}]  & = \Tr_{B_{\rm o}}[\rho_{ -,  A_{\rm o}B_{\rm o}  C_{\rm i}}]   =:  \sigma_{A_{\rm o}  C_{\rm i}}
\end{align} 
Let us now define the channel $\map N   \in    \Chan  (   A_{\rm i} B_{\rm i}  ,   A_{\rm o} B_{\rm o}  C_{\rm i} )$ that measures the  systems $A_{\rm i}B_{\rm i}$ and prepares the  systems $A_{\rm o}B_{\rm o}C_{\rm i}$ in either the state $\rho_{ +}$ or in the state $\rho_-$, depening on the outcome of a measurement with projectors $P_+$ and $P_-$, respectively.  Explicitly, the action of the channel $\map N$ is 
\begin{align}
\map N (\rho)   =  \Tr[P_+  \, \rho ] \,  \rho_+  +   \Tr[P_-  \, \rho ] \,  \rho_- \, .       
\end{align}
The channel $\map N$ has Choi operator  
\begin{align}
{\sf Choi}_{\map N}   =   P_{+,  A_{\rm i} B_{\rm i}}  \otimes   \rho_{+,  A_{\rm o}B_{\rm o}C_{\rm i}}     +  P_{-,  A_{\rm i} B_{\rm i}}  \otimes   \rho_{-,  A_{\rm o}B_{\rm o}C_{\rm i}}  \, , 
\end{align}
and  direct inspection shows that one has the matrix inequality 
\begin{align}
{\sf Choi}_{\map N}  \otimes I_{C_{\rm o}}     \ge   \frac{6}{5}  \, \Omega  \,. 
\end{align}
Eq.~(\ref{fisomega}) then yields the bound
\begin{align}
\nonumber F  &  =  \Tr  [  S\,   \Omega]  \\
  &  \le \frac{5}{6} \,   \Tr [S  \,   ({\sf Choi}_{\map N}  \otimes I_{C_{\rm o}})] \, .\label{bounboun}
\end{align}

We now show that the factor inside the trace is equal to 1.   To this purpose, we observe that the channel $\map N$ satisfies the conditions 
\begin{align}
\nonumber \Tr_{  A_{\rm o}  }  [\map N (\rho)]   & =   \sigma_{B_{\rm o}  C_{\rm i}}     \\
\Tr_{  B_{\rm o}  }  [\map N (\rho)]   & =   \sigma_{A_{\rm o}  C_{\rm i}}     \quad  \forall \rho  \in  L(\spc H_{A_{\rm i}} \otimes \spc H_{B_{\rm i}})  \, , 
\end{align}
meaning that the marginals of the output state on systems $B_{\rm o}C_{\rm i}$ and $A_{\rm o}  C_{\rm i}$ are independent of the input state $\rho$.    In particular, these condidions imply that the channel $\map N$ is no-signalling with respect to the tripartition  $(  A_{\rm i},A_{\rm o} )  \, , (B_{\rm i},B_{\rm o}) \, ,  (   C_*,  C_{\rm i})$, where $C_*$ is a fictitious one-dimensional system, serving as input for the $C$-part of the tripartition.     Thanks to the no-signalling property, the channel $\map N$ can be decomposed as an affine combination of product channels, namely 
\begin{align}\label{akbkck}
\map N  =  \sum_{k=1}^K  \,  r_k  \,    \map A_k\otimes \map B_k \otimes \map C_k \,,
\end{align} where  $\map A_k  \in  \Chan (A_{\rm i},A_{\rm o})$,  $\map B_k  \in \Chan  (B_{\rm i},B_{\rm o})$, and $\map C_k  \in  \Chan  (C_*,C_{\rm i})$ are quantum channels,  and $(r_k)_{k=1}^K$ are real coefficients summing up to 1.    Note that, since the system $C_*$ is trivial, the ``channel'' $\map C_k  \in  \Chan (C_*, C_{\rm i})$ is just a quantum state of system $C_{\rm i}$. Such state will be denoted by $\gamma_k$ in the following.   

The decomposition~(\ref{akbkck}) implies that the Choi operator of the channel $\map N$ can be decomposed as 
\begin{align}
{\sf Choi}_{\map N}   =   \sum_k \,  r_k \,   \Choi_{\map A_k}  \otimes \Choi_{\map B_k}  \otimes \gamma_k  \, . 
\end{align} 
Hence, we have 
\begin{align}
 \nonumber  & \Tr [S  \,   ({\sf Choi}_{\map N}  \otimes I_{C_{\rm o}})]   \\
 \nonumber  &=  \sum_k \,r_k\,   \Tr [S \,   ( \Choi_{\map A_k}  \otimes \Choi_{\map B_k}  \otimes \gamma_k\otimes I_{C_{\rm o}} )\,  ]\\
 \nonumber  &  =  \sum_k   r_k\,    \Tr[  \map S   ( \overline {\map A}_k,  \overline {\map B}_k)  \,  (\gamma_k)]  \\
 \nonumber    &  =  \sum_k  \, r_k  \\
     &  =  1 \, ,  \label{unouno}
\end{align}
where the second equality follows from Eq.~(\ref{SAB}), by defining $\overline{\map A}_k$ and $\overline{\map B}_k$, respectively. Then, the third inequality follows from the fact that the map $\map S  (\overline{\map A}_k ,  \overline{\map B}_k)$ is a quantum channel, and therefore is trace-preserving.    

Finally, inserting Eq.~(\ref{unouno}) into Eq.~(\ref{bounboun}) we obtain the bound $F  \le  5/6$. \qed }

\medskip  
 
\section{Bound on the error probability for strategies with definite time direction
}\label{app:bound}  

{
\subsection{Numerical bound for arbitrary strategies}

Here we consider the game defined in the main text: a player is given access to two black boxes, implementing two unknown gates $U$ and $V$, respectively.  The problem is to determine whether a given pair of gates $(U,V)$ belongs to the set  
\begin{align}\label{splus} 
\set S_+ &= \{ (U, V) :    UV^T   =  U^TV   \} \, ,
\end{align}
or to the set  
\begin{align}\label{sminus} 
\set S_- &= \{ (U, V) :    UV^T   = - U^TV   \} \, ,
\end{align}
where $U$ and $V$ are generic elements of ${\sf U}(d)$, the set of $d \times d$ unitary matrices.
  
In the following we will show that  every player who uses the two black boxes in a definite time direction    
 will make errors with probability of at least $11.2\%$.   This bound on the probability of error holds for every strategy in which the two gates are accessed in the same time direction (either both in the forward direction, or both in the backward direction), even if the relative order of the two black boxes is indefinite.    

Measurement strategies with indefinite causal order were defined in Ref.~\cite{chiribella2019quantum}, where they were called {\em indefinite testers}  (see also the recent work \cite{ bavaresco2021strict}).    Mathematically, an indefinite tester is a linear map from the set of no-signalling channels to the set of probability distributions over a given set of outcomes  $\set X$.   Since we are interested in measurements on a pair of qubit channels $\map A:  \rho  \mapsto  U\rho  U^\dag$ and $\map B  :  \rho \mapsto  V\rho  V^\dag$, here we will focus on the case of bipartite no-signalling channels in the set $ \NS  (A_{\rm i}  ,  A_{\rm o}|  B_{\rm i}  , B_{\rm o})$, with  $\spc H_{A_{\rm i}} \simeq \spc H_{A_{\rm o}}  \simeq \spc H_{B_{\rm i}}  \simeq \spc H_{B_{\rm o}} \simeq  \C^d$.    In the Choi representation, the tester is described by a set of positive operators  $(T_x)_{x\in\set X}$ where each operator $T_x$ acts on the Hilbert space $\spc H_{A_{\rm i}}  \otimes  \spc H_{A_{\rm o}}  \otimes  \spc H_{B_{\rm i}}  \otimes\spc H_{B_{\rm o}}$.
When the test is performed on a pair of channels $(\map A, \map B)$,  the probability of the outcome $x$ is given by the generalised Born rule  
\begin{align}\label{borntobewild}p_x  =  \Tr [  T_x    (\Choi (\map A)  \otimes  \Choi (\map B))] \,.
\end{align}    The normalisation of the tester is expressed by the condition  
\begin{align}
\nonumber &\sum_{x\in\set X}       \Tr [  T_x    (\Choi (\map A)  \otimes  \Choi (\map B))^T]   =   1   \\
 &\qquad~~~~~      \forall \map A  \in  \Chan  (A_{\rm i},  A_{\rm o}) \, ,
 \, \forall \map B  \in  \Chan  (B_{\rm i},  B_{\rm o})\,. \label{genborn}
\end{align}
 Equivalently, this means that the positive operator $T:  = \sum_{x\in\set X}$ satisfies the condition   
\begin{align}
\nonumber &      \Tr [  T    (\Choi (\map A)  \otimes  \Choi (\map B))^T]   =   1   \\
 &\qquad~~~~~      \forall \map A  \in  \Chan  (A_{\rm i},  A_{\rm o}) \, ,
 \, \forall \map B  \in  \Chan  (B_{\rm i},  B_{\rm o})\,. \label{genborn2}
\end{align}
This condition shows that the operator $T$ is a process matrix, in the sense of Ref.~\cite{oreshkov2012quantum}.  In the notation of our paper, the above  conditions is equivalent to the linear constraints 
    \begin{eqnarray}
     \nonumber        T        &=  & \frac { I_{A_{\rm o}}\otimes  \Tr_{A_{\rm o} }   [T]    +  I_{B_{\rm o}} \otimes \Tr_{ B_{\rm o}}[T]}d      \\
   \nonumber          & &  - \frac{   I_{A_{\rm o}}  \otimes   I_{B_{\rm o}}  \otimes \Tr_{A_{\rm o}  B_{\rm o}}  [T] }{d^2}  \, , \\
   \nonumber          \Tr_{A_{\rm i}  A_{\rm o}}[  T] &=&  I_{B_{\rm o}}  \otimes  \frac{\Tr_{A_{\rm i}  A_{\rm o}  B_{\rm o}  }[  T]}{d}\, , \\
  \nonumber           \Tr_{B_{\rm i}  B_{\rm o}}[  T] &=  &   I_{Q_{\rm o}}  \otimes  \frac{\Tr_{A_{\rm o}  B_{\rm i}  B_{\rm o}  }[  T]}d\, , \\
                  \Tr[T] &=& d^2 \, .  \label{testernorm}
     \end{eqnarray}

We now  give a numerical bound of the minimum probability of error in distinguishing between two generic elements of the sets $\set S_+$ and $\set S_-$ defined in Eqs.~(\ref{splus}) and~(\ref{sminus}), respectively.  For this purpose, we consider an indefinite tester with binary outcome set $\set X  =  \{+,  -\}$ and  tester operators are denoted as  $(T_+,   T_-)$. 

     To obtain our bound, we consider  two  subsets of $\set S_+$ and $\set S_-$, denoted by $\set S_0'$ and $\set S_1'$, respectively, and we show that  these two subsets  not be distinguished perfectly by any indefinite tester.  The subsets are defined as follows: 
    \begin{align}
      \nonumber       \set S'_0 &= \Big\{(I, I),\, (I, X),\, (I, Z),\, \\
      \nonumber                 &(X, I),\, (X, X),\, (X, Z), \\
    \nonumber                   &(Z, I),\, (Z, X),\, (Z, Z), \\
    \nonumber                   &\left(\frac{X - Y}{\sqrt 2}, \frac{X + Y}{\sqrt 2}\right),\, \left(\frac{X + Y}{\sqrt 2}, \frac{X - Y}{\sqrt 2}\right), \\
                      &\left(\frac{Z - Y}{\sqrt 2}, \frac{Z + Y}{\sqrt 2}\right),\, \left(\frac{Z + Y}{\sqrt 2}, \frac{Z - Y}{\sqrt 2}\right) \Big\} \, ,
                      \end{align}
                      and 
                      \begin{align}
    \nonumber         \set S'_1 &= \Big\{(Y, I),\, (Y, X),\, (Y, Z),\, \\
   \nonumber                    &(I, Y),\, (X, Y),\, (Z, Y), \\
                      &\left(\frac{I + iY}{\sqrt 2}, \frac{I - iY}{\sqrt 2}\right),\, \left(\frac{I - iY}{\sqrt 2}, \frac{I + iY}{\sqrt 2}\right)\Big\} \, .
        \end{align}

The worst-case probability in distinguishing between the sets $\set S'_0$ or $\set S'_1$ is
   \begin{equation}
        \max(\{e_{0,i}\} \cup \{e_{1,j}\})
    \end{equation}
    with
    \begin{equation}
        \begin{split}
            &e_{0,i} = \Tr[T_-  \,  (|V_i\kk\bb V_i| \otimes |U_i\kk\bb U_i|)^T], \quad (U_i, V_i) \in \set S'_0 \, ,\\
            &e_{1,j} = \Tr[T_+\,  (|V_j\kk\bb V_j| \otimes |U_j\kk\bb U_j|)^T], \quad (U_j, V_j) \in \set S'_1 \, .
        \end{split}
    \end{equation}

    Hence, the minimum worst-case error probability is given by the following program:
      \begin{alignat}{2}
      \nonumber   & \text{minimize} \quad && \max(\{e_{0,i}\} \cup \{e_{1,j}\}) \\
        & \text{subject to~Eq.~(\ref{testernorm})} \, . 
    \end{alignat}

Numerical calculation by MATLAB CVX \cite{cvx,gb08} and QETLAB~\cite{qetlab}  then yields the optimal value 0.112149.}

\section{Linearity of input-output inversion
}\label{app:linear}  

Here we show that every input-output inversion defined on a convex subset $\sf B$ of quantum channels can be extended to a linear supermap on the (complex) linear space spanned by $\sf B$.

 \begin{prop}\label{prop:linear}
 Every input-output inversion $\Theta$, defined on a convex subset ${\sf B}  \subseteq \Chan (\spc H)$ and satisfying Requirement 4 in the main text, can be uniquely  extended to a  linear  supermap  $\Gamma$ on the  vector space ${\sf Span}  ({\sf B})$  spanned by the quantum channels in $\sf B$.  In other words, there exists a linear supermap $\Gamma:  {\sf Span}  (\sf B)  \to  {\sf Span}  ({\sf B})  $ such that $\Gamma  (\map C)  =  \Theta  (\map C)$ for every channel $\map C\in  \sf B$. 
   \end{prop} 

The proof is   somewhat lengthy, and can be skipped at a first reading.  
  
  \medskip 
   
\Proof   Let  $\map M$ be a generic element of ${\sf Span}  ({\sf B})$, written as 
\begin{align}\label{combin}
\map M  = \sum_j  \, c_j  \, \map C_j
\end{align} for some complex numbers $\{c_j\}  \subset \C$ and some quantum channels $\{\map C_j\} \subset  \sf B$.   {  Note that, if the input-output inversion $\Theta$ can be extended to a linear supermap  $\Gamma$ on  ${\sf Span}  ({\sf B})$, then such extension is necessarily unique: indeed, the action of the supermap $\Gamma$ is uniquely fixed by the linearity condition  
\begin{align}\label{wtheta}
\Gamma  \left(  \map M\right)  := \sum_j   c_j\,  \Theta (\map C_j) \, ,
\end{align}   
which defines $\Gamma$ on every element of ${\sf Span}  ({\sf B})$.}

We now show that the above definition is independent of the way in which $\map M$ is represented as a linear combination.  That is, if  $\map M =  \sum_k  \, c'_k  \, \map C'_k$ for some other set of  complex numbers $\{c_k'\}  \subset \C$, and  for some other set of quantum channels $\{\map C_k'\} \subset  \sf B$,    then one has  
\begin{align}\label{wellposed}
\sum_j  c_j\,  \Theta (\map C_j)   =  \sum_k   c_k'\,  \Theta (\map C'_k) \, . 
\end{align}  

To prove Equation (\ref{wellposed}), we start from the  special case where the numbers $\{c_j\}$ and $\{c_k'\}$ are probabilities, so that (\ref{combin}) is a convex combination.   In this case, the map $\map M$ is a quantum channel and Condition 3 in the main text implies  $\sum_j  c_j\,  \Theta (\map C_j)   =  \Theta (\map M)  =  \sum_k  c_k'  \,  \Theta  (\map C_k')$.    Hence, the definition (\ref{wtheta})   is well-defined on convex combinations.

Consider now the case where the numbers $\{c_j\}$ and $\{c_k'\}$ are  non-negative, so that (\ref{combin}) is a conic combination.  In this case, the trace-preserving property of the channels $\{\map C_j\}$ and  $\{\map C_k'\}$  implies 
\begin{align}
\nonumber \sum_j  c_j     &= \sum_j   \,  c_j  \,    \Tr[\map C_j  (|0\>\<0|)  ]  \\
 \nonumber  &    =    \Tr [\map M(|0\>\<0|)]   \\
 \nonumber &   =     \sum_k   \,  c_k'  \,    \Tr[\map C_k'  (|0\>\<0|)  ]  \\
  &  = \sum_k \, c_k'   =:  \lambda  \, .
\end{align}
Define the probabilities $p_j:   =  c_j/\lambda$ and $p_k'  : =  c_k'/\lambda$, and note that they satisfy the condition   
$\sum_j   p_j \, \map C_j  =  \map M/\lambda  =  \sum_k   p_k'\,  \map C_k' $.       Hence, one has    
\begin{align}
\nonumber \sum_j  c_j\,  \Theta (\map C_j)    &=   \lambda  \,  \sum_j   p_j  \Theta ( \map C_j)   \qquad &   \\
\nonumber  &   =    \lambda  \,  \Theta   ( \widetilde{ \map M} )   \qquad      & \widetilde{\map M}  :  =   \map M/\lambda  \\ 
\nonumber &   = \lambda   \,  \sum_j   p_k'  \Theta ( \map C_k')   \qquad & \\
 \label{conic}   &  =\sum_k  c_k\,  \Theta (\map C_k')   \, ,  &
\end{align}
where the third and fifth equalities follow from the condition $  \sum_j  p_j \,\map C_j    =   \sum_k  \, p_k'  \, \map C_k'$ and from the fact that $\Theta$ is well-defined on convex combinations.   Summarizing, Equation (\ref{conic}) shows that the definition (\ref{wtheta})  is well-posed on conic combinations.  

We now show that $\widetilde {\Theta}$ is well-defined on real-valued combinations. When  the coefficients $\{c_j\}$ and $\{c_k'\}$ are real, they can be partitioned into positive (negative) subsets, denoted by $\{c_j\}_{j\in\set S_+}$   ($\{c_j\}_{j\in\set S_-}$)    and $\{c_k'\}_{k\in\set S'_+}$   ($\{c_k'\}_{k\in\set S_-'}$), respectively.       Define the maps $\map M_{\pm}:  =  \sum_{j \in  \set S_\pm}  |c_j| \, \map C_j$ and $\map M_{\pm}':  =  \sum_{k \in  \set S'_\pm}  |c_k'| \, \map C_k'$.   By construction, we have $\map M_+ -  \map M_-   =  \map M_+'  -  \map M_-' $, and equivalently, $\map M_+  +  \map M_-'  =   \map M_+'  +  \map M_-$.     Since $\widetilde{\Theta}$ is well-defined on conic combinations, we have 
\begin{align}
\nonumber &\sum_{j\in  S_+}   |c_j| \,  \Theta  (\map C_j)   + \sum_{k\in  S_-'}  |c_k'|   \, \Theta (\map C_k')   \\
&   \qquad \qquad =\sum_{k\in  S_+'}   |c_k'| \,  \Theta  (\map C_k')   + \sum_{j\in  S_-}  |c_j|   \, \Theta (\map C_j) \, , 
\end{align}
and therefore, 
\begin{align}
\nonumber &\sum_{j\in  S_+}   |c_j|   \,  \Theta  (\map C_j)   -    \sum_{j\in  S_-}  |c_j|\,   \Theta (\map C_j)  \\
&   \qquad \qquad =\sum_{k\in  S_+'}   |c_k'| \,  \Theta  (\map C_k')   -   \sum_{k\in  S_-'}  |c_k'|   \, \Theta (\map C_k')     \, , 
\end{align}
which is equivalent to  $\sum_j  c_j\,  \Theta (\map C_j)   =  \sum_k   c_k'\,  \Theta (\map C'_k)$.  Hence, we conclude that the definition (\ref{wtheta}) is well-posed on linear combinations with real coefficients.  

Finally, consider  linear combinations with arbitrary complex coefficients.  In this case, the map $\map M$ in Equation (\ref{combin}) can be decomposed as  $\map M  =   \map A  +  i \map B$,  with 
\begin{align}
\nonumber \map A    =   \sum_j    {\sf Re}   (c_j)  \,     \map C_j         =   \sum_k   {\sf Re}   (c_k')  \,     \map C_k'  \\
   \map B    =   \sum_j    {\sf Im}   (c_j)  \,     \map C_j         =   \sum_k   {\sf Im}   (c_k')  \,     \map C_k'   \,,
\end{align}
where ${\sf Re}(z)$ and ${\sf Im}(z)$ denote the real and imaginary part of a generic complex number $z\in \C$, respectively. 
Since the definition (\ref{wtheta}) is well-posed on linear combinations with real coefficients, we have the equalities 
\begin{align}
\nonumber 
\sum_j    {\sf Re}   (c_j)  \,   \Theta (  \map C_j   )      &=   \sum_k   {\sf Re}   (c_k')  \,     \Theta (\map C_k') \\
  \sum_j    {\sf Im}   (c_j)  \,    \Theta ( \map C_j)         &=   \sum_k   {\sf Im}   (c_k')  \,    \Theta  ( \map C_k')   \,  ,  
\end{align}
which, summed up, yield the desired equality  $\sum_j   c_j  \,   \Theta (  \map C_j   )      =   \sum_k  c_k'  \,     \Theta (\map C_k')$.  Hence, the definition  (\ref{wtheta}) is well-posed on arbitrary linear combinations. \qed

\section{Proof of Theorem 1 in the main text
}\label{app:bisto}  
    
Here we  provide a constructive proof of the fact  that 
 the unitary channels are a spanning set for the linear space spanned by bistochastic channels \cite{mendl2009unital}.  Our proof provides an explicit way to decompose a given bistochastic channel into a linear (in fact, affine) combination of unitary channels.  

  Hereafter we will denote by  
$\Map  (\spc H)$ the set of linear maps from $L(\spc H)$ to itself, and by $\Chan  (\spc H)\subset  \Map (\spc H)$ the subset of quantum channels (completely positive trace-preserving maps).

The proof makes use of a one-to-one correspondence between linear maps in  $\Map  (\spc H)$  and vectors  in $\spc H\otimes \spc H\otimes \spc H\otimes\spc H$. 
The correspondence associates the linear map $\map M$  to the vector  $|{\rm Vec}  (\map M)\>$ defined as 
\begin{align}\label{vectorrep}
|{\rm Vec}  (\map M)\>  : =   \sum_{j,k,l}    \, \big[ \map M (  |j\>\<k|)|l\> \big]  \otimes |j\>   \otimes |l\> \otimes |k\>   \, .
\end{align}

For a completely positive map with Kraus representation $  \map M  (\cdot)  =  \sum_i   M_i  \cdot  M_i^\dag$,  the vector
$|{\rm Vec}  (\map M)\>$ has the simple form 
\begin{align}
|{\rm Vec}  (\map M)\>    =    \sum_i  \,  |M_i\kk   |\overline M_i\kk    \, ,
\end{align} 
where we used the double ket notation in Eq. (\ref{doubleket}).    In particular, the unitary channels $\map U (\cdot)   =  U\cdot U^\dag$ correspond to vectors of the form $|{\rm Vec}  (\map U)\>    =      |  U\kk   |\overline  U\kk$.

We now show that the span of the vectors of the form $|U\kk|\overline U\kk $ coincides with the span of the vectors of the form $|{\rm Vec}  (\map B)\>$, where $\map B$ is a generic bistochastic channel.  
To this purpose, we use the fact that the linear span of a set of vectors   $\{  |v_i\>\}$ is equal to the support of their frame operator 
\begin{align}F :  =  \sum_i  |v_i\>\<v_i| \,,
\end{align} and every vector $|v\rangle$ in the linear span can be expanded as 
\begin{align}\label{frame}
|v\rangle =  \sum_j     \< v_j  | F^{-1}  |v\rangle  ~  |v_j\rangle \,  \, ,    
\end{align}    
{ where $F^{-1}$ denotes the inverse of $F$ on its support, also known as the Moore-Penrose pseudo-inverse} (see {\em e.g.}  \cite{casazza2000art}).    

For the vectors $|U\kk  |\overline  U\kk/d$, the frame operator can be defined as 
\begin{align}\label{F}
F   =   \frac 1 {d^2}\int  \d  U  \,   |U\kk \bb U| \otimes  |\overline  U\kk  \bb \overline U|  \, ,\end{align}
where  $  \d U$ is the normalized Haar measure on $\grp{ SU}  (d)$. 

The integral can be computed with the methods of representation theory, which give rise to the following
\begin{lemma}\label{lem:Fcomputed}
The frame operator (\ref{F}) is given by  
\begin{align}\label{Fcomputed}
F    = \frac{  E_{13} \otimes E_{24}  }{d^2} +  \left  ( 1  -  \frac 1{d^2} \right)    \frac{  E^\perp_{13}  \otimes E^\perp_{24}  }{(d^2 -1)^2} \, ,  
\end{align}
where   $E$ and $E^\perp$ are the projectors $E:  =  |I\kk  \bb I|/d$ and $E^\perp  :  =   I\otimes I  -  E$, and the subscripts $13$ and $24$ specify  the Hilbert spaces on which the operators act. 
\end{lemma}

\Proof    Note that the vectors $|U\kk  |\overline  U\kk$  can be expressed as  $(  U_1\otimes  I_2\otimes \overline U_3 \otimes I_4)  ( |I\kk_{12}  |I\kk_{34})$.   The product $U\otimes  \overline U$ defines a representation of $\grp {SU}  (d)$ that can be decomposed into two irreducible representations: one is the trivial representation, acting on the one-dimensional subspace spanned by the vector $|I\kk$, and the other is its orthogonal complement, acting on the $(d^2-1)$-dimensional subspace orthogonal to $|I\kk$.  Hence, Schur's lemmas imply the relation  
\begin{align}
\int   \d U    \,  (U\otimes \overline U)  A  (U\otimes \overline U)^\dag   = \Tr[ E  A ]    \,  E    +  \Tr[   E^\perp  A]  \,  \frac{  E_\perp}{d^2-1} \, .  
\end{align}
Inserting this relation into the definition of $F$,  one obtains Equation (\ref{Fcomputed}). \qed 
 
 Let    $\BiChan (\spc H)$ be the set of bistochastic channels mapping density matrices on $\spc H$ into density matrices on $\spc H$.   We have the following:   
 
\begin{prop}\label{prop:span}
Every bistochastic channel is a linear combination of  unitary channels. 
\end{prop}  

\Proof   Let  $\map B\in \BiChan (\spc H)$ be a generic bistochastic channel, and let $\map B    (\cdot)   =  \sum_i  B_i \cdot  B_i^\dag$ be a Kraus representation for $\map B$. Then, the vector representation of $\map B$ is 
$  |{\rm Vec}  (  \map B) \> :  =  \sum_i  |B_i  \kk  |\overline B_i  \kk$.   
  We will now show that the vector $|{\rm Vec}  (\map B)\>$ is contained in the linear span of the vectors $|U\kk |\overline U\kk$, which coincides the the support of the frame operator   $F$ in Equation (\ref{F}).    Using Lemma (\ref{lem:Fcomputed}), the projector on the support of $F$ can be expressed as   
  \begin{align}
   P     &=     E_{13} \otimes  E_{24}   +    E^\perp_{13}  \otimes  E^\perp_{24}  \, ,
  \end{align}
where $E$ and $E^\perp$ are as in Lemma \ref{lem:Fcomputed}. 

Note that one has the relation   $P  =   I_{13} \otimes I_{24}    +  2    E_{13} \otimes  E_{24}    -    E_{13}  \otimes  I_{24}   -   I_{13} \otimes E_{24} $. Using this relation, one obtains
\begin{align}
\nonumber P   |{\rm Vec}  (\map B)\>   &  =|{\rm Vec}  (\map B)\>    \\
\nonumber  & \quad    +   \frac 2{d^2}   \, \sum_i   \Tr[  B_i^\dag B_i] \,  |I\kk_{13} |  I  \kk_{24}\\
\nonumber 
&\quad -\frac 1d \sum_i  |  B_iB_i^\dag  \kk_{13} |I\kk_{24}  \\
 &\quad -\frac 1d \sum_i   |I\kk_{13}  |  B_i^T  \overline  B_i  \kk_{24}  \, .\label{decomp}
  \end{align}
  Since the channel $\map B$ is bistochastic, it satisfies the conditions $\sum_i  B_i  B_i^\dag =  I$ and $\sum_i  B_i^T \overline B_i  = I  $.  Hence, one obtains $P  |{\rm Vec}  (\map B)  \kk    = |{\rm Vec}  (\map B)  \kk$, meaning that the vector $|{\rm Vec}  (\map B)  \kk$ is contained in the support of the frame operator $F$.  Equivalently, this means that the bistochastic channel $\map B$ is contained in the linear span of the unitary channels.  \qed

\medskip  

Note that Equation (\ref{frame}), combined with the vector representation (\ref{vectorrep}),  also provides an explicit way to decompose every bistochastic channel as a linear combination of unitary channels.  Explicitly, one has 
\begin{align}
\nonumber   &  |{\rm Vec} (\map B) \>  \\
\nonumber   &   = \frac 1 {d^2} \,  \int \d U~  (  \bb U|  \bb \overline U|  ) F^{-1}    |{\rm Vec} (\map B) \> ~ |U\kk|\overline U\kk  \\
\nonumber &=  \frac 1 {d^2} \,  \int \d U~  \Big\{   d^2  ( \bb U|  \bb \overline U| )    (   E_{13}\otimes E_{24}) |{\rm Vec} (\map B) \>      \\
\nonumber &\quad  +  d^2 (d^2-1)   ( \bb U|  \bb \overline U| )   (  E^\perp_{13}\otimes E^\perp_{24}) |{\rm Vec} (\map B) \>       \Big\} ~  |U\kk  | \overline U \kk\\
\nonumber  &  =   \int \d U~  \left\{ (d^2-1)        \sum_i  \left|  \Tr [U^\dag B_i]\right|^2  
 \right.\\
 &   \quad  - \left.    \frac{d^2-2}{d} \,  \left(   \sum_i  \Tr[  B_i^\dag B_i]  \right)  \right\} \, ~ |U\kk|\overline U\kk \,, 
\end{align}
or equivalently, 
\begin{align}
\nonumber \map B     &=   \int \d U~  \Big\{  (d^2-1)     \Tr[{\sf Choi}  (\map U)  \, {\sf Choi}  (\map B)]   
 \\
 &   \quad  -     \frac{d^2-2}{d} \,     \Tr [\map B (I) ]    \Big\} \, ~\map U \,, 
\end{align}
where { $ {\rm Choi}({\map M})      =   (\map M\otimes \map I)  (|I\kk\bb I|)$ is the Choi operator of  a generic linear map $\map M$ (cf. Eq.~(\ref{choief}) for the explicit definition).} 

Since every unitary channel is trivially bistochastic, Proposition \ref{prop:span} implies Theorem 2 in the main text. 

\section{Local action of the input-output inversion
}\label{app:CP}

Let $\Theta_S$ be the input-output inversion for evolutions of system $S$, defined on the set  $\BiChan (\spc H_S)$ of bistochastic channels.   As shown in Supplementary Note 3, $\Theta_S$ can be extended to a linear supermap acting  on the linear span of the set of bistochastic channels. In the following, we will denote the linear span by $\SpanBiChan (\spc H_S)$, and we will call its the maps $\SpanBiChan (\spc H_S)$   {\em bistochastic maps}.  

Here we ask whether it is possible to define the local action of the input-output inversion $\Theta_S$ on joint evolutions  of a composite system $S\otimes E$.  Let us first specify the properties that the local action is required to satisfy.  
Let $\map C$ be a quantum channel on system $S\otimes E$, with the property that the reduced evolution of system $S$ is well-defined, meaning that one has
\begin{align}
\Tr_{E} \, \circ \, \map C     =    \map C_S \otimes \Tr_E \, ,
\end{align}
where $\Tr_E$ is the partial trace over the Hilbert space of system $E$, and $\map C_S$ is a channel on system $S$.     If the channel $\map C_S$ is bistochastic, one may want to apply the input-output inversion locally on the $S$-part of the joint  evolution $\map C$, without changing the $E$-part.  Mathematically, this means extending the map $\Theta_S$ to a  linear supermap $\Gamma_S$ acting on the whole space of linear maps $\Map (\spc H_S)$, instead of just the space of bistochastic  channels:  
\begin{defi}
Let $\Theta_S  :  \SpanBiChan (\spc H_S) \to  \SpanBiChan (\spc H_S)$ be a linear supermap defined on the space of bistochastic maps. An extension of $\Theta_S$ is a linear supermap  $\Gamma_S  :  \Map (\spc H_S) \to  \Map (\spc H_S)$ such that 
\begin{align}
\Gamma_S  (\map C)  = \Theta_S (\map C)  \qquad \forall \map C \in  \SpanBiChan (\spc H_S) \, . 
\end{align}
\end{defi}   

The local action  is then given by the linear supermap  $\Gamma_S\otimes \map I_E$, where $\map I_E$ is the identity supermap on $\Map (\spc H_E)$, and the  supermap  $\Gamma_S\otimes \map I_E$ is uniquely defined by the condition $(\Gamma_S\otimes \map I_E)  (\map A\otimes \map B) :  =  \Gamma_S (\map A)  \otimes \map B $, for every pair of maps $\map A \in \Map (\spc H_S)$ and $\map B  \in \Map  (\spc H_E)$. 

Now, a crucial requirement for the local action $\Gamma_S$ is that the extended supermap $\Gamma_S\otimes \map I_E$  should transform  quantum operations (completely positive trace non-increasing maps) into quantum operations \cite{chiribella2008transforming,chiribella2013quantum}.   This requirement implies in particular that $\Gamma_S\otimes \map I_E$  should  be {\em complete positivity  preserving (CP-preserving)}, that is, it should transform completely positive maps into completely positive maps.

In the main text, we showed that, up to unitary equivalences, the input-output inversion supermap $\Theta_S$ is  either the transpose  or the adjoint.  These two supermaps have two natural extensions to the set $\Map  (\spc H_S)$.  For a generic map $\map M \in \Map (\spc H_S) $, the transpose  map $\map M^T$ is defined by the relation 
\begin{align}\label{transpmap}
\Tr\left[  A^T   \,  \map M (\rho)\right]   =  \Tr\left[    \left(  \map M^T (A) \right)^T  \rho \right]   \qquad \forall \rho ,\forall A \, , 
\end{align}
and the adjoint map $\map M^\dag$ is defined by the relation 
\begin{align}
\Tr\left[  A^\dag   \,  \map M (\rho)\right]   =  \Tr\left[    \left(  \map M^\dag (A) \right)^\dag  \rho \right] \qquad\forall \rho, \forall A \, . 
\end{align}
For a completely positive map $\map C:  \rho \mapsto  \map C(\rho)  =    \sum_i  C_i  \rho  C_i^\dag$, the transpose and adjoint are given by  $\map C^T  :   A \mapsto \map C^T (A)  =   \sum_i  C_i^T  A  \overline C_i$  and $\map C^\dag  :  A  \mapsto  \map C^\dag (A)  =  \sum_i  C_i^\dag  A  C_i$, respectively.

For the transpose supermap $\Gamma_S^{\rm trans}  :  \map M\mapsto  \map M^T$, a  proof of CP-preservation comes from the Choi representation. 
In this representation,   the map $\Gamma_S$ is  represented by  a map $\widehat \Gamma_S$, uniquely defined by the relation 
  \begin{align}\label{choimap}
  \widehat \Gamma_S  (  {\rm Choi}  (\map M))   =  {\rm Choi}  (\Gamma_S  ( \map M)) \, .
  \end{align} 
  The map $\Gamma_S$ is CP-preserving if and only if the  map $\widehat \Gamma_S$  is completely positive  \cite{chiribella2008transforming}.

\begin{prop}\label{prop:trans}
Let $\Gamma_S^{\rm trans}  :  \map M\mapsto  \map M^T$  be the transpose supermap, and let $\widehat \Gamma_S^{\rm trans}$ be its Choi map, defined as in Equation (\ref{choimap}).   The Choi map  has the form 
\begin{align}\label{swap}
\widehat \Gamma_S^{\rm trans}   (X)    =    {\tt SWAP} \, X \,  {\tt SWAP}^\dag   \qquad \forall X \in  L(\spc H_S\otimes \spc H_S) \, ,
\end{align}
where ${\tt SWAP}$ is the swap operator, defined by the condition ${\tt SWAP}  (|\phi\>\otimes  |\psi\>  )  =   |\psi\>\otimes |\phi\>$, $\forall |\phi\>,  |\psi\> \in \spc H_S$. 
\end{prop}

\Proof 
For an arbitrary completely positive map $\map C  (\cdot)  = \sum_i  C_i\cdot C_i^\dag$, one has 
\begin{align}
\nonumber {\rm Choi}  ( \map M)    &  =  \sum_i  (C_i\otimes I) |  I\kk \bb   I|   (C_i^\dag\otimes I)  \\
 \nonumber &   =  \sum_i  (I\otimes C_i^T) |  I\kk \bb   I|   (I\otimes \overline C_i)  \\  
 \nonumber &  =  (\map I \otimes \map M^T)  (|I\kk \bb I|)  \\
 \nonumber  &   =  {\tt SWAP} \,   {\rm Choi}  (\map M^T)  \, {\tt SWAP} \\
 \nonumber
 &  =  {\tt SWAP} \, {\rm Choi}  ( \Gamma^{\rm trans}_S  (\map M) )  \,  {\tt SWAP} \\
 &  =  {\tt SWAP} \,   \widehat \Gamma^{\rm trans}_S  ({\rm Choi} (\map M))  \,  {\tt SWAP}  \, ,\label{aa}
\end{align} 
where the second equality follows from  the property  $|A\kk   =  (A\otimes I)  |I \kk  =  (I \otimes A^T)|I\kk$, while the last equality follows from Equation (\ref{choimap}).  

Applying the  ${\tt SWAP}$  on both sides of  Equation (\ref{aa}), we obtain the relation 
\begin{align}
\widehat \Gamma_S^{\rm trans}  ({\rm Choi}  ( \map M))  =  {\tt SWAP} \, {\rm Choi}  ( \map M) \,{\tt SWAP} \,.
\end{align} 
Hence, Equation (\ref{swap}) holds whenever $X$ is the Choi operator of a completely positive map.  Since every positive semidefinite operator is the Choi operator of a completely positive map, and since every operator is a linear combination of positive semidefinite operators, Equation (\ref{swap}) holds for every operator $X$. 
 \qed   
    
CP-preservation of the transpose supermap  is then immediate:  

\begin{cor}
The transpose supermap $\Gamma_S^{\rm trans}  :   \map M\mapsto  \map M^T$  is CP-preserving.   
\end{cor}
\Proof Since the map $\widehat \Gamma_S^{\rm trans}$  has a Kraus representation, it follows that it is completely positive.  Since  $\widehat \Gamma_S^{\rm trans}$ is completely positive, $\Gamma_S^{\rm trans}$
 is CP preserving. \qed  
  
We now show that the adjoint supermap $\Gamma_S^{\rm adj}  :    \map M\mapsto  \map M^\dag$ is not CP-preserving for every dimension $d_S  >1$.    

\begin{prop}
Let $\Gamma_S^{\rm adj}  :     \map M\mapsto  \map M^\dag$  be the adjoint supermap, and let $\widehat \Gamma_S^{\rm adj}$ be its Choi map, defined as in Equation (\ref{choimap}).   The Choi map  has the form 
\begin{align}\label{swaptrans}
\widehat \Gamma_S^{\rm adj}   (X)    =    {\tt SWAP} \, X^T \,  {\tt SWAP}^\dag   \qquad \forall X \in  L(\spc H_S\otimes \spc H_S) \, .
\end{align}
\end{prop}

\Proof  The proof has the same structure of the proof of Proposition \ref{prop:trans}.  For an arbitrary completely positive map $\map C  (\cdot)  = \sum_i  C_i\cdot C_i^\dag$, one has 
\begin{align}
\nonumber {\rm Choi}  ( \map M)    &  =  \sum_i  (C_i\otimes I) |  I\kk \bb   I|   (C_i^\dag\otimes I)  \\
 \nonumber &   =  \sum_i  (I\otimes C_i^T) |  I\kk \bb   I|   (I\otimes \overline C_i)  \\  
 \nonumber &   =      \left[  \sum_i  (I\otimes C_i^\dag) |  I\kk \bb   I|   (I\otimes C_i) \right]^T \\  
 \nonumber &  =  \big[  (\map I \otimes \map M^\dag)  (|I\kk \bb I|)\big]^T  \\
 \nonumber  &   =\big[  {\tt SWAP} \,   {\rm Choi}  (\map M^\dag)  \, {\tt SWAP} \big]^T\\
 \nonumber
 &  =  \big[  {\tt SWAP} \, {\rm Choi}  ( \Gamma^{\rm adj}_S  (\map M) )  \,  {\tt SWAP} \big]^T\\
 &  = \big[ {\tt SWAP} \,   \widehat \Gamma^{\rm adj}_S  ({\rm Choi} (\map M))  \,  {\tt SWAP}\big]^T  \, ,\label{bb}
\end{align} 
where the second equality follows from  the property  $|A\kk   =  (A\otimes I)  |I \kk  =  (I \otimes A^T)|I\kk$, while the last equality follows from Equation (\ref{choimap}).

Equation (\ref{bb}) implies 
\begin{align}
\widehat \Gamma_S^{\rm adj}  ({\rm Choi}  ( \map M))  =  {\tt SWAP} \, {\rm Choi}  ( \map M)^T \,{\tt SWAP} \,,
\end{align} 
which in turn implies Equation (\ref{swaptrans}). \qed

\begin{cor}
The adjoint supermap $\Gamma_S^{\rm adj}  :     \map M\mapsto  \map M^\dag$  is not CP-preserving. 
\end{cor}
\Proof   The Choi map   $\widehat \Gamma^{\rm adj}_S$ is unitarily equivalent to the transpose, which is known to be positive but not completely positive.  Since $\widehat \Gamma_S^{\rm adj}$ is not completely positive, $ \Gamma_S^{\rm adj}$ is not CP-preserving. 
\qed

We have shown that the adjoint supermap is not CP-preserving. We conclude with a strengthening of this result, showing that, in dimension $d_S>2$, no CP-preserving supermap can act as the adjoint supermap on the set of bistochastic channels. In other words, every extension of the adjoint supermap from the set of  bistochastic maps to the set of all maps will necessarily be non-CP-preserving. In dimension $d_S  =  2$, instead, the adjoint and the transpose are unitarily equivalent on bistochastic channels, and therefore the adjoint has a positive extension to the set of all maps.

\begin{prop}
Let $\Theta_S^{\rm adj}  :    \SpanBiChan (\spc H_S)  \to \SpanBiChan (\spc H_S)   \, , \map C \mapsto  \map C^\dag$ be the adjoint supermap on the space of bistochastic channels.  
Then,  
\begin{enumerate}
\item for $d_S  = 2$, $\Theta_S^{\rm adj}$ has a unique CP-preserving extension $\Gamma_S^{\rm adj}  :    \Map (\spc H_S)  \to \Map (\spc H_S)$, defined by the relation 
\begin{align} 
\widehat  \Gamma_S^{\rm adj}    (X)  :=  G  X G^\dag  \qquad  \forall X  \in  L(\spc H_S \otimes \spc H_S)     
\end{align} with
\begin{align}
G =    I\otimes I  -  2  |\Phi^+\>\<\Phi^+| \, ,   
\end{align}  
where $|\Phi^+\>   :=  |I\kk/\sqrt 2$ is the canonical maximally entangled state. 
 \item for $d_S  >2$, $\Theta_S^{\rm adj}$ has no  CP-preserving extension.
\end{enumerate}
\end{prop}

 \Proof   Let us start from the $d_S=2$  case.   Let $  U $ be a generic element of $\grp{SU}  (2)$, parametrised as $U =  \cos (\theta/2)  \, I  -  i \sin (\theta/2)  \, \st n\cdot \bs \sigma$, where $\theta \in  [0,2\pi)$ is the rotation angle, $\st n  =  (n_x,n_y,n_z)  \in \R^3$ is the rotation axis, and $\st n\cdot \bs \sigma  =  n_x  \sigma_x  +  n_y \sigma_y + n_z\sigma_z$, $\{\sigma_k\}_{k\in  \{x,y,z\}}$ being the three Pauli matrices.  

Then, one has 
\begin{align}
\nonumber G |U\kk  &=  \cos (\theta/2)  \,  G| I \kk   -  i \sin (\theta/2)  \, G |\st n\cdot \bs \sigma  \kk\\
\nonumber &=  - \cos (\theta/2)  \,  | I \kk   -  i \sin (\theta/2)  \,  |\st n\cdot \bs \sigma  \kk\\
&  =  -  |U^\dag\kk  \, .
\end{align}
Hence, we obtained the relation 
\begin{align}
\widehat  \Gamma_S^{\rm adj}  (|U\kk\bb U|)    =   |U^\dag\kk \bb U^\dag|  \, ,   
\end{align}
and, in turn, the relation
\begin{align}
\Gamma_S^{\rm adj}  (\map U)    = \map U^\dag  =  \Theta_S^{\rm adj}  (\map U)\, ,   
\end{align}
valid for every unitary channel $\map U$.  Since the unitary channels span the space of bistochastic maps  (Theorem 2 in the main text),  the above relation implies 
\begin{align}
\Gamma_S^{\rm adj}  (\map M)  = \Theta_S^{\rm adj}  (\map M) \qquad \forall \map M \in \SpanBiChan (\spc H_S)  \, .   
\end{align}
In summary, we proved that $\Gamma_S^{\rm adj}$ is an extension of  $\Theta_S^{\rm adj}$, and  is CP-preserving (because $\widehat \Gamma_S^{\rm adj}$ has a Kraus representation).  Note that   $\Gamma_S^{\rm adj}$ does not act as the adjoint outside the space of bistochastic maps.

Let consider now the $d_S  > 2$ case. We now prove by contradiction that the map $\Theta_S^{\rm adj}$ admits no CP-preserving extension  $\Theta_S^{\rm adj}$. To get the contradiction, we suppose that the CP-preserving extension exist, and we let  $\widehat  \Gamma_S$ be the  corresponding completely positive map. 
Since  $\widehat \Gamma_S$ is completely positive, it has a Kraus representation 
  \begin{align}\label{superkraus}
  \Gamma_S  (C)  =  \sum_j  G_j  C G_j^\dag 
  \end{align}
  for some suitable set of operators $\{G_j\}$.  

The condition that  $\Gamma_S$  extends  $\Theta_S$ can be expressed as
\begin{align}\label{unitaryact}
\widehat  \Gamma_S(|U\kk\bb U|)  = |U^\dag\kk\bb U^\dag|)  \qquad \forall U \in \grp {SU} (d_S) \, .
\end{align}

Combining Equations (\ref{superkraus}) and  (\ref{unitaryact}),  we obtain the condition  
 \begin{align}\label{superkraus2}
  \sum_j  G_j  |U\kk \bb U|  G_j^\dag  =  |U^\dag\kk \bb  U^\dag |  \qquad \forall U \in \grp {SU} (d_S) \, .
  \end{align}
Since the left-hand side is a positive semidefinite operator, and the right-hand side is a rank-one operator, we must have   
\begin{align}\label{basta}
G_j  |U\kk  =  c_{j,U}  \,  |U^\dag\kk  \, , \qquad \forall j, \, \forall U \in \grp{SU}  (d_S)\, ,
\end{align}
for some suitable coefficients $\{c_{j,U}\}$ satisfying $\sum_j  |c_{j,U}|^2  = 1$ for every $U$.

We now show that it is impossible to satisfy condition  (\ref{basta})   in dimension $d_S  > 2$.  For simplicity, we will first illustrate the argument in dimension $d_S  =  3$.  
Consider the  unitary gates 
$U_{1\pm}    =  |1\>\<1|   \pm   i   |2\>\<2|    \mp i |3\>\<3|$. 
In this case, Equation (\ref{basta}) becomes  
\begin{align}
G_j  |  U_{1 \pm}\kk  =  c_{j,1 \pm} \, |U_{1, \mp}\kk 
\end{align} 
for some coefficients $c_{j,1\pm}$.   

Note that the operator $  V_1  =  (  U_{1+}   +  i    \,U_{1-} )/\sqrt 2$ is also  unitary, and therefore one must have $G_j  |V_1\kk  =   c_{j, 1}  \,  |V_1^\dag\kk$ for some coefficient $c_{j,1}$. 

Combining the above relations, one obtains  
\begin{align}
\nonumber \frac {c_{j,1+}    |U_{ 1-} \kk    + i  \,  c_{j,1-} \, | U_{1+}  \kk  }{\sqrt 2}   &  =  G_j     |V_1\kk  \\
 \nonumber  &  =  c_{j,1}  \, |V_1^\dag\kk\\
  &  = c_{j,1}  \, \frac {    |U_{ 1-} \kk    - i  \, | U_{1+}  \kk  }{\sqrt 2}   \, ,
\end{align}
from which we obtain $c_{j,1\pm}  =    \pm c_{j,  1}$.

Now, note the relation 
\begin{align}
\nonumber G_j  \,  |1\> |1\>   &  =  G_j  \frac {  | U_{1+}  \kk +  | U_{1-} \kk}2  \\
 \nonumber  &  =  c_{j,1}   \,  \frac { | U_{1-}\kk  - | U_{1+}\kk}2  \\
  &  = - i \,  c_{j,1}  \,   ( |2\>|2\> -|3\>|3\>) \, .   
\end{align}

Similarly, one can define the unitaries $  U_{2 \pm} :  =  |2\>\<2|   \pm  i  |3\>\<3|  \mp i |1\>\<1| $ and $U_{3\pm}  :  =  |3\>\<3|   \pm   i |1\>\<1|  \mp  i  |2\>\<2|$, and prove the relations  
\begin{align}
\nonumber 
G_j  \,  |2\> |2\>  &= -i \,  c_{j,2}  \,   ( |3\>|3\> -|1\>|1\>)\\
G_j  \,  |3\> |3\>  &= -i \,  c_{j,3}  \,   ( |1\>|1\> -|2\>|2\>) \, .   
\end{align}
All together, these relations imply
\begin{align}
\nonumber G_j  \, |I\kk  &  =  G_{j} \,  (  |1\>|1\>  +|2\>|2\>  +|3\>|3\> )\\
 \nonumber &  = i   (c_{j,1}-  c_{j,2})\,  |3\>|3\>\\
 \nonumber    &\quad +  i   (c_{j,2}-  c_{j,3})\,  |1\>|1\> \\  
   &\quad +  i   (c_{j,3}-  c_{j,1})\,  |2\>|2\>   
\end{align} 
On the other hand, one must have 
\begin{align}
\nonumber G_j  |I\kk    &  =  c_{j,  I}   |I\kk\\
  &  =c_{j,  I}   \,  (  |1\>|1\>  +|2\>|2\>  +|3\>|3\> )    \, ,
  \end{align}
  for some coefficient $c_{j,I}$.    Comparing the above equations, we obtain the relation $c_{j, I}  =  i(c_{j,1}-  c_{j,2})   = i(c_{j,2}-  c_{j,3})  = i(c_{j,3}-  c_{j,1})$, which implies $c_{j,1}=c_{j,2}=  c_{j,3}$ and $c_{j,I}  =  0$.  
  
Since this relation should hold for every $j$, we obtained the contradiction 
\begin{align}
\nonumber |I\kk \bb I|    &=  \widehat \Theta_S  (  |I\kk  \bb I|  ) \\
\nonumber &    =\widehat \Gamma_S  (|I\kk\bb I|)   \\
\nonumber & =  \sum_j  \,  G_j |I\kk \bb  I|G_j^\dag   \\
& =  0 \, .
\end{align}  
For $d_S  \ge 3$, the same argument can be applied to the unitary gates $U_{mn \pm}   :  =    (I-|m\>\<m|  -  |n\>\<n|)   \pm   i |m\>\<m|  \mp   i |n\>\<n|$, obtaining the relations 
\begin{align}
G_j  |U_{mn\pm}  \>    = \pm   c_{j, mn}  \,  |U_{mn  \mp}\kk \ 
\end{align} 
and 
\begin{align}
\nonumber G_j     (  |I  \kk  -   |m\>|m\>  -  |n\>|n\>)  &  =  G_j  \frac{  |U_{mn+} \kk +  |U_{mn-}\kk}2  \\
  \nonumber & = c_{j,mn}  \,  \frac{ |U_{mn-} \kk   -  |U_{mn  +}\kk }2\\
  &  = -i  c_{j,mn}   \,   ( |m\>|m\>  -  |n\>|n\>)\, .  
\end{align}
To conclude the proof, note that one has 
\begin{align}
\nonumber \bb I | G_j  |I\kk  &  =\bb I|   G_j \frac {\sum_{l=1}^d  |I \kk  -  |l\>|l\>  -  |l\oplus 1  \>  |l\oplus 1\>  }{d-2}   \\
\nonumber &   =        \frac {\sum_{l=1}^d    -  i\,   c_{j,  l \,l+1}   \,  \bb I| \, ( |l\>|l\>  -  |l\oplus 1\>|l\oplus1\>)  }{d-2} \\
& =  0  \, ,
\end{align}
where $\oplus$ denotes the addition modulo $d$, and we define $|0\>:  =  |d\>$.     Hence, we obtained the contradiction 
\begin{align}
\nonumber d_S^2     &=  |  \bb I  |  I  \kk|^2 \\
 \nonumber  &  = \bb I|  \, \widehat \Gamma_S^{\rm adj}  (|I   \kk \bb I|  )  \,  |I  \kk  \\
 \nonumber  &  =    \sum_j  |\bb  I  |  G_j   |I  \kk |^2  \\
  &=  0  \, .     
\end{align} \qed 
\section{Proof of Theorem 2 in the main text
}\label{app:allchannels}
Here we prove Theorem 2
of the main text: for every channel admitting a input-output inversion satisfying Requirements (1)-(4), the input-output inversion is bistochastic.    
\subsection{The space of bistochastic channels}

We start by providing a few properties of the subset of bistochastic channels.  The set   $\BiChan(\spc H)  \subset \Chan  (\spc H)$ can be equivalently be defined as the set of linear maps $\map B $  that are completely positive and satisfy the conditions $\map B(I)   = I$ and $\map B^T  (I)   =  I$, where  $\map B^T$ is the transpose of the map $\map B$, defined in Eq. (\ref{transpmap}). 

The following proposition provides a characterization of the linear  space spanned by the bistochastic channels.

\begin{prop}\label{prop:spanbichan}
The space of bistochastic maps  $\SpanBiChan (\spc H)$   consists of maps $\map M  \in  \Map (\spc H)$ with the property that the operators $\map M (I)$ and $ \map M^T  (I)$ are both proportional to the identity.  
\end{prop}
\Proof Let $\map M  $  be a generic element of $\SpanBiChan (\spc H)$, written as $\map M  = \sum_i  \, c_i \,  \map B_i$ where $\{c_i\}$  are complex numbers and $\{\map B_i\}$ are bistochastic channels.  By definition, one has $\map B_i(  I)   =  \map B_i^T  (I)   =  I$ for every $i$. Hence, one has $\map M  (I)     =  \map M^T  (I)   = ( \sum_i  c_i  )\,     I$.

Conversely, suppose that a map $\map M$ is such that $\map M(I)$ and $\map M^T  (I)$ are both proportional to the identity.  Note that, since $\Tr  [   \map M (I)]    =  \Tr[\map M^T (I)]$, the proportionality constant should be the same. Hence, we can write $\map M (I)    =  \map M^T (I)   =  c\, I$ for some $c\in  \C$.   

Now, $\map M$ can be decomposed as $\map M  =  \map A    +  i  \,  \map B$, with $\map A:  =  ( \map M   +  \widetilde {\map M} )/2 $ and $\map B : =    ( \map M   -  \widetilde{ \map M} )/{2i}$, where $ \widetilde{ \map M} $ is the map defined by  $ \widetilde{ \map M}    (  X)   =       [ \map M (X^\dag)]^\dag$, for a generic operator $X\in L(\spc H)$.  Note that one has $\map A  (I)   = \map A^T  (I)   = a  \,  I$ and $\map B  (I)   = \map B^T  (I)   = b  \,  I$ where $a$ and $b$ are the real and imaginary part of $c$, respectively. 
 
We now show that $\map A$ and $\map B$ are linear combinations of bistochastic channels.    The maps $\map A$ and $\map B$ are Hermitian-preserving, meaning that they map Hermitian operators into Hermitian operators. 
 Consider a generic  Hermitian-preserving map $\map G  \not  =  0 $ satisfying the conditions $  \map G  (I)   =   \map G^T  (I)   =  g   \,  I$ for some real number $g \in \R$.  In the following we will show that $\map G$ is a linear combination of bistochastic channels.  

{  Let $G$ be the Choi operator of the map $\map G$. In the Choi representation, the condition that the map $\map G$ be Hermitian-preserving is equivalent to the condition that the operator $G$ be Hermitian. Hence, $G$ has real eigenvalues and can be decomposed into a positive part and a negative part, namely 
\begin{align}\label{Gpm}
G=  G_+  -  G_- \, ,  
\end{align} with $G_+\ge 0$ and $G_-\ge 0$.   

The conditions $  \map G  (I)   =   \map G^T  (I)   =  g   \,  I$  become  $\Tr_{o}  [G]=  g \,  I_i$ and $\Tr_i  [  G]   =  g\,  I_o$, where the subscripts $i$ and $o$ refer to the input and output of the map $\map G$, respectively.  In turn, these conditions are equivalent to 
\begin{align}
\Tr_{o}  [G_+]   &=  \Tr_o [G_-]  +  g\,  I_i   \label{summeG1} \\
\Tr_{i}  [G_+]   &=  \Tr_i [G_-]  +  g\,  I_o \, .   
 \label{summeG2}
\end{align}  
Equation~( \ref{summeG1}) implies that the operators $\Tr_o  [G_+]$ and $\Tr_o[G_-]$ commute, and therefore are  diagonal in the same basis.  Also, the equation implies that the maximum eigenvalues of $\Tr_o[G_+]$ and $\Tr_o[G_-]$, denoted by $\gamma_{i,+}$ and $\gamma_{i,-}$, respectively, satisfy the condition  
\begin{align}
\label{gammai}\gamma_{i,+}   =\gamma_{i,-}   +  g \, .  
\end{align}  
 Similarly, Equation~(\ref{summeG2})  implies that  the maximum eigenvalues 
 of the operators $\Tr_i [G_\pm]$, denoted by $  \gamma_{o,\pm}$,
 satisfy the condition  
 \begin{align}
\label{gammao} \gamma_{o,+}   &=\gamma_{o,-}   +  g \, ,  
\end{align}  

Now, define the  operators 
\begin{align}
\nonumber G_\pm'     :   =    &G_\pm    +   \frac {I_i}{d}  \otimes   (  \gamma_{o,\pm}   \,  I_o  -  \Tr_i  [  G_\pm]  )   \\
 &  ~\,\quad +     (  \gamma_{i,\pm}   \,  I_i  -  \Tr_o  [  G_\pm]  )  \otimes  \frac{I_o}{d} \, . 
\end{align}  
 Note that the operators $G_\pm'$ are positive by construction.  Moreover, they satisfy the condition $G_+'  -  G_-' =  G$.   Indeed, one has 
 \begin{align}\nonumber
 G_+'  -  G_-'  =   &  G_+  -  G_-   \\
 \nonumber &   +      \frac {I_i}{d}  \otimes   (  \gamma_{o,+}   -  \gamma_{o,-}  )     \,  I_o \\
\nonumber &   -      \frac {I_i}{d}  \otimes   (  \Tr_i[G_+] - \Tr_i[G_-]  )      \\  
 \nonumber &   +         (  \gamma_{i,+}   -  \gamma_{i,-}  )     \,  I_i \otimes \frac {I_o}{d}   \\
\nonumber &   -   (  \Tr_i[G_+] - \Tr_i[G_-]  )     \otimes     \frac {I_i}{d} \\
= &  G \, , 
 \end{align}
 where the second equality follows from Eqs.~(\ref{Gpm}),~(\ref{summeG1}),~(\ref{summeG2}),~(\ref{gammai}), and~~(\ref{gammao}).  Finally, note that $G_\pm$ are proportional to the Choi operators of two bistochastic channels: indeed,  one has 
 \begin{align}
\nonumber \Tr_o[G_\pm' ]         =    &\Tr_o[G_\pm]    +   \frac {I_i}{d}    (  \gamma_{o,\pm}   \, d  -  \Tr  [  G_\pm]  )   \\
\nonumber  &  ~\,\quad +       \gamma_{i,\pm}   \,  I_i  -  \Tr_o  [  G_\pm] \\
 &  =  \left( \gamma_{o,\pm}   +  \gamma_{i,\pm} - \frac{\Tr[G_\pm]}d \right) I_i    
 \end{align}
 and  
  \begin{align}
\nonumber \Tr_i[G_\pm' ]         =    &\Tr_i[G_\pm]    +     \gamma_{o,\pm}   \, I_o -  \Tr_i  [  G_\pm]      \\
\nonumber  &  ~\,\quad +       (  \gamma_{i,\pm}   \, d  -  \Tr  [  G_\pm]  )     \otimes \frac {I_o}{d}  
    \\
 &  =  \left( \gamma_{i,\pm}   +  \gamma_{o,\pm} - \frac{\Tr[G_\pm]}d  \right) I_o    
 \end{align}
In conclusion, we have shown that the Choi operator of the map $\map G$ can be decomposed as $G  =  G_+ - G_-$, where $G_\pm$ are proportional to the Choi operators of two bistochastic channels.  Hence, $\map G$ is a linear combination of bistochastic channels.  \qed   }

\subsection{Projection on the space of bistochastic  channels}

Here we define a supermap   that projects   onto the space of bistochastic maps.

\begin{defi}
A projection on the space of bistochastic maps is a linear supermap $\Pi:  \Map  ( \spc H ) \to \Map (\spc H) $  satisfying the conditions $\Pi  (\map M)  \in  \SpanBiChan  (\spc H)$,  $\forall \map M\in \Map (\spc H)$,     and $\Pi  (\map B)  =  \map B$, $\forall \map B  \in  \SpanBiChan (\spc H)$. 
\end{defi}

A projection  on the space of bistochastic channels can be constructed as follows:  
\begin{prop}
The supermap $\Pi:  \Map  ( \spc H ) \to \Map (\spc H) $  defined by 
\begin{align}
\nonumber \Pi  (\map M)  (\rho)     &:=   \map M (\rho)     +    2     \Tr \left[\map M \left( I \right)\right]  ~  \frac{\Tr[\rho]}d\,   \frac   I d   \,   \\
 \nonumber  &    \quad -   \map M \left( I \right)   \,  \frac{\Tr[\rho] }d  \\
 \label{Pi} &\quad  -  { \Tr \left[\map M (\rho)\right]} \,  \frac I d  \, .             
\end{align}
is a projection on the space of bistochastic maps. 
\end{prop}

\Proof   First, let us show that $\Pi  (\map M)$ belongs to $\SpanBiChan  (\spc H)$.   Thanks to Proposition \ref{prop:spanbichan}, we just need to check that $   \Pi  (\map M)  (I) $ and $[\Pi  (\map M)]^T  (I)$ are proportional to the identity.  
Indeed, one has  
\begin{align}
\nonumber \Pi  (\map M)  (  I)     &=   \map M (I)     +    2     \Tr \left[\map M \left( I \right)\right]  ~   \frac I d  \\
 \nonumber  &    \quad -   \map M \left( I\right)   \,   \\
 \nonumber  &\quad  -   \,  \Tr \left[\map M  \left( I    \right) \right] \frac I d  \\
  & =  \Tr \left[\map M \left( I \right)\right]  ~   \frac I d \, ,             
\end{align}
and, for every $X  \in L(\spc H)$,  
\begin{align}
\nonumber \Tr\Big\{ [ \Pi  (   \map M )]^T  (I)    \,  X  \Big\} 
 &  =  \Tr[  I^T  \,       \Pi  (\map M)  (X^T) ]\\
\nonumber  &   =  \Tr[ \map M (X^T)]     +    2     \Tr \left[\map M \left( I \right)\right]  ~  \frac{\Tr[X^T]}d\,     \\
 \nonumber  &    \quad -   \Tr[\map M \left( I \right) ]  \,  \frac{\Tr[X^T] }d  \\
 \nonumber  &\quad  -  { \Tr \left[\map M  \left(X^T \right)  \right]}  \\
  \nonumber  &   =  
    \Tr \left[\map M \left( I \right)\right]  ~  \frac{\Tr[X]}d\,     \\
  &   =        \Tr \left\{\,   \Big [   \Tr\left[\map M \left( I \right)\right]    \frac{ I}d\,    \Big  ]\,  X\right\}   \, ,
\end{align} 
 which implies 
 \begin{align} [ \Pi  (   \map M )]^T  (I)     = \Tr\left[\map M \left( I \right)\right]    \frac{ I}d \,.
 \end{align} 
Then, it remains to show that $\Pi$ maps every  element of $ \SpanBiChan  (\spc H)$ into itself.   Recall that a generic element  $\map B  \in \SpanBiChan  (\spc H)$ satisfies the condition $\map B(I)   = \map B^T  (I)  = b\, I$, for some $b\in  \C$.   { The second condition implies 
\begin{align}
\nonumber \Tr[  \map B  (\rho)]  & =  \Tr[  I^T  \,  \map B(\rho)]  \\
\nonumber &  =  \Tr[ ( \map B^T (I))^T  \rho  ]  \\
\nonumber &  =  \Tr [  (b\,  I)^T  \, \rho  ]\\
&  =  b \, \Tr[\rho]\,.
\end{align}
Hence, one has  
\begin{align}
\nonumber \Pi  (\map B)  (\rho)     &:=   \map B (\rho)     +    2    b   ~  \frac{\Tr[\rho]}d\,    I    \,   \\
 \nonumber  &    \quad -   b   I \,\frac{ \Tr[\rho] }d  \\
 \nonumber &\quad  -   b\,   \Tr \left[ \rho\right] \,  \frac I d \\
   &  =  \map B(\rho)   \, .             
\end{align}}
Hence, $\Pi$ is a projection on the space of bistochastic maps.   \qed

\subsection{Decomposition of arbitrary quantum channels}  

We now show that any arbitrary quantum channel $\map C$ can be decomposed as a linear combination of its projection on the space of bistochastic maps and of  two constant channels.  

Specifically, we prove the following proposition:  
 \begin{prop}\label{prop:decomp}
Let $\map M \in  \Map (\spc H) $ be a generic linear map.  Then, $\map M$ can be decomposed as 
\begin{align}\label{decomp1}
\nonumber \map M  (\rho)    &=    \Pi  (\map M)  (\rho)    \\    
\nonumber   &  \quad  +      \frac 1d  \, \left(     \map   M(I)    -  \Tr[\map M(I)] \, \frac I d    \right)   \,  \Tr [\rho]  \\      
&  \quad +  \frac Id      \,  \Tr  \left[ \left(     \map   M^T(I)    -  \Tr[\map M^T(I)] \, \frac I d    \right)   \, \rho^T   \right] \, . 
\end{align}
where $\Pi $ is the projection defined in Equation (\ref{Pi}). 
    In particular, when $\map M$ is trace-preserving, the decomposition takes the simpler form   
   \begin{align}\label{decomp3}
\map M   = \Pi  ( \map M)     +     \map K_{  \map C(I/d)}   -  \map K_{I/d} \, ,
\end{align}
where   $\Pi$ is the projection defined in Equation (\ref{Pi}), and, for every density matrix $\rho_0$,  $\map K_{\rho_0}$ is the constant channel defined by 
\begin{align}
\map K_{\rho_0}  (\rho) :=  \rho_0  \Tr [  \rho] \, ,\qquad \forall \rho  \in L(\spc H) \, .
\end{align}
\end{prop}
 
 \Proof    Equation (\ref{decomp1}) follows immediately from the definition of $\Pi$ in Equation (\ref{Pi}).  
 When $\map M$ is trace-preserving, one has  $\Tr[\map M(I)]   =d$ and $\map M^T  (I)  =  I$.   Using  these two relations, the decomposition (\ref{decomp1}) reduces to (\ref{decomp3}).      \qed 
 
 

\subsection{Input-output inversion of constant channels}
  
  Here we show that  Requirements 1-4 in the main text imply that the input-output inversion of every constant channel is the completely depolarizing channel.  This result implies in particular that the input-output inversion cannot be a one-to-one map on the  set of all quantum channels.

 \begin{prop}\label{prop:noninvertible}
Let $\Theta$ be an arbitrary input-output inversion satisfying Requirements 1-4 in the main text, and let $\map K_{\rho_0} \in \sf B$ be an arbitrary constant channel, defined by the condition $\map  K_{\rho_0} (\rho)   =  \rho_0   \forall \rho \in L(\spc H)$ where $\rho_0 \in L(\spc H)$ is a fixed (but otherwise arbitrary) density matrix.  Then, one has  $\Theta  (\map K_{\rho_0})   =   \map K_{I/d} $ for every   $\rho_0$.  
\end{prop}

The proof uses Lemma \ref{lem:unitarypreserving} and the following

\begin{lemma}\label{lem:depolarizing}
Let $\Theta$ be a input-output inversion satisfying Requirements  1-4 in the main text.  Every input-output inversion $\Theta$ leaves the completely depolarizing channel  invariant.
\end{lemma}

\Proof The completely depolarizing channel $\map C_{I/d}$ can be expressed as a random mixture of $d^2$ unitary channels: for every set of  unitary operators  $\{  U_i\}_{i=1}^{D^2}$ satisfying the Hilbert-Schmidt orthogonality condition $\Tr [  U_i^\dag U_j ]   =  d  \,  \delta_{i,j}$, one has $\map C_{I/d}   =\sum_i  \map U_i  / d^2$. 

Requirement 4  (preservation of random mixtures) implies that the input-output inversion of $\map C_{I/d}$ can be expressed as $\Theta  (\map C_{I/d})   = \sum_i   \Theta (\map U_i)/d^2$. 

Then, discussion after the proof of Lemma \ref{lem:unitarypreserving} shows that the input-output inversion of each unitary channel $\map U_i$ can be expressed as $\Theta  (\map U_i)  (\cdot)     =     \theta (U_i)  (\cdot) \theta  (U_i)^\dag$, where  $\theta$ is an input-output inversion on the special unitary group.  

Under Requirements 1-3, we know that   the input-output inversion on the special unitary group is either the adjoint or the transpose (Lemma~\ref{lem:unitary}), we conclude that one has either $\Theta  (\map U_i)  (\cdot)     =     U_i^\dag  (\cdot)  U_i$ or $\Theta  (\map U_i)  (\cdot)     =     U_i^T  (\cdot)  \overline U_i$. In either case, the set $\{ \theta  (U_i)\}$ satisfies the orthogonality condition $\Tr [  \theta  (U_i)^\dag  U_j]  =  d \, \delta_{i,j}$.  Hence,  we obtain the equality $\Theta  (\map C_{I/d})   = \sum_i   \Theta (\map U_i)/d^2  =  \map C_{I/d}$. \qed

\medskip  

{\bf Proof of Proposition \ref{prop:noninvertible}. }  For every constant channel $\map C_{\rho_0}$ one has the relation $\map  C_{\rho_0}   =  \map C_{\rho_0}  \map C_{I/d}$.  Hence, the input-output inversion satisfies the condition $ \Theta  (\map C_{\rho_0})   =   \Theta  (\map C_{I/d})  \Theta  (\map C_{\rho_0})    =    \map C_{I/d}  \,   \Theta  (\map C_{\rho_0}) =   \map C_{I/d} $, the second equality following from Lemma \ref{lem:depolarizing} \qed

\smallskip  {  An immediate consequence of Proposition \ref{prop:noninvertible} is that the set of bidirectional quantum channels can only contain one constant channel, namely the completely depolarising channel. In other words, the set of bidirectional quantum channels contains only one  way to re-set the system to a fixed state.}

\subsection{The input-output inversion of a generic quantum channel is a bistochastic quantum channel}   

Here we prove Theorem 2 in the main text:  if a quantum channel admits an input-output inversion, then its input-output inversion is a bistochastic channel.   

\begin{prop}\label{prop:bisto}
Let $\Theta$ be a input-output inversion satisfying Requirements 1-4  in the main text, and let $\map C\in \sf B$ be a generic bidirectional channel, decomposed as in Equation (\ref{decomp3}). Then, one has $\Theta (\map C)   = \Theta  (\Pi  (\map C))$.  
\end{prop}

\Proof  The decomposition (\ref{decomp3}) implies  $\Theta  (\map C)  =  \Theta  (\Pi  (\map C)+   \map K_{  \map C(I)/d}   -  \map K_{I/d}  )$.   By the linearity of the input-output inversion  (Proposition \ref{prop:linear}),  we obtain the condition 
 $\Theta  (\map C)  =  \Theta  (\Pi  (\map C))      +  \Theta  (   \map K_{  \map C(I)/d} )  -\Theta  (  \map K_{I/d}  )$.   Since $\Theta$ maps all constant channels into the completely depolarizing channel (Proposition \ref{prop:noninvertible}), the last two terms coincide, and one has $\Theta (\map C)  =  \Theta  (\Pi  (\map C))$. \qed 
  
\medskip 

In summary, we have shown that the input-output inversion of a given quantum channel depends only on its projection on the space of bistochastic channels.  In particular, this implies Theorem 2 in the main text:  



\medskip 

{\bf Proof  of Theorem 2 in the main text.}     By Proposition \ref{prop:bisto}, one has $\Theta (\map C)   =  \Theta  (\Pi (\map C))$.  Since the unitary channels are a spanning set for the space of bistochastic maps, one has $\Pi  (\map C)  = \sum_i   c_i\,   \map U_i$, for suitable coefficients $\{c_i\}$ and suitable unitary channels  $\{ \map  U_i\}$.     By linearity of the input-output inversion, we have $\Theta  (\map C )=  \sum_i  c_i\,  \Theta (\map U_i)$.  
  Since $\Theta$ maps unitary channels into unitary channels,  the channel $\Theta (\map C)$ is a linear combination of unitary channels.  But we know that any such channel is bistochastic (Theorem 1 of the main text). \qed 


\section{Relaxing the notion of input-output inversion
}\label{app:nobij}  

Here we explore a relaxed notion of input-output inversion, which is not required to be  injective.   We ask whether, under such relaxation, it is possible to define a  non-trivial input-output inversion for every quantum channel.  As it turns out, the answer is negative in dimension $d>2$, and affirmative for $d=2$.    

\subsection{Positivity of the input-output inversion supermap} 

Let $\Theta:  \Chan (\spc H) \to \Chan (\spc H)$  be a  supermap defined on all channels and satisfying Requirements 1 (order reversal), 2 (identity presevation), and 4 (compatibility with random mixtures) in the main text.  A trivial choice for the map $\Theta $ is to transform every channel $\map C$ into the identity channel $\map I$. The question is whether any other nontrivial choice exists.  

Here we show that, for every non-trivial $\Theta$,   the restriction of $\Theta$ to the space of bistochastic maps  must map completely positive maps into completely positive maps: 
\begin{prop}\label{prop:bistopositive}
Let $\Theta:  \Chan (\spc H) \to \Chan (\spc H) $ be a non-trivial supermap satisfying Requirements 1, 2, and 4 in the main text.  Then, for every bistochastic map $\map M \in  \SpanBiChan(\spc H)$, $\map M$ is completely positive if and only if $\Theta (\map M)$ is completely positive.  
\end{prop} 
\Proof 
By Lemma~\ref{lem:unitarypreserving} of this Supplemental Material,  Requirements 1 and 2 imply that the supermap $\Theta$ maps unitary channels into unitary channels, and therefore induces a map $\theta$ on the special unitary group.   Since the supermap $\Theta$ is non-trivial,  the map $\theta$ cannot be $\theta  (U)  =  I,\, \forall U$.  Hence, the proof of  Lemma~\ref{lem:unitary} shows that the action of  $\Theta$ on unitary channels is unitarily equivalent either to the transpose, or to the adjoint.   

Now, Requirement 4   implies that the action of the supermap  $\Theta$  is uniquely defined on the set of bistochastic channels (which is the linear span of the set of unitary channels, cf. by Theorem 1 in the main text).  Hence, the input-output inversion of bistochastic channel must be unitarily equivalent to the transpose or to the adjoint. The action of these maps in the Choi representation is provided by Eqs.  (\ref{swap}) and Eqs. (\ref{swaptrans}).  Both maps are involutions and send positive operators into positive operators. Hence, a generic bistochastic map  $\map M \in  \SpanBiChan(\spc H)$ is completely positive if and only if  $\Theta (\map M) $ is completely positive.\qed

\subsection{(Non)-positivity of the projection on the set of bistochastic maps} 
By Proposition \ref{prop:bisto}, the action of any non-trivial supermap $\Theta$ satisfying Requirements 1,2 and 4 on a generic channel $\map C$ satisfies the condition 
\begin{align}\label{thetapi}
\Theta  (\map C)  =   \Theta  (\Pi  (\map C)) \, ,
\end{align}  
where  $\Pi$ is the projection on the set of bistochastic maps. 
 We now  observe that the supermap $\Pi$ is generally not positive, in the sense that it may  not map completely positive maps into completely positive maps.  

\begin{prop}\label{prop:d>2}  
For $d>2$, there exist channels $\map C  \in  \Chan  (\spc  H)$ such that the map $\Pi (\map C)$ is not positive.
\end{prop}

\Proof   A counterexample can be found among the classical channels of the form  $\map C   (\rho)   =  \sum_{x,y}   p(  y|x)  \,   |y\>\<y|    \<  x|  \rho  |x\> $, where $p(y|x)$ is a conditional probability distribution over the variables $x,y  \in   \{1,\dots,  d\}$.    In particular,  
 consider the probability distribution  defined by  
 \begin{align}
 p(y|x)    = 
 \left\{ 
 \begin{array}{ll} 
 \frac{  (d-2) }{2(d-1)}    \qquad    &y=1  \, ,  x = 1  \\
  1     &   y  = 1  \, ,  x  \not = 1 \\
 \frac 1 {d-1} \, \left [ 1-  \frac{  (d-2) }{2(d-1)} \right] &   y  \not = 1  \, ,  x  = 1 \\
 \end{array}
 \right.   
 \end{align}
  
In this case, one has 
\begin{align}
\nonumber \<1|  \Pi  (\map  C)  (|1\>\<1|) |1\>     &  = \<  1  |  \,  \left[  \map C (|1\>\<1|)   +  \frac I d   -   \map C  \left(\frac Id \right) \, \right] \,  |1\>\\
\nonumber &    =        \frac{  (d-2) }{2(d-1)}      +  \frac 1 d  -  \frac  1d\left [  \frac{  (d-2) }{2(d-1)}        +  (d-1) \right ]
 \\
\nonumber &    =  \left( 1-  \frac 1d \right) \, \frac{  (d-2) }{2(d-1)}      -   \frac{(d-2)}d  \\
&  =   - \frac{  (d-2) }{2 d}  \,.
\end{align}
Hence, the map $\Pi (\map C)$ is not positive for every $d>2$.   \qed

\begin{cor}
For $d>2$, it is impossible to find a non-trivial supermap $\Theta$  defined on the set of all quantum channels and satisfying Requirements 1,2, and 4.   
\end{cor}
\Proof    Equation  (\ref{thetapi}) implies that  the input-output inversion  $\Theta  (\map C)$ of a channel $\map C$ is completely positive if and only if the map $\Theta (\Pi (\map C))$ is completely positive.  In turn,  Proposition \ref{prop:bistopositive} implies that the map $\Theta (\Pi  (\map C))$ is completely positive if and only if the map $\Pi (\map C)$ is  completely positive. For $d>2$,  Proposition \ref{prop:d>2} shows that there exist channels for which the map $\Pi (\map C)$ is not positive.  Hence, the supermap $\Theta$ cannot be defined on these channels. \qed

In the special case $d=2$, instead, the supermap $\Pi$ is guaranteed to be positive, and an input-output inversion satisfying Requirements 1, 2 and 4  can be defined on every quantum channel. The positivity of the supermap $\Pi$ follows from the fact that the projection on the set of bistochastic channels transforms completely positive maps into completely positive maps: 
\begin{prop}\label{prop:d=2}
Let $  \Pi   $ be the projection on the vector space spanned by the bistochastic qubit channels. For every completely positive qubit map $\map M$ the map $\Pi  (\map M)$ is completely positive.   
\end{prop}

\Proof   The proof is done in the Choi representation.  Let $|\Psi\>  $ be a unit vector. When applied to  a completely positive map with Choi operator $|\Psi\>\<\Psi|$, the projection $\Pi$  yields a map with Choi operator  
\begin{align}
 A  =   |\Psi\>\<\Psi|     +      \frac {  I\otimes  I}{2}   -  \rho_1 \otimes \frac I 2  -  \frac I 2 \otimes \rho_2 \, ,  \end{align} 
with $\rho_1  = \Tr_2  [|\Psi\>\<\Psi|]$   and $\rho_2  = \Tr_1  [|\Psi\>\<\Psi|]$.    

We now show that the operator $A$ is positive for every $|\Psi\>$. Let  $|\Psi\>  =  \sqrt  p  \, |\alpha_1\>  |\beta_1\>  +  \sqrt{1-p}  \,  |\alpha_2\>  |\beta_2\>$ be  Schmidt decomposition of $|\Psi\>$.  
Hence, $A$ can be rewritten as 
{\begin{align}
\nonumber A    &  =      \frac 12    |\alpha_1\>\<\alpha_1| \otimes |\beta_1\>\<\beta_1  |   +   \frac 12  |\alpha_2\>\<\alpha_2|  \otimes |\beta_2\>\<\beta_2|   \\
\nonumber 
&  \quad +  \sqrt{  p(1-p)    }   \,    |\alpha_1\>\<\alpha_2| \otimes |\beta_1\>\<\beta_2  |    \\
\nonumber &  \quad +  \sqrt{  p(1-p)    }   \,    |\alpha_2\>\<\alpha_1| \otimes |\beta_2\>\<\beta_1  | \\
&  =  \frac 12 \,  |\Psi\>\<\Psi|   +  \frac 12  |\Psi'\>\<\Psi'|    \,
 ,\end{align}}
with $|\Psi'\>    :  =  \sqrt{1-p} \,  |\alpha_1\>|\beta_1\>  +  \sqrt{p}  \,  |\alpha_2\>|\beta_2\> \, .$ \qed 

\smallskip 

Building on the above result, one can define an input-output inversion $\Theta$ on the set of all qubit channels by first projecting on the subspace of bistochastic  channels, and then applying one of the input-output inversions of bistochastic channels defined earlier in the paper.

\bibliography{references}

\begin{thebibliography}{75}%
\makeatletter
\providecommand \@ifxundefined [1]{%
 \@ifx{#1\undefined}
}%
\providecommand \@ifnum [1]{%
 \ifnum #1\expandafter \@firstoftwo
 \else \expandafter \@secondoftwo
 \fi
}%
\providecommand \@ifx [1]{%
 \ifx #1\expandafter \@firstoftwo
 \else \expandafter \@secondoftwo
 \fi
}%
\providecommand \natexlab [1]{#1}%
\providecommand \enquote  [1]{``#1''}%
\providecommand \bibnamefont  [1]{#1}%
\providecommand \bibfnamefont [1]{#1}%
\providecommand \citenamefont [1]{#1}%
\providecommand \href@noop [0]{\@secondoftwo}%
\providecommand \href [0]{\begingroup \@sanitize@url \@href}%
\providecommand \@href[1]{\@@startlink{#1}\@@href}%
\providecommand \@@href[1]{\endgroup#1\@@endlink}%
\providecommand \@sanitize@url [0]{\catcode `\\12\catcode `\$12\catcode
  `\&12\catcode `\#12\catcode `\^12\catcode `\_12\catcode `\%12\relax}%
\providecommand \@@startlink[1]{}%
\providecommand \@@endlink[0]{}%
\providecommand \url  [0]{\begingroup\@sanitize@url \@url }%
\providecommand \@url [1]{\endgroup\@href {#1}{\urlprefix }}%
\providecommand \urlprefix  [0]{URL }%
\providecommand \Eprint [0]{\href }%
\providecommand \doibase [0]{http://dx.doi.org/}%
\providecommand \selectlanguage [0]{\@gobble}%
\providecommand \bibinfo  [0]{\@secondoftwo}%
\providecommand \bibfield  [0]{\@secondoftwo}%
\providecommand \translation [1]{[#1]}%
\providecommand \BibitemOpen [0]{}%
\providecommand \bibitemStop [0]{}%
\providecommand \bibitemNoStop [0]{.\EOS\space}%
\providecommand \EOS [0]{\spacefactor3000\relax}%
\providecommand \BibitemShut  [1]{\csname bibitem#1\endcsname}%
\let\auto@bib@innerbib\@empty
\bibitem [{\citenamefont {L{\"u}ders}(1954)}]{luders1954equivalence}%
  \BibitemOpen
  \bibfield  {author} {\bibinfo {author} {\bibfnamefont {G.}~\bibnamefont
  {L{\"u}ders}},\ }\href@noop {} {\bibfield  {journal} {\bibinfo  {journal}
  {Dan. Mat. Fys. Medd.}\ }\textbf {\bibinfo {volume} {28}},\ \bibinfo {pages}
  {1} (\bibinfo {year} {1954})}\BibitemShut {NoStop}%
\bibitem [{\citenamefont {Pauli}(1955)}]{pauli1955niels}%
  \BibitemOpen
  \bibfield  {author} {\bibinfo {author} {\bibfnamefont {W.}~\bibnamefont
  {Pauli}},\ }\href@noop {} {\emph {\bibinfo {title} {Niels Bohr and the
  development of physics}}},\ edited by\ \bibinfo {editor} {\bibfnamefont
  {W.}~\bibnamefont {Pauli}}, \bibinfo {editor} {\bibfnamefont
  {L.}~\bibnamefont {Rosenfeld}}, \ and\ \bibinfo {editor} {\bibfnamefont
  {V.}~\bibnamefont {Weisskopf}},\ Vol.\ \bibinfo {volume} {129}\ (\bibinfo
  {publisher} {McGraw-Hill},\ \bibinfo {year} {1955})\BibitemShut {NoStop}%
\bibitem [{\citenamefont {Halliwell}\ \emph {et~al.}(1996)\citenamefont
  {Halliwell}, \citenamefont {P{\'e}rez-Mercader},\ and\ \citenamefont
  {Zurek}}]{halliwell1996physical}%
  \BibitemOpen
  \bibfield  {author} {\bibinfo {author} {\bibfnamefont {J.~J.}\ \bibnamefont
  {Halliwell}}, \bibinfo {author} {\bibfnamefont {J.}~\bibnamefont
  {P{\'e}rez-Mercader}}, \ and\ \bibinfo {author} {\bibfnamefont {W.~H.}\
  \bibnamefont {Zurek}},\ }\href@noop {} {\emph {\bibinfo {title} {Physical
  origins of time asymmetry}}}\ (\bibinfo  {publisher} {Cambridge University
  Press},\ \bibinfo {year} {1996})\BibitemShut {NoStop}%
\bibitem [{\citenamefont {Wald}(2006)}]{wald2006arrow}%
  \BibitemOpen
  \bibfield  {author} {\bibinfo {author} {\bibfnamefont {R.~M.}\ \bibnamefont
  {Wald}},\ }\href@noop {} {\bibfield  {journal} {\bibinfo  {journal} {Studies
  in History and Philosophy of Science Part B: Studies in History and
  Philosophy of Modern Physics}\ }\textbf {\bibinfo {volume} {37}},\ \bibinfo
  {pages} {394} (\bibinfo {year} {2006})}\BibitemShut {NoStop}%
\bibitem [{\citenamefont {Maccone}(2009)}]{maccone2009quantum}%
  \BibitemOpen
  \bibfield  {author} {\bibinfo {author} {\bibfnamefont {L.}~\bibnamefont
  {Maccone}},\ }\href@noop {} {\bibfield  {journal} {\bibinfo  {journal}
  {Physical Review Letters}\ }\textbf {\bibinfo {volume} {103}},\ \bibinfo
  {pages} {080401} (\bibinfo {year} {2009})}\BibitemShut {NoStop}%
\bibitem [{\citenamefont {Rovelli}(2017)}]{rovelli2017time}%
  \BibitemOpen
  \bibfield  {author} {\bibinfo {author} {\bibfnamefont {C.}~\bibnamefont
  {Rovelli}},\ }in\ \href@noop {} {\emph {\bibinfo {booktitle} {The Philosophy
  of Cosmology}}}\ (\bibinfo  {publisher} {Cambridge University Press},\
  \bibinfo {year} {2017})\ pp.\ \bibinfo {pages} {285--296}\BibitemShut
  {NoStop}%
\bibitem [{\citenamefont {Di~Biagio}\ \emph {et~al.}(2020)\citenamefont
  {Di~Biagio}, \citenamefont {Don{\`a}},\ and\ \citenamefont
  {Rovelli}}]{di2020quantum}%
  \BibitemOpen
  \bibfield  {author} {\bibinfo {author} {\bibfnamefont {A.}~\bibnamefont
  {Di~Biagio}}, \bibinfo {author} {\bibfnamefont {P.}~\bibnamefont {Don{\`a}}},
  \ and\ \bibinfo {author} {\bibfnamefont {C.}~\bibnamefont {Rovelli}},\
  }\href@noop {} {\bibfield  {journal} {\bibinfo  {journal} {Preprint at
  arXiv:2010.05734}\ } (\bibinfo {year} {2020})}\BibitemShut {NoStop}%
\bibitem [{\citenamefont {Hardy}(2021)}]{hardy2021time}%
  \BibitemOpen
  \bibfield  {author} {\bibinfo {author} {\bibfnamefont {L.}~\bibnamefont
  {Hardy}},\ }\href@noop {} {\bibfield  {journal} {\bibinfo  {journal}
  {Preprint at arXiv:2104.00071}\ } (\bibinfo {year} {2021})}\BibitemShut
  {NoStop}%
\bibitem [{\citenamefont {Aharonov}\ \emph {et~al.}(1964)\citenamefont
  {Aharonov}, \citenamefont {Bergmann},\ and\ \citenamefont
  {Lebowitz}}]{aharonov1964time}%
  \BibitemOpen
  \bibfield  {author} {\bibinfo {author} {\bibfnamefont {Y.}~\bibnamefont
  {Aharonov}}, \bibinfo {author} {\bibfnamefont {P.~G.}\ \bibnamefont
  {Bergmann}}, \ and\ \bibinfo {author} {\bibfnamefont {J.~L.}\ \bibnamefont
  {Lebowitz}},\ }\href@noop {} {\bibfield  {journal} {\bibinfo  {journal}
  {Physical Review}\ }\textbf {\bibinfo {volume} {134}},\ \bibinfo {pages}
  {B1410} (\bibinfo {year} {1964})}\BibitemShut {NoStop}%
\bibitem [{\citenamefont {Aharonov}\ \emph {et~al.}(1990)\citenamefont
  {Aharonov}, \citenamefont {Anandan}, \citenamefont {Popescu},\ and\
  \citenamefont {Vaidman}}]{aharonov1990superpositions}%
  \BibitemOpen
  \bibfield  {author} {\bibinfo {author} {\bibfnamefont {Y.}~\bibnamefont
  {Aharonov}}, \bibinfo {author} {\bibfnamefont {J.}~\bibnamefont {Anandan}},
  \bibinfo {author} {\bibfnamefont {S.}~\bibnamefont {Popescu}}, \ and\
  \bibinfo {author} {\bibfnamefont {L.}~\bibnamefont {Vaidman}},\ }\href@noop
  {} {\bibfield  {journal} {\bibinfo  {journal} {Physical Review Letters}\
  }\textbf {\bibinfo {volume} {64}},\ \bibinfo {pages} {2965} (\bibinfo {year}
  {1990})}\BibitemShut {NoStop}%
\bibitem [{\citenamefont {Aharonov}\ and\ \citenamefont
  {Vaidman}(2002)}]{aharonov2002two}%
  \BibitemOpen
  \bibfield  {author} {\bibinfo {author} {\bibfnamefont {Y.}~\bibnamefont
  {Aharonov}}\ and\ \bibinfo {author} {\bibfnamefont {L.}~\bibnamefont
  {Vaidman}},\ }in\ \href@noop {} {\emph {\bibinfo {booktitle} {Time in quantum
  mechanics}}}\ (\bibinfo  {publisher} {Springer},\ \bibinfo {year} {2002})\
  pp.\ \bibinfo {pages} {369--412}\BibitemShut {NoStop}%
\bibitem [{\citenamefont {Abramsky}\ and\ \citenamefont
  {Coecke}(2004)}]{abramsky2004categorical}%
  \BibitemOpen
  \bibfield  {author} {\bibinfo {author} {\bibfnamefont {S.}~\bibnamefont
  {Abramsky}}\ and\ \bibinfo {author} {\bibfnamefont {B.}~\bibnamefont
  {Coecke}},\ }in\ \href@noop {} {\emph {\bibinfo {booktitle} {Proceedings of
  the 19th Annual IEEE Symposium on Logic in Computer Science, 2004.}}}\
  (\bibinfo {organization} {IEEE},\ \bibinfo {year} {2004})\ pp.\ \bibinfo
  {pages} {415--425}\BibitemShut {NoStop}%
\bibitem [{\citenamefont {Hardy}(2007)}]{hardy2007towards}%
  \BibitemOpen
  \bibfield  {author} {\bibinfo {author} {\bibfnamefont {L.}~\bibnamefont
  {Hardy}},\ }\href@noop {} {\bibfield  {journal} {\bibinfo  {journal} {Journal
  of Physics A: Mathematical and Theoretical}\ }\textbf {\bibinfo {volume}
  {40}},\ \bibinfo {pages} {3081} (\bibinfo {year} {2007})}\BibitemShut
  {NoStop}%
\bibitem [{\citenamefont {Oeckl}(2008)}]{oeckl2008general}%
  \BibitemOpen
  \bibfield  {author} {\bibinfo {author} {\bibfnamefont {R.}~\bibnamefont
  {Oeckl}},\ }\href@noop {} {\bibfield  {journal} {\bibinfo  {journal}
  {Advances in Theoretical and Mathematical Physics}\ }\textbf {\bibinfo
  {volume} {12}},\ \bibinfo {pages} {319} (\bibinfo {year} {2008})}\BibitemShut
  {NoStop}%
\bibitem [{\citenamefont {Svetlichny}(2011)}]{svetlichny2011time}%
  \BibitemOpen
  \bibfield  {author} {\bibinfo {author} {\bibfnamefont {G.}~\bibnamefont
  {Svetlichny}},\ }\href@noop {} {\bibfield  {journal} {\bibinfo  {journal}
  {International Journal of Theoretical Physics}\ }\textbf {\bibinfo {volume}
  {50}},\ \bibinfo {pages} {3903} (\bibinfo {year} {2011})}\BibitemShut
  {NoStop}%
\bibitem [{\citenamefont {Lloyd}\ \emph {et~al.}(2011)\citenamefont {Lloyd},
  \citenamefont {Maccone}, \citenamefont {Garcia-Patron}, \citenamefont
  {Giovannetti}, \citenamefont {Shikano}, \citenamefont {Pirandola},
  \citenamefont {Rozema}, \citenamefont {Darabi}, \citenamefont {Soudagar},
  \citenamefont {Shalm} \emph {et~al.}}]{lloyd2011closed}%
  \BibitemOpen
  \bibfield  {author} {\bibinfo {author} {\bibfnamefont {S.}~\bibnamefont
  {Lloyd}}, \bibinfo {author} {\bibfnamefont {L.}~\bibnamefont {Maccone}},
  \bibinfo {author} {\bibfnamefont {R.}~\bibnamefont {Garcia-Patron}}, \bibinfo
  {author} {\bibfnamefont {V.}~\bibnamefont {Giovannetti}}, \bibinfo {author}
  {\bibfnamefont {Y.}~\bibnamefont {Shikano}}, \bibinfo {author} {\bibfnamefont
  {S.}~\bibnamefont {Pirandola}}, \bibinfo {author} {\bibfnamefont {L.~A.}\
  \bibnamefont {Rozema}}, \bibinfo {author} {\bibfnamefont {A.}~\bibnamefont
  {Darabi}}, \bibinfo {author} {\bibfnamefont {Y.}~\bibnamefont {Soudagar}},
  \bibinfo {author} {\bibfnamefont {L.~K.}\ \bibnamefont {Shalm}},  \emph
  {et~al.},\ }\href@noop {} {\bibfield  {journal} {\bibinfo  {journal}
  {Physical Review Letters}\ }\textbf {\bibinfo {volume} {106}},\ \bibinfo
  {pages} {040403} (\bibinfo {year} {2011})}\BibitemShut {NoStop}%
\bibitem [{\citenamefont {Genkina}\ \emph {et~al.}(2012)\citenamefont
  {Genkina}, \citenamefont {Chiribella},\ and\ \citenamefont
  {Hardy}}]{genkina2012optimal}%
  \BibitemOpen
  \bibfield  {author} {\bibinfo {author} {\bibfnamefont {D.}~\bibnamefont
  {Genkina}}, \bibinfo {author} {\bibfnamefont {G.}~\bibnamefont {Chiribella}},
  \ and\ \bibinfo {author} {\bibfnamefont {L.}~\bibnamefont {Hardy}},\
  }\href@noop {} {\bibfield  {journal} {\bibinfo  {journal} {Physical Review
  A}\ }\textbf {\bibinfo {volume} {85}},\ \bibinfo {pages} {022330} (\bibinfo
  {year} {2012})}\BibitemShut {NoStop}%
\bibitem [{\citenamefont {Oreshkov}\ and\ \citenamefont
  {Cerf}(2015)}]{oreshkov2015operational}%
  \BibitemOpen
  \bibfield  {author} {\bibinfo {author} {\bibfnamefont {O.}~\bibnamefont
  {Oreshkov}}\ and\ \bibinfo {author} {\bibfnamefont {N.~J.}\ \bibnamefont
  {Cerf}},\ }\href@noop {} {\bibfield  {journal} {\bibinfo  {journal} {Nature
  Physics}\ }\textbf {\bibinfo {volume} {11}},\ \bibinfo {pages} {853}
  (\bibinfo {year} {2015})}\BibitemShut {NoStop}%
\bibitem [{\citenamefont {Silva}\ \emph {et~al.}(2017)\citenamefont {Silva},
  \citenamefont {Guryanova}, \citenamefont {Short}, \citenamefont {Skrzypczyk},
  \citenamefont {Brunner},\ and\ \citenamefont
  {Popescu}}]{silva2017connecting}%
  \BibitemOpen
  \bibfield  {author} {\bibinfo {author} {\bibfnamefont {R.}~\bibnamefont
  {Silva}}, \bibinfo {author} {\bibfnamefont {Y.}~\bibnamefont {Guryanova}},
  \bibinfo {author} {\bibfnamefont {A.~J.}\ \bibnamefont {Short}}, \bibinfo
  {author} {\bibfnamefont {P.}~\bibnamefont {Skrzypczyk}}, \bibinfo {author}
  {\bibfnamefont {N.}~\bibnamefont {Brunner}}, \ and\ \bibinfo {author}
  {\bibfnamefont {S.}~\bibnamefont {Popescu}},\ }\href@noop {} {\bibfield
  {journal} {\bibinfo  {journal} {New Journal of Physics}\ }\textbf {\bibinfo
  {volume} {19}},\ \bibinfo {pages} {103022} (\bibinfo {year}
  {2017})}\BibitemShut {NoStop}%
\bibitem [{\citenamefont {Chiribella}\ \emph
  {et~al.}(2009{\natexlab{a}})\citenamefont {Chiribella}, \citenamefont
  {D’Ariano}, \citenamefont {Perinotti},\ and\ \citenamefont
  {Valiron}}]{chiribella2009beyond}%
  \BibitemOpen
  \bibfield  {author} {\bibinfo {author} {\bibfnamefont {G.}~\bibnamefont
  {Chiribella}}, \bibinfo {author} {\bibfnamefont {G.}~\bibnamefont
  {D’Ariano}}, \bibinfo {author} {\bibfnamefont {P.}~\bibnamefont
  {Perinotti}}, \ and\ \bibinfo {author} {\bibfnamefont {B.}~\bibnamefont
  {Valiron}},\ }\href@noop {} {\bibfield  {journal} {\bibinfo  {journal}
  {Preprint at arXiv:0912.0195}\ } (\bibinfo {year}
  {2009}{\natexlab{a}})}\BibitemShut {NoStop}%
\bibitem [{\citenamefont {Oreshkov}\ \emph {et~al.}(2012)\citenamefont
  {Oreshkov}, \citenamefont {Costa},\ and\ \citenamefont
  {Brukner}}]{oreshkov2012quantum}%
  \BibitemOpen
  \bibfield  {author} {\bibinfo {author} {\bibfnamefont {O.}~\bibnamefont
  {Oreshkov}}, \bibinfo {author} {\bibfnamefont {F.}~\bibnamefont {Costa}}, \
  and\ \bibinfo {author} {\bibfnamefont {{\v{C}}.}~\bibnamefont {Brukner}},\
  }\href@noop {} {\bibfield  {journal} {\bibinfo  {journal} {Nature
  Communications}\ }\textbf {\bibinfo {volume} {3}},\ \bibinfo {pages} {1}
  (\bibinfo {year} {2012})}\BibitemShut {NoStop}%
\bibitem [{\citenamefont {Chiribella}\ \emph {et~al.}(2013)\citenamefont
  {Chiribella}, \citenamefont {D{’}Ariano}, \citenamefont {Perinotti},\ and\
  \citenamefont {Valiron}}]{chiribella2013quantum}%
  \BibitemOpen
  \bibfield  {author} {\bibinfo {author} {\bibfnamefont {G.}~\bibnamefont
  {Chiribella}}, \bibinfo {author} {\bibfnamefont {G.~M.}\ \bibnamefont
  {D{’}Ariano}}, \bibinfo {author} {\bibfnamefont {P.}~\bibnamefont
  {Perinotti}}, \ and\ \bibinfo {author} {\bibfnamefont {B.}~\bibnamefont
  {Valiron}},\ }\href@noop {} {\bibfield  {journal} {\bibinfo  {journal}
  {Physical Review A}\ }\textbf {\bibinfo {volume} {88}},\ \bibinfo {pages}
  {022318} (\bibinfo {year} {2013})}\BibitemShut {NoStop}%
\bibitem [{\citenamefont {Heinosaari}\ and\ \citenamefont
  {Ziman}(2011)}]{heinosaari2011mathematical}%
  \BibitemOpen
  \bibfield  {author} {\bibinfo {author} {\bibfnamefont {T.}~\bibnamefont
  {Heinosaari}}\ and\ \bibinfo {author} {\bibfnamefont {M.}~\bibnamefont
  {Ziman}},\ }\href@noop {} {\emph {\bibinfo {title} {The mathematical language
  of quantum theory: from uncertainty to entanglement}}}\ (\bibinfo
  {publisher} {Cambridge University Press},\ \bibinfo {year}
  {2011})\BibitemShut {NoStop}%
\bibitem [{\citenamefont {Wigner}(1959)}]{wigner1959group}%
  \BibitemOpen
  \bibfield  {author} {\bibinfo {author} {\bibfnamefont {E.~P.}\ \bibnamefont
  {Wigner}},\ }\href@noop {} {\emph {\bibinfo {title} {Group theory and its
  application to the quantum mechanics of atomic spectra}}}\ (\bibinfo
  {publisher} {Academic Press},\ \bibinfo {year} {1959})\BibitemShut {NoStop}%
\bibitem [{\citenamefont {Messiah}(1965)}]{messiah1965quantum}%
  \BibitemOpen
  \bibfield  {author} {\bibinfo {author} {\bibfnamefont {A.}~\bibnamefont
  {Messiah}},\ }\href@noop {} {\emph {\bibinfo {title} {Quantum mechanics}}}\
  (\bibinfo  {publisher} {North-Holland Publishing Company Amsterdam},\
  \bibinfo {year} {1965})\BibitemShut {NoStop}%
\bibitem [{\citenamefont {Campisi}\ \emph {et~al.}(2011)\citenamefont
  {Campisi}, \citenamefont {H{\"a}nggi},\ and\ \citenamefont
  {Talkner}}]{campisi2011colloquium}%
  \BibitemOpen
  \bibfield  {author} {\bibinfo {author} {\bibfnamefont {M.}~\bibnamefont
  {Campisi}}, \bibinfo {author} {\bibfnamefont {P.}~\bibnamefont {H{\"a}nggi}},
  \ and\ \bibinfo {author} {\bibfnamefont {P.}~\bibnamefont {Talkner}},\
  }\href@noop {} {\bibfield  {journal} {\bibinfo  {journal} {Reviews of Modern
  Physics}\ }\textbf {\bibinfo {volume} {83}},\ \bibinfo {pages} {771}
  (\bibinfo {year} {2011})}\BibitemShut {NoStop}%
\bibitem [{\citenamefont {Landau}\ and\ \citenamefont
  {Streater}(1993)}]{landau1993Birkhoff}%
  \BibitemOpen
  \bibfield  {author} {\bibinfo {author} {\bibfnamefont {L.}~\bibnamefont
  {Landau}}\ and\ \bibinfo {author} {\bibfnamefont {R.}~\bibnamefont
  {Streater}},\ }\href@noop {} {\bibfield  {journal} {\bibinfo  {journal}
  {Linear Algebra and Its Applications}\ }\textbf {\bibinfo {volume} {193}},\
  \bibinfo {pages} {107} (\bibinfo {year} {1993})}\BibitemShut {NoStop}%
\bibitem [{\citenamefont {Mendl}\ and\ \citenamefont
  {Wolf}(2009)}]{mendl2009unital}%
  \BibitemOpen
  \bibfield  {author} {\bibinfo {author} {\bibfnamefont {C.~B.}\ \bibnamefont
  {Mendl}}\ and\ \bibinfo {author} {\bibfnamefont {M.~M.}\ \bibnamefont
  {Wolf}},\ }\href@noop {} {\bibfield  {journal} {\bibinfo  {journal}
  {Communications in Mathematical Physics}\ }\textbf {\bibinfo {volume}
  {289}},\ \bibinfo {pages} {1057} (\bibinfo {year} {2009})}\BibitemShut
  {NoStop}%
\bibitem [{\citenamefont {Chiribella}\ \emph {et~al.}(2008)\citenamefont
  {Chiribella}, \citenamefont {D'Ariano},\ and\ \citenamefont
  {Perinotti}}]{chiribella2008transforming}%
  \BibitemOpen
  \bibfield  {author} {\bibinfo {author} {\bibfnamefont {G.}~\bibnamefont
  {Chiribella}}, \bibinfo {author} {\bibfnamefont {G.~M.}\ \bibnamefont
  {D'Ariano}}, \ and\ \bibinfo {author} {\bibfnamefont {P.}~\bibnamefont
  {Perinotti}},\ }\href@noop {} {\bibfield  {journal} {\bibinfo  {journal} {EPL
  (Europhysics Letters)}\ }\textbf {\bibinfo {volume} {83}},\ \bibinfo {pages}
  {30004} (\bibinfo {year} {2008})}\BibitemShut {NoStop}%
\bibitem [{\citenamefont {Chiribella}\ \emph
  {et~al.}(2009{\natexlab{b}})\citenamefont {Chiribella}, \citenamefont
  {D’Ariano},\ and\ \citenamefont {Perinotti}}]{chiribella2009theoretical}%
  \BibitemOpen
  \bibfield  {author} {\bibinfo {author} {\bibfnamefont {G.}~\bibnamefont
  {Chiribella}}, \bibinfo {author} {\bibfnamefont {G.~M.}\ \bibnamefont
  {D’Ariano}}, \ and\ \bibinfo {author} {\bibfnamefont {P.}~\bibnamefont
  {Perinotti}},\ }\href@noop {} {\bibfield  {journal} {\bibinfo  {journal}
  {Physical Review A}\ }\textbf {\bibinfo {volume} {80}},\ \bibinfo {pages}
  {022339} (\bibinfo {year} {2009}{\natexlab{b}})}\BibitemShut {NoStop}%
\bibitem [{\citenamefont {Bisio}\ and\ \citenamefont
  {Perinotti}(2019)}]{bisio2019theoretical}%
  \BibitemOpen
  \bibfield  {author} {\bibinfo {author} {\bibfnamefont {A.}~\bibnamefont
  {Bisio}}\ and\ \bibinfo {author} {\bibfnamefont {P.}~\bibnamefont
  {Perinotti}},\ }\href@noop {} {\bibfield  {journal} {\bibinfo  {journal}
  {Proceedings of the Royal Society A}\ }\textbf {\bibinfo {volume} {475}},\
  \bibinfo {pages} {20180706} (\bibinfo {year} {2019})}\BibitemShut {NoStop}%
\bibitem [{\citenamefont {Ara{\'u}jo}\ \emph {et~al.}(2015)\citenamefont
  {Ara{\'u}jo}, \citenamefont {Branciard}, \citenamefont {Costa}, \citenamefont
  {Feix}, \citenamefont {Giarmatzi},\ and\ \citenamefont
  {Brukner}}]{araujo2015witnessing}%
  \BibitemOpen
  \bibfield  {author} {\bibinfo {author} {\bibfnamefont {M.}~\bibnamefont
  {Ara{\'u}jo}}, \bibinfo {author} {\bibfnamefont {C.}~\bibnamefont
  {Branciard}}, \bibinfo {author} {\bibfnamefont {F.}~\bibnamefont {Costa}},
  \bibinfo {author} {\bibfnamefont {A.}~\bibnamefont {Feix}}, \bibinfo {author}
  {\bibfnamefont {C.}~\bibnamefont {Giarmatzi}}, \ and\ \bibinfo {author}
  {\bibfnamefont {{\v{C}}.}~\bibnamefont {Brukner}},\ }\href@noop {} {\bibfield
   {journal} {\bibinfo  {journal} {New Journal of Physics}\ }\textbf {\bibinfo
  {volume} {17}},\ \bibinfo {pages} {102001} (\bibinfo {year}
  {2015})}\BibitemShut {NoStop}%
\bibitem [{\citenamefont {Oreshkov}\ and\ \citenamefont
  {Giarmatzi}(2016)}]{oreshkov2016causal}%
  \BibitemOpen
  \bibfield  {author} {\bibinfo {author} {\bibfnamefont {O.}~\bibnamefont
  {Oreshkov}}\ and\ \bibinfo {author} {\bibfnamefont {C.}~\bibnamefont
  {Giarmatzi}},\ }\href@noop {} {\bibfield  {journal} {\bibinfo  {journal} {New
  Journal of Physics}\ }\textbf {\bibinfo {volume} {18}},\ \bibinfo {pages}
  {093020} (\bibinfo {year} {2016})}\BibitemShut {NoStop}%
\bibitem [{\citenamefont {Oi}(2003)}]{oi2003interference}%
  \BibitemOpen
  \bibfield  {author} {\bibinfo {author} {\bibfnamefont {D.~K.}\ \bibnamefont
  {Oi}},\ }\href@noop {} {\bibfield  {journal} {\bibinfo  {journal} {Physical
  Review Letters}\ }\textbf {\bibinfo {volume} {91}},\ \bibinfo {pages}
  {067902} (\bibinfo {year} {2003})}\BibitemShut {NoStop}%
\bibitem [{\citenamefont {Chiribella}\ and\ \citenamefont
  {Ebler}(2019)}]{chiribella2019quantum}%
  \BibitemOpen
  \bibfield  {author} {\bibinfo {author} {\bibfnamefont {G.}~\bibnamefont
  {Chiribella}}\ and\ \bibinfo {author} {\bibfnamefont {D.}~\bibnamefont
  {Ebler}},\ }\href@noop {} {\bibfield  {journal} {\bibinfo  {journal} {Nature
  communications}\ }\textbf {\bibinfo {volume} {10}},\ \bibinfo {pages} {1}
  (\bibinfo {year} {2019})}\BibitemShut {NoStop}%
\bibitem [{\citenamefont {Abbott}\ \emph {et~al.}(2020)\citenamefont {Abbott},
  \citenamefont {Wechs}, \citenamefont {Horsman}, \citenamefont {Mhalla},\ and\
  \citenamefont {Branciard}}]{abbott2020communication}%
  \BibitemOpen
  \bibfield  {author} {\bibinfo {author} {\bibfnamefont {A.~A.}\ \bibnamefont
  {Abbott}}, \bibinfo {author} {\bibfnamefont {J.}~\bibnamefont {Wechs}},
  \bibinfo {author} {\bibfnamefont {D.}~\bibnamefont {Horsman}}, \bibinfo
  {author} {\bibfnamefont {M.}~\bibnamefont {Mhalla}}, \ and\ \bibinfo {author}
  {\bibfnamefont {C.}~\bibnamefont {Branciard}},\ }\href@noop {} {\bibfield
  {journal} {\bibinfo  {journal} {Quantum}\ }\textbf {\bibinfo {volume} {4}},\
  \bibinfo {pages} {333} (\bibinfo {year} {2020})}\BibitemShut {NoStop}%
\bibitem [{\citenamefont {Dong}\ \emph {et~al.}(2019)\citenamefont {Dong},
  \citenamefont {Nakayama}, \citenamefont {Soeda},\ and\ \citenamefont
  {Murao}}]{dong2019controlled}%
  \BibitemOpen
  \bibfield  {author} {\bibinfo {author} {\bibfnamefont {Q.}~\bibnamefont
  {Dong}}, \bibinfo {author} {\bibfnamefont {S.}~\bibnamefont {Nakayama}},
  \bibinfo {author} {\bibfnamefont {A.}~\bibnamefont {Soeda}}, \ and\ \bibinfo
  {author} {\bibfnamefont {M.}~\bibnamefont {Murao}},\ }\href@noop {}
  {\bibfield  {journal} {\bibinfo  {journal} {Preprint at arXiv:1911.01645}\ }
  (\bibinfo {year} {2019})}\BibitemShut {NoStop}%
\bibitem [{\citenamefont {Chiribella}\ and\ \citenamefont
  {Ebler}(2016)}]{chiribella2016optimal}%
  \BibitemOpen
  \bibfield  {author} {\bibinfo {author} {\bibfnamefont {G.}~\bibnamefont
  {Chiribella}}\ and\ \bibinfo {author} {\bibfnamefont {D.}~\bibnamefont
  {Ebler}},\ }\href@noop {} {\bibfield  {journal} {\bibinfo  {journal} {New
  Journal of Physics}\ }\textbf {\bibinfo {volume} {18}},\ \bibinfo {pages}
  {093053} (\bibinfo {year} {2016})}\BibitemShut {NoStop}%
\bibitem [{\citenamefont {Quintino}\ \emph {et~al.}(2019)\citenamefont
  {Quintino}, \citenamefont {Dong}, \citenamefont {Shimbo}, \citenamefont
  {Soeda},\ and\ \citenamefont {Murao}}]{quintino2019probabilistic}%
  \BibitemOpen
  \bibfield  {author} {\bibinfo {author} {\bibfnamefont {M.~T.}\ \bibnamefont
  {Quintino}}, \bibinfo {author} {\bibfnamefont {Q.}~\bibnamefont {Dong}},
  \bibinfo {author} {\bibfnamefont {A.}~\bibnamefont {Shimbo}}, \bibinfo
  {author} {\bibfnamefont {A.}~\bibnamefont {Soeda}}, \ and\ \bibinfo {author}
  {\bibfnamefont {M.}~\bibnamefont {Murao}},\ }\href@noop {} {\bibfield
  {journal} {\bibinfo  {journal} {Physical Review A}\ }\textbf {\bibinfo
  {volume} {100}},\ \bibinfo {pages} {062339} (\bibinfo {year}
  {2019})}\BibitemShut {NoStop}%
\bibitem [{\citenamefont {Bennett}\ \emph {et~al.}(1993)\citenamefont
  {Bennett}, \citenamefont {Brassard}, \citenamefont {Cr{\'e}peau},
  \citenamefont {Jozsa}, \citenamefont {Peres},\ and\ \citenamefont
  {Wootters}}]{bennett1993teleporting}%
  \BibitemOpen
  \bibfield  {author} {\bibinfo {author} {\bibfnamefont {C.~H.}\ \bibnamefont
  {Bennett}}, \bibinfo {author} {\bibfnamefont {G.}~\bibnamefont {Brassard}},
  \bibinfo {author} {\bibfnamefont {C.}~\bibnamefont {Cr{\'e}peau}}, \bibinfo
  {author} {\bibfnamefont {R.}~\bibnamefont {Jozsa}}, \bibinfo {author}
  {\bibfnamefont {A.}~\bibnamefont {Peres}}, \ and\ \bibinfo {author}
  {\bibfnamefont {W.~K.}\ \bibnamefont {Wootters}},\ }\href@noop {} {\bibfield
  {journal} {\bibinfo  {journal} {Physical Review Letters}\ }\textbf {\bibinfo
  {volume} {70}},\ \bibinfo {pages} {1895} (\bibinfo {year}
  {1993})}\BibitemShut {NoStop}%
\bibitem [{\citenamefont {Chiribella}(2012)}]{chiribella2012perfect}%
  \BibitemOpen
  \bibfield  {author} {\bibinfo {author} {\bibfnamefont {G.}~\bibnamefont
  {Chiribella}},\ }\href@noop {} {\bibfield  {journal} {\bibinfo  {journal}
  {Physical Review A}\ }\textbf {\bibinfo {volume} {86}},\ \bibinfo {pages}
  {040301} (\bibinfo {year} {2012})}\BibitemShut {NoStop}%
\bibitem [{\citenamefont {Nakayama}\ \emph {et~al.}(2014)\citenamefont
  {Nakayama}, \citenamefont {Soeda},\ and\ \citenamefont
  {Murao}}]{nakayama2014universal}%
  \BibitemOpen
  \bibfield  {author} {\bibinfo {author} {\bibfnamefont {S.}~\bibnamefont
  {Nakayama}}, \bibinfo {author} {\bibfnamefont {A.}~\bibnamefont {Soeda}}, \
  and\ \bibinfo {author} {\bibfnamefont {M.}~\bibnamefont {Murao}},\ }in\
  \href@noop {} {\emph {\bibinfo {booktitle} {AIP Conference Proceedings}}},\
  Vol.\ \bibinfo {volume} {1633}\ (\bibinfo {organization} {American Institute
  of Physics},\ \bibinfo {year} {2014})\ pp.\ \bibinfo {pages}
  {183--185}\BibitemShut {NoStop}%
\bibitem [{\citenamefont {Ara{\'u}jo}\ \emph {et~al.}(2014)\citenamefont
  {Ara{\'u}jo}, \citenamefont {Feix}, \citenamefont {Costa},\ and\
  \citenamefont {Brukner}}]{araujo2014quantum}%
  \BibitemOpen
  \bibfield  {author} {\bibinfo {author} {\bibfnamefont {M.}~\bibnamefont
  {Ara{\'u}jo}}, \bibinfo {author} {\bibfnamefont {A.}~\bibnamefont {Feix}},
  \bibinfo {author} {\bibfnamefont {F.}~\bibnamefont {Costa}}, \ and\ \bibinfo
  {author} {\bibfnamefont {{\v{C}}.}~\bibnamefont {Brukner}},\ }\href@noop {}
  {\bibfield  {journal} {\bibinfo  {journal} {New Journal of Physics}\ }\textbf
  {\bibinfo {volume} {16}},\ \bibinfo {pages} {093026} (\bibinfo {year}
  {2014})}\BibitemShut {NoStop}%
\bibitem [{\citenamefont {Thompson}\ \emph {et~al.}(2018)\citenamefont
  {Thompson}, \citenamefont {Modi}, \citenamefont {Vedral},\ and\ \citenamefont
  {Gu}}]{thompson2018quantum}%
  \BibitemOpen
  \bibfield  {author} {\bibinfo {author} {\bibfnamefont {J.}~\bibnamefont
  {Thompson}}, \bibinfo {author} {\bibfnamefont {K.}~\bibnamefont {Modi}},
  \bibinfo {author} {\bibfnamefont {V.}~\bibnamefont {Vedral}}, \ and\ \bibinfo
  {author} {\bibfnamefont {M.}~\bibnamefont {Gu}},\ }\href@noop {} {\bibfield
  {journal} {\bibinfo  {journal} {New Journal of Physics}\ }\textbf {\bibinfo
  {volume} {20}},\ \bibinfo {pages} {013004} (\bibinfo {year}
  {2018})}\BibitemShut {NoStop}%
\bibitem [{\citenamefont {Zhou}\ \emph {et~al.}(2011)\citenamefont {Zhou},
  \citenamefont {Ralph}, \citenamefont {Kalasuwan}, \citenamefont {Zhang},
  \citenamefont {Peruzzo}, \citenamefont {Lanyon},\ and\ \citenamefont
  {O'Brien}}]{zhou2011adding}%
  \BibitemOpen
  \bibfield  {author} {\bibinfo {author} {\bibfnamefont {X.-Q.}\ \bibnamefont
  {Zhou}}, \bibinfo {author} {\bibfnamefont {T.~C.}\ \bibnamefont {Ralph}},
  \bibinfo {author} {\bibfnamefont {P.}~\bibnamefont {Kalasuwan}}, \bibinfo
  {author} {\bibfnamefont {M.}~\bibnamefont {Zhang}}, \bibinfo {author}
  {\bibfnamefont {A.}~\bibnamefont {Peruzzo}}, \bibinfo {author} {\bibfnamefont
  {B.~P.}\ \bibnamefont {Lanyon}}, \ and\ \bibinfo {author} {\bibfnamefont
  {J.~L.}\ \bibnamefont {O'Brien}},\ }\href@noop {} {\bibfield  {journal}
  {\bibinfo  {journal} {Nature Communications}\ }\textbf {\bibinfo {volume}
  {2}},\ \bibinfo {pages} {1} (\bibinfo {year} {2011})}\BibitemShut {NoStop}%
\bibitem [{\citenamefont {Friis}\ \emph {et~al.}(2014)\citenamefont {Friis},
  \citenamefont {Dunjko}, \citenamefont {D{\"u}r},\ and\ \citenamefont
  {Briegel}}]{friis2014implementing}%
  \BibitemOpen
  \bibfield  {author} {\bibinfo {author} {\bibfnamefont {N.}~\bibnamefont
  {Friis}}, \bibinfo {author} {\bibfnamefont {V.}~\bibnamefont {Dunjko}},
  \bibinfo {author} {\bibfnamefont {W.}~\bibnamefont {D{\"u}r}}, \ and\
  \bibinfo {author} {\bibfnamefont {H.~J.}\ \bibnamefont {Briegel}},\
  }\href@noop {} {\bibfield  {journal} {\bibinfo  {journal} {Physical Review
  A}\ }\textbf {\bibinfo {volume} {89}},\ \bibinfo {pages} {030303} (\bibinfo
  {year} {2014})}\BibitemShut {NoStop}%
\bibitem [{\citenamefont {Gour}\ \emph {et~al.}(2015)\citenamefont {Gour},
  \citenamefont {M{\"u}ller}, \citenamefont {Narasimhachar}, \citenamefont
  {Spekkens},\ and\ \citenamefont {Halpern}}]{gour2015resource}%
  \BibitemOpen
  \bibfield  {author} {\bibinfo {author} {\bibfnamefont {G.}~\bibnamefont
  {Gour}}, \bibinfo {author} {\bibfnamefont {M.~P.}\ \bibnamefont
  {M{\"u}ller}}, \bibinfo {author} {\bibfnamefont {V.}~\bibnamefont
  {Narasimhachar}}, \bibinfo {author} {\bibfnamefont {R.~W.}\ \bibnamefont
  {Spekkens}}, \ and\ \bibinfo {author} {\bibfnamefont {N.~Y.}\ \bibnamefont
  {Halpern}},\ }\href@noop {} {\bibfield  {journal} {\bibinfo  {journal}
  {Physics Reports}\ }\textbf {\bibinfo {volume} {583}},\ \bibinfo {pages} {1}
  (\bibinfo {year} {2015})}\BibitemShut {NoStop}%
\bibitem [{\citenamefont {Chiribella}\ and\ \citenamefont
  {Scandolo}(2017)}]{chiribella2017microcanonical}%
  \BibitemOpen
  \bibfield  {author} {\bibinfo {author} {\bibfnamefont {G.}~\bibnamefont
  {Chiribella}}\ and\ \bibinfo {author} {\bibfnamefont {C.~M.}\ \bibnamefont
  {Scandolo}},\ }\href@noop {} {\bibfield  {journal} {\bibinfo  {journal} {New
  Journal of Physics}\ }\textbf {\bibinfo {volume} {19}},\ \bibinfo {pages}
  {123043} (\bibinfo {year} {2017})}\BibitemShut {NoStop}%
\bibitem [{\citenamefont {Krumm}\ \emph {et~al.}(2017)\citenamefont {Krumm},
  \citenamefont {Barnum}, \citenamefont {Barrett},\ and\ \citenamefont
  {Müller}}]{krumm2017thermodynamics}%
  \BibitemOpen
  \bibfield  {author} {\bibinfo {author} {\bibfnamefont {M.}~\bibnamefont
  {Krumm}}, \bibinfo {author} {\bibfnamefont {H.}~\bibnamefont {Barnum}},
  \bibinfo {author} {\bibfnamefont {J.}~\bibnamefont {Barrett}}, \ and\
  \bibinfo {author} {\bibfnamefont {M.~P.}\ \bibnamefont {Müller}},\ }\href
  {\doibase 10.1088/1367-2630/aa68ef} {\bibfield  {journal} {\bibinfo
  {journal} {New Journal of Physics}\ }\textbf {\bibinfo {volume} {19}},\
  \bibinfo {pages} {043025} (\bibinfo {year} {2017})}\BibitemShut {NoStop}%
\bibitem [{\citenamefont {Rubino}\ \emph {et~al.}(2020)\citenamefont {Rubino},
  \citenamefont {Manzano},\ and\ \citenamefont {Brukner}}]{rubino2020time}%
  \BibitemOpen
  \bibfield  {author} {\bibinfo {author} {\bibfnamefont {G.}~\bibnamefont
  {Rubino}}, \bibinfo {author} {\bibfnamefont {G.}~\bibnamefont {Manzano}}, \
  and\ \bibinfo {author} {\bibfnamefont {{\v{C}}.}~\bibnamefont {Brukner}},\
  }\href@noop {} {\bibfield  {journal} {\bibinfo  {journal} {Preprint at
  arXiv:2008.02818}\ } (\bibinfo {year} {2020})}\BibitemShut {NoStop}%
\bibitem [{\citenamefont {Felce}\ and\ \citenamefont {Vedral}(2020)}]{Felce20}%
  \BibitemOpen
  \bibfield  {author} {\bibinfo {author} {\bibfnamefont {D.}~\bibnamefont
  {Felce}}\ and\ \bibinfo {author} {\bibfnamefont {V.}~\bibnamefont {Vedral}},\
  }\href {\doibase 10.1103/physrevlett.125.070603} {\bibfield  {journal}
  {\bibinfo  {journal} {Physical Review Letters}\ }\textbf {\bibinfo {volume}
  {125}} (\bibinfo {year} {2020}),\ 10.1103/physrevlett.125.070603}\BibitemShut
  {NoStop}%
\bibitem [{\citenamefont {Guha}\ \emph {et~al.}(2020)\citenamefont {Guha},
  \citenamefont {Alimuddin},\ and\ \citenamefont {Parashar}}]{Guha20}%
  \BibitemOpen
  \bibfield  {author} {\bibinfo {author} {\bibfnamefont {T.}~\bibnamefont
  {Guha}}, \bibinfo {author} {\bibfnamefont {M.}~\bibnamefont {Alimuddin}}, \
  and\ \bibinfo {author} {\bibfnamefont {P.}~\bibnamefont {Parashar}},\ }\href
  {\doibase 10.1103/physreva.102.032215} {\bibfield  {journal} {\bibinfo
  {journal} {Physical Review A}\ }\textbf {\bibinfo {volume} {102}} (\bibinfo
  {year} {2020}),\ 10.1103/physreva.102.032215}\BibitemShut {NoStop}%
\bibitem [{\citenamefont {Simonov}\ \emph {et~al.}(2022)\citenamefont
  {Simonov}, \citenamefont {Francica}, \citenamefont {Guarnieri},\ and\
  \citenamefont {Paternostro}}]{Simonov21}%
  \BibitemOpen
  \bibfield  {author} {\bibinfo {author} {\bibfnamefont {K.}~\bibnamefont
  {Simonov}}, \bibinfo {author} {\bibfnamefont {G.}~\bibnamefont {Francica}},
  \bibinfo {author} {\bibfnamefont {G.}~\bibnamefont {Guarnieri}}, \ and\
  \bibinfo {author} {\bibfnamefont {M.}~\bibnamefont {Paternostro}},\ }\href
  {\doibase 10.1103/PhysRevA.105.032217} {\bibfield  {journal} {\bibinfo
  {journal} {Phys. Rev. A}\ }\textbf {\bibinfo {volume} {105}},\ \bibinfo
  {pages} {032217} (\bibinfo {year} {2022})}\BibitemShut {NoStop}%
\bibitem [{\citenamefont {Fulton}\ and\ \citenamefont
  {Harris}(2013)}]{fulton2013representation}%
  \BibitemOpen
  \bibfield  {author} {\bibinfo {author} {\bibfnamefont {W.}~\bibnamefont
  {Fulton}}\ and\ \bibinfo {author} {\bibfnamefont {J.}~\bibnamefont
  {Harris}},\ }\href@noop {} {\emph {\bibinfo {title} {Representation theory: a
  first course}}},\ Vol.\ \bibinfo {volume} {129}\ (\bibinfo  {publisher}
  {Springer Science \& Business Media},\ \bibinfo {address} {New York},\
  \bibinfo {year} {2013})\BibitemShut {NoStop}%
\bibitem [{\citenamefont {Uhlmann}(2016)}]{Uhl16}%
  \BibitemOpen
  \bibfield  {author} {\bibinfo {author} {\bibfnamefont {A.}~\bibnamefont
  {Uhlmann}},\ }\href@noop {} {\bibfield  {journal} {\bibinfo  {journal}
  {Science China Physics, Mechanics and Astronomy}\ }\textbf {\bibinfo {volume}
  {59}},\ \bibinfo {pages} {630301} (\bibinfo {year} {2016})}\BibitemShut
  {NoStop}%
\bibitem [{\citenamefont {Weinberg}(1995)}]{weinberg1995quantum}%
  \BibitemOpen
  \bibfield  {author} {\bibinfo {author} {\bibfnamefont {S.}~\bibnamefont
  {Weinberg}},\ }\href@noop {} {\emph {\bibinfo {title} {The quantum theory of
  fields}}},\ Vol.~\bibinfo {volume} {2}\ (\bibinfo  {publisher} {Cambridge
  University Press},\ \bibinfo {year} {1995})\BibitemShut {NoStop}%
\bibitem [{\citenamefont {Roberts}(2017)}]{roberts2017three}%
  \BibitemOpen
  \bibfield  {author} {\bibinfo {author} {\bibfnamefont {B.~W.}\ \bibnamefont
  {Roberts}},\ }\href@noop {} {\bibfield  {journal} {\bibinfo  {journal}
  {Philosophy of Science}\ }\textbf {\bibinfo {volume} {84}},\ \bibinfo {pages}
  {315} (\bibinfo {year} {2017})}\BibitemShut {NoStop}%
\bibitem [{\citenamefont {Albert}(2000)}]{albert2000time}%
  \BibitemOpen
  \bibfield  {author} {\bibinfo {author} {\bibfnamefont {D.~Z.}\ \bibnamefont
  {Albert}},\ }\href@noop {} {\emph {\bibinfo {title} {Time and chance}}}\
  (\bibinfo  {publisher} {Harvard University Press},\ \bibinfo {year}
  {2000})\BibitemShut {NoStop}%
\bibitem [{\citenamefont {Callender}(2000)}]{callender2000time}%
  \BibitemOpen
  \bibfield  {author} {\bibinfo {author} {\bibfnamefont {C.}~\bibnamefont
  {Callender}},\ }in\ \href@noop {} {\emph {\bibinfo {booktitle} {Proceedings
  of the Aristotelian Society (Hardback)}}},\ Vol.\ \bibinfo {volume} {100}\
  (\bibinfo {organization} {Wiley Online Library},\ \bibinfo {year} {2000})\
  pp.\ \bibinfo {pages} {247--269}\BibitemShut {NoStop}%
\bibitem [{\citenamefont {Skotiniotis}\ \emph {et~al.}(2013)\citenamefont
  {Skotiniotis}, \citenamefont {Toloui}, \citenamefont {Durham},\ and\
  \citenamefont {Sanders}}]{skotiniotis2013quantum}%
  \BibitemOpen
  \bibfield  {author} {\bibinfo {author} {\bibfnamefont {M.}~\bibnamefont
  {Skotiniotis}}, \bibinfo {author} {\bibfnamefont {B.}~\bibnamefont {Toloui}},
  \bibinfo {author} {\bibfnamefont {I.~T.}\ \bibnamefont {Durham}}, \ and\
  \bibinfo {author} {\bibfnamefont {B.~C.}\ \bibnamefont {Sanders}},\
  }\href@noop {} {\bibfield  {journal} {\bibinfo  {journal} {Physical review
  letters}\ }\textbf {\bibinfo {volume} {111}},\ \bibinfo {pages} {020504}
  (\bibinfo {year} {2013})}\BibitemShut {NoStop}%
\bibitem [{\citenamefont {Sachs}(1987)}]{sachs1987physics}%
  \BibitemOpen
  \bibfield  {author} {\bibinfo {author} {\bibfnamefont {R.~G.}\ \bibnamefont
  {Sachs}},\ }\href@noop {} {\emph {\bibinfo {title} {The physics of time
  reversal}}}\ (\bibinfo  {publisher} {University of Chicago Press},\ \bibinfo
  {year} {1987})\BibitemShut {NoStop}%
\bibitem [{\citenamefont {Crooks}(2008)}]{crooks2008quantum}%
  \BibitemOpen
  \bibfield  {author} {\bibinfo {author} {\bibfnamefont {G.~E.}\ \bibnamefont
  {Crooks}},\ }\href@noop {} {\bibfield  {journal} {\bibinfo  {journal}
  {Physical Review A}\ }\textbf {\bibinfo {volume} {77}},\ \bibinfo {pages}
  {034101} (\bibinfo {year} {2008})}\BibitemShut {NoStop}%
\bibitem [{\citenamefont {Aurell}\ \emph {et~al.}(2015)\citenamefont {Aurell},
  \citenamefont {Zakrzewski},\ and\ \citenamefont
  {{\.Z}yczkowski}}]{aurell2015time}%
  \BibitemOpen
  \bibfield  {author} {\bibinfo {author} {\bibfnamefont {E.}~\bibnamefont
  {Aurell}}, \bibinfo {author} {\bibfnamefont {J.}~\bibnamefont {Zakrzewski}},
  \ and\ \bibinfo {author} {\bibfnamefont {K.}~\bibnamefont {{\.Z}yczkowski}},\
  }\href@noop {} {\bibfield  {journal} {\bibinfo  {journal} {Journal of Physics
  A: Mathematical and Theoretical}\ }\textbf {\bibinfo {volume} {48}},\
  \bibinfo {pages} {38FT01} (\bibinfo {year} {2015})}\BibitemShut {NoStop}%
\bibitem [{\citenamefont {Chiribella}\ \emph {et~al.}(2020)\citenamefont
  {Chiribella}, \citenamefont {Aurell},\ and\ \citenamefont
  {{\.Z}yczkowski}}]{chiribella2020symmetries}%
  \BibitemOpen
  \bibfield  {author} {\bibinfo {author} {\bibfnamefont {G.}~\bibnamefont
  {Chiribella}}, \bibinfo {author} {\bibfnamefont {E.}~\bibnamefont {Aurell}},
  \ and\ \bibinfo {author} {\bibfnamefont {K.}~\bibnamefont {{\.Z}yczkowski}},\
  }\href@noop {} {\bibfield  {journal} {\bibinfo  {journal} {manuscript in
  preparation}\ } (\bibinfo {year} {2020})}\BibitemShut {NoStop}%
\bibitem [{\citenamefont {Petz}(1988)}]{petz1988sufficiency}%
  \BibitemOpen
  \bibfield  {author} {\bibinfo {author} {\bibfnamefont {D.}~\bibnamefont
  {Petz}},\ }\href@noop {} {\bibfield  {journal} {\bibinfo  {journal} {The
  Quarterly Journal of Mathematics}\ }\textbf {\bibinfo {volume} {39}},\
  \bibinfo {pages} {97} (\bibinfo {year} {1988})}\BibitemShut {NoStop}%
\bibitem [{\citenamefont {Hayden}\ \emph {et~al.}(2004)\citenamefont {Hayden},
  \citenamefont {Jozsa}, \citenamefont {Petz},\ and\ \citenamefont
  {Winter}}]{hayden2004structure}%
  \BibitemOpen
  \bibfield  {author} {\bibinfo {author} {\bibfnamefont {P.}~\bibnamefont
  {Hayden}}, \bibinfo {author} {\bibfnamefont {R.}~\bibnamefont {Jozsa}},
  \bibinfo {author} {\bibfnamefont {D.}~\bibnamefont {Petz}}, \ and\ \bibinfo
  {author} {\bibfnamefont {A.}~\bibnamefont {Winter}},\ }\href@noop {}
  {\bibfield  {journal} {\bibinfo  {journal} {Communications in mathematical
  physics}\ }\textbf {\bibinfo {volume} {246}},\ \bibinfo {pages} {359}
  (\bibinfo {year} {2004})}\BibitemShut {NoStop}%
\bibitem [{\citenamefont {Choi}(1975)}]{choi1975completely}%
  \BibitemOpen
  \bibfield  {author} {\bibinfo {author} {\bibfnamefont {M.-D.}\ \bibnamefont
  {Choi}},\ }\href@noop {} {\bibfield  {journal} {\bibinfo  {journal} {Linear
  algebra and its applications}\ }\textbf {\bibinfo {volume} {10}},\ \bibinfo
  {pages} {285} (\bibinfo {year} {1975})}\BibitemShut {NoStop}%
\bibitem [{\citenamefont {Royer}(1991)}]{royer1991wigner}%
  \BibitemOpen
  \bibfield  {author} {\bibinfo {author} {\bibfnamefont {A.}~\bibnamefont
  {Royer}},\ }\href@noop {} {\bibfield  {journal} {\bibinfo  {journal}
  {Physical Review A}\ }\textbf {\bibinfo {volume} {43}},\ \bibinfo {pages}
  {44} (\bibinfo {year} {1991})}\BibitemShut {NoStop}%
\bibitem [{\citenamefont {D'Ariano}\ \emph {et~al.}(2000)\citenamefont
  {D'Ariano}, \citenamefont {Presti},\ and\ \citenamefont
  {Sacchi}}]{dariano2000bell}%
  \BibitemOpen
  \bibfield  {author} {\bibinfo {author} {\bibfnamefont {G.}~\bibnamefont
  {D'Ariano}}, \bibinfo {author} {\bibfnamefont {P.~L.}\ \bibnamefont
  {Presti}}, \ and\ \bibinfo {author} {\bibfnamefont {M.}~\bibnamefont
  {Sacchi}},\ }\href@noop {} {\bibfield  {journal} {\bibinfo  {journal}
  {Physics Letters A}\ }\textbf {\bibinfo {volume} {272}},\ \bibinfo {pages}
  {32} (\bibinfo {year} {2000})}\BibitemShut {NoStop}%
\bibitem [{\citenamefont {Dong}\ \emph {et~al.}(2021)\citenamefont {Dong},
  \citenamefont {Quintino}, \citenamefont {Soeda},\ and\ \citenamefont
  {Murao}}]{dong2021quantum}%
  \BibitemOpen
  \bibfield  {author} {\bibinfo {author} {\bibfnamefont {Q.}~\bibnamefont
  {Dong}}, \bibinfo {author} {\bibfnamefont {M.~T.}\ \bibnamefont {Quintino}},
  \bibinfo {author} {\bibfnamefont {A.}~\bibnamefont {Soeda}}, \ and\ \bibinfo
  {author} {\bibfnamefont {M.}~\bibnamefont {Murao}},\ }\href@noop {}
  {\bibfield  {journal} {\bibinfo  {journal} {arXiv preprint arXiv:2106.00034}\
  } (\bibinfo {year} {2021})}\BibitemShut {NoStop}%
\bibitem [{\citenamefont {Bavaresco}\ \emph {et~al.}(2021)\citenamefont
  {Bavaresco}, \citenamefont {Murao},\ and\ \citenamefont
  {Quintino}}]{bavaresco2021strict}%
  \BibitemOpen
  \bibfield  {author} {\bibinfo {author} {\bibfnamefont {J.}~\bibnamefont
  {Bavaresco}}, \bibinfo {author} {\bibfnamefont {M.}~\bibnamefont {Murao}}, \
  and\ \bibinfo {author} {\bibfnamefont {M.~T.}\ \bibnamefont {Quintino}},\
  }\href@noop {} {\bibfield  {journal} {\bibinfo  {journal} {Physical review
  letters}\ }\textbf {\bibinfo {volume} {127}},\ \bibinfo {pages} {200504}
  (\bibinfo {year} {2021})}\BibitemShut {NoStop}%
\bibitem [{\citenamefont {Grant}\ and\ \citenamefont {Boyd}(2014)}]{cvx}%
  \BibitemOpen
  \bibfield  {author} {\bibinfo {author} {\bibfnamefont {M.}~\bibnamefont
  {Grant}}\ and\ \bibinfo {author} {\bibfnamefont {S.}~\bibnamefont {Boyd}},\
  }\href@noop {} {\enquote {\bibinfo {title} {{CVX}: Matlab software for
  disciplined convex programming, version 2.1},}\ }\bibinfo {howpublished}
  {\url{http://cvxr.com/cvx}} (\bibinfo {year} {2014})\BibitemShut {NoStop}%
\bibitem [{\citenamefont {Grant}\ and\ \citenamefont {Boyd}(2008)}]{gb08}%
  \BibitemOpen
  \bibfield  {author} {\bibinfo {author} {\bibfnamefont {M.}~\bibnamefont
  {Grant}}\ and\ \bibinfo {author} {\bibfnamefont {S.}~\bibnamefont {Boyd}},\
  }in\ \href@noop {} {\emph {\bibinfo {booktitle} {Recent Advances in Learning
  and Control}}},\ \bibinfo {series and number} {Lecture Notes in Control and
  Information Sciences},\ \bibinfo {editor} {edited by\ \bibinfo {editor}
  {\bibfnamefont {V.}~\bibnamefont {Blondel}}, \bibinfo {editor} {\bibfnamefont
  {S.}~\bibnamefont {Boyd}}, \ and\ \bibinfo {editor} {\bibfnamefont
  {H.}~\bibnamefont {Kimura}}}\ (\bibinfo  {publisher} {Springer-Verlag
  Limited},\ \bibinfo {year} {2008})\ pp.\ \bibinfo {pages} {95--110},\
  \bibinfo {note} {\url{http://stanford.edu/~boyd/graph_dcp.html}}\BibitemShut
  {NoStop}%
\bibitem [{\citenamefont {Johnston}(2016)}]{qetlab}%
  \BibitemOpen
  \bibfield  {author} {\bibinfo {author} {\bibfnamefont {N.}~\bibnamefont
  {Johnston}},\ }\href {\doibase 10.5281/zenodo.44637} {\enquote {\bibinfo
  {title} {{QETLAB}: A {MATLAB} toolbox for quantum entanglement, version
  0.9},}\ }\bibinfo {howpublished} {\url{http://qetlab.com}} (\bibinfo {year}
  {2016})\BibitemShut {NoStop}%
\bibitem [{\citenamefont {Casazza}\ \emph {et~al.}(2000)\citenamefont {Casazza}
  \emph {et~al.}}]{casazza2000art}%
  \BibitemOpen
  \bibfield  {author} {\bibinfo {author} {\bibfnamefont {P.~G.}\ \bibnamefont
  {Casazza}} \emph {et~al.},\ }\href@noop {} {\bibfield  {journal} {\bibinfo
  {journal} {Taiwanese Journal of Mathematics}\ }\textbf {\bibinfo {volume}
  {4}},\ \bibinfo {pages} {129} (\bibinfo {year} {2000})}\BibitemShut {NoStop}%
\end{thebibliography}%
    
\end{document}